\documentclass[iop,apj]{emulateapj}
\slugcomment{{\sc Received} 2017 November 1; {\sc Revised} 2018 March 19; {\sc Accepted to ApJ} 2018 March 19; {\sc Published in ApJ} 2018 April 24} 
\shorttitle{Stellar Populations of COSMOS BCGs since $\MakeLowercase{z}$ $\sim$ 1.0}
\shortauthors{Cooke \& Fogarty et al. }

\usepackage{natbib}
\usepackage{graphicx} 
\usepackage{textcomp}
\usepackage{gensymb}
\usepackage{epsfig}%

\usepackage{verbatim}
\usepackage[outdir=./]{epstopdf}
\usepackage{indentfirst}
\usepackage{hyperref}
\usepackage{amssymb}
\usepackage{amsmath}
\usepackage{amsfonts}%
\usepackage{fancyhdr}

\begin{document}

\title{\textbf{Stellar Mass and 3.4\micron \;M/L Ratio Evolution of Brightest Cluster Galaxies in COSMOS since $\MakeLowercase{z}$ $\sim$ 1.0}}

\author{Kevin C. Cooke$^{1, \dagger}$, Kevin Fogarty$^{2, 5, \dagger}$, Jeyhan S. Kartaltepe$^{1}$,John Moustakas$^{3}$,Christopher P. O'Dea$^{4}$,Marc Postman$^{5}$}
\affil{$^1$School of Physics and Astronomy, Rochester Institute of Technology, Rochester, NY 14623, USA\\
$^2$Department of Physics and Astronomy, Johns Hopkins University, 3400 North Charles Street, Baltimore, MD 21218, USA\\
$^3$Department of Physics and Astronomy, Siena College, 515 Loudon Road, Loudonville, NY 12211, USA\\
$^4$Department of Physics and Astronomy, University of Manitoba, Winnipeg, MB R3T 2N2, Canada\\
$^5$Space Telescope Science Institute, 3700 San Martin Drive, Baltimore, MD 21208, USA}
\email{$^1$ kcc7952@rit.edu, $^2$ kfogarty@stsci.edu\\
$\dagger$ First two authors are considered co-first-authors}
\begin{abstract}
We investigate the evolution of star formation rates (SFRs), stellar masses, and M/L$_{3.4 \micron}$ ratios of brightest cluster galaxies (BCGs) in the COSMOS survey since $ z \sim 1$ to determine the contribution of star formation to the growth-rate of BCG stellar mass over time.   Through the spectral energy distribution (SED) fitting of the \emph{GALEX}, CFHT, Subaru, \emph{Vista}, \emph{Spitzer}, and \emph{Herschel} photometric data available in the COSMOS2015 catalog, we estimate the stellar mass and SFR of each BCG.  We use a modified version of the {\tt iSEDfit} package to fit the SEDs of our sample with both stellar and dust emission models, as well as constrain the impact of star formation history assumptions on our results. We find that in our sample of COSMOS BCGs, star formation evolves similarly to that in BCGs in samples of more massive galaxy clusters.  However, compared to the latter, the magnitude of star formation in our sample is lower by $\sim$ 1 dex. Additionally, we find an evolution of BCG baryonic mass-to-light ratio ($M/L_{3.4 \mu m}$) with redshift which is consistent with a passively aging stellar population. We use this to build upon \citeauthor{Wen:2013aa}'s low-redshift $\nu L_{3.4 \micron}-M_{Stellar}$ relation, quantifying a correlation between $\nu L_{3.4 \micron}$ and M$_{Stellar}$ to z $\sim$ 1. By comparing our results to BCGs in Sunyaev--Zel'dovich and X-ray-selected samples of galaxy clusters, we find evidence that the normalization of star formation evolution in a cluster sample is driven by the mass range of the sample and may be biased upwards by cool cores.
\end{abstract}


\section{\textbf{Introduction}}\label{sec:intro}

Dominating the luminosity and stellar mass in the central region of galaxy clusters, and influencing their evolution, are brightest cluster galaxies (BCGs).  These massive ellipticals occupy a narrower distribution of position-velocity space in relation to other cluster members \citep{Lauer:2014aa}, indicating a relaxed state within the parent cluster.  However, they do not represent the same population \citep{Von-Der-Linden:2007aa} as other ellipticals in the cluster or the field.  One unique characteristic of BCGs is their extended light profiles \citep[e.g.][]{Oemler:1976aa, Graham:1996aa}, indicating a rich merger history \citep[e.g][]{Bernardi:2007aa, Liu:2008aa}.  Additionally, BCGs exhibit larger sizes and luminosities than predicted from cluster luminosity functions \citep[e.g.][]{Loh:2006aa, Shen:2014aa}.

Recent observations of star-forming BCGs has led to a revision of their formation scenario over the past decade.  The original model derived from theoretical predictions \citep[e.g.][]{Merritt:1984aa} and observations \citep[e.g.][]{Stott:2010aa, Stott:2011aa} postulates a BCG formation mechanism in which the original gas and stellar content of a BCG formed in the initial matter density peaks \citep{Treu:2005aa}. The rest of its constituent stars form via in situ processes rapidly before $z = 1.5$, which evolve passively to the present day.  This is contrasted by semi-analytical models in which BCGs grow in stellar mass by a factor of three from $z = 1$ to the present day \citep{De-Lucia:2007aa}.  The newest models exhibit a factor of two growth over the same redshift range \citep[e.g.][]{Shankar:2015aa}, in which BCG stellar mass is accumulated through many minor mergers \citep[e.g.][]{Naab:2009aa,Edwards:2012aa}, which has led to greater agreement between observations and models \citep{Lidman:2012aa,Lin:2013aa}.  Whether this merger-driven era dominates mass growth is still under investigation, as minor mergers are not 100\% efficient in the delivery of their gas supply or stars \citep{Liu:2009aa,Burke:2013aa,Liu:2015aa}. 
 
One important problem is to constrain the role of star formation in BCG growth, since this process may contribute significant stellar mass to BCGs in the case of wet mergers, or in the case of cool-core clusters, if condensed intracluster medium (ICM) gas fuels a significant buildup of stellar mass \citep{ODea:2008aa}. While the mass growth of BCGs at $z < 1$ is believed to be substantial (with estimates placing stellar mass growth in BCGs doubling or tripling), there is considerable disagreement over how much of that mass growth is ongoing at $z \lesssim 0.5$  \citep[e.g.][]{McIntosh:2008aa, Tonini:2012aa, Bai:2014aa, Oliva-altamirano:2014aa, Inagaki:2015aa, Bellstedt:2016aa}, and whether new formed stars are contributing substantially to that growth \citep{Gozaliasl:2016aa}. Selection effects due to different selection criteria (e.g. whether the sample is optically or X-ray limited) may play a role in driving this disparity of results\citep{Burke:2000aa}, as well as different observed wavelengths, or assumptions such as selection of initial mass function (IMF) or evolution models to constrain stellar population. 
 
In this paper, we seek to constrain the contribution of BCG stellar mass growth made by in situ star formation in a sample of X-ray-selected, low-mass (M$_{500}$ = $\sim 10^{13}-10^{14}$ M$_{solar}$) galaxy clusters in the COSMOS field. We compare our findings with BCGs in higher-mass cluster samples obtained using both X-ray and Sunyaev--Zel'dovich (SZ) effect selection criteria to better understand the impact of cluster environment and sample selection on the estimation of the evolution of BCG star formation. By modeling BCG spectral energy distributions (SEDs) using far-UV (FUV) to far-IR (FIR) observations, we can better constrain the old and young stellar populations in a self-consistent manner that considers stellar populations obscured by IR-emitting dust. We also can consider several stellar population models, and constrain the impact of these models on our results.  By estimating specific star formation rates (sSFRs) in BCGs over a wide redshift range, we better constrain when star formation took place and whether or not star formation in different cluster samples evolves similarly. 
 
 In \textsection \ref{sec:sample} we discuss target selection criteria and sample completeness.  We review the archival data used in the SED fitting procedure in \textsection \ref{sec:inst} and discuss the data reduction and SED fitting software in \textsection \ref{sec:analysis}.  We discuss the results in \textsection \ref{sec:disc}. Finally, we summarize our results and future prospects in \textsection \ref{sec:conc}. We use the $\Lambda$CDM standard cosmological parameters of  $H_{0}$ = 70 Mpc$^{-1}$ km s$^{-1}$, $\Omega_M$ = 0.3, and $\Omega_{vac}$ = 0.7.

\section{\textbf{Sample Selection}}\label{sec:sample}

\begin{figure}[h]
\begin{center}
\begin{tabular}{c}
\includegraphics[width=0.45\textwidth]{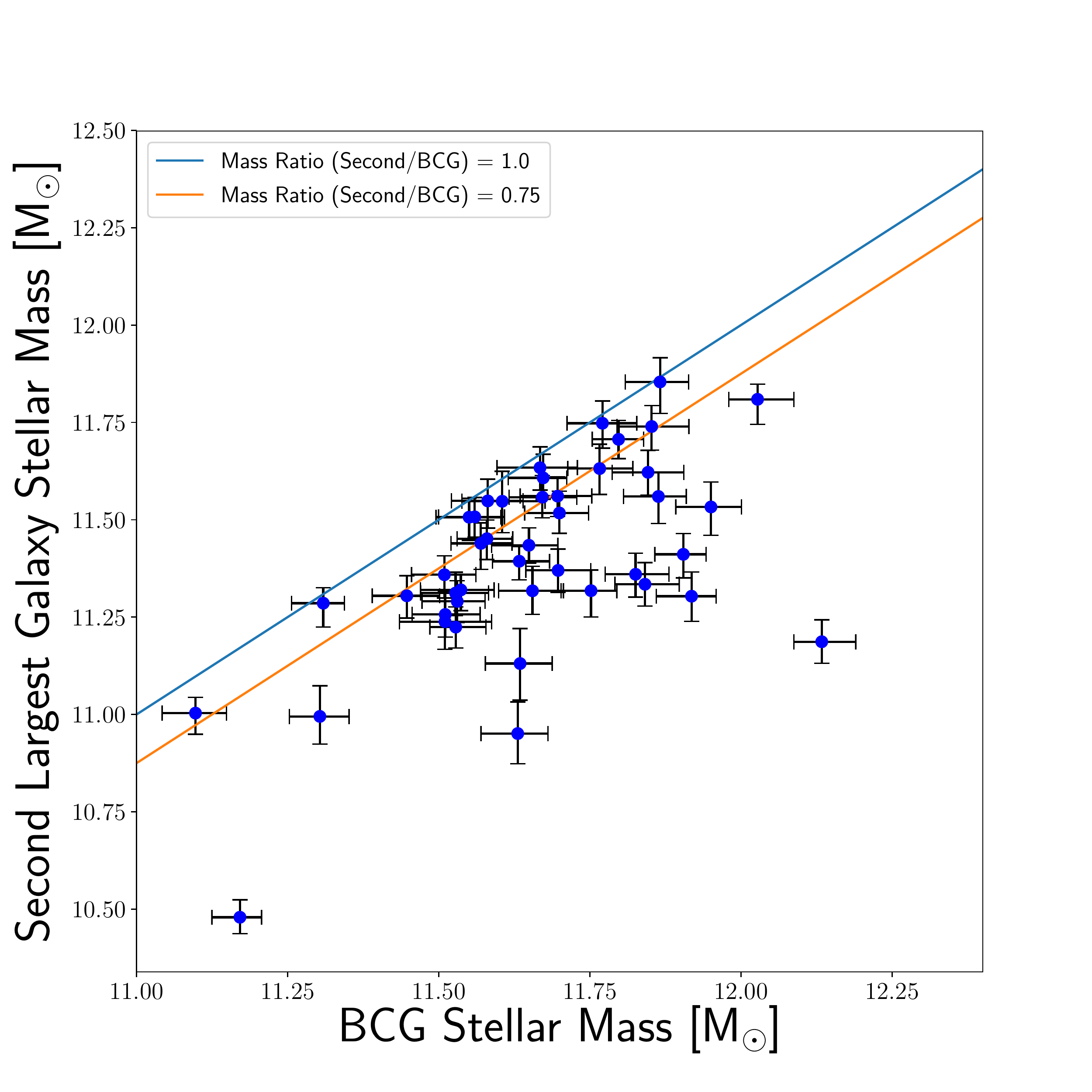}
\end{tabular}
\end{center}
\caption
{ \label{fig:BCGDetection} Ratio of BCG stellar mass to second most massive member of the same galaxy group.  Many targets are within errors with their next most massive neighbor, and the difference between them becomes indistinguishable by 10$^{11}$ M$_{\odot}$.}
\end{figure}

In order to build our sample from uniform observations and selection method, we select clusters in the COSMOS X-ray group and cluster catalog \citep{Finoguenov:2007aa,George:2011aa} between $0 < z < 1$, with greater than 30 spectroscopically confirmed or high probability ($\mathcal{P}_{MEM}$ $>$ 0.5) cluster members.  We choose a cutoff of 30 members to select well-detected clusters as have been previously studied in COSMOS \citep{Delaye:2014aa}, and are more populated and massive than the known protoclusters in COSMOS which have 5-10 members \citep[e.g.][]{Diener:2015aa}. These redshift and angular size limits correspond to a contiguous comoving volume of 7 $\times$ 10$^6$ Mpc$^3$.  To identify the BCG within each cluster, we estimate the stellar mass of all group members with $\mathcal{P}_{MEM}$ $>$ 0.5 using the SED fitting procedure described in Section \ref{sec:analysis} and select the group member with the highest estimated stellar mass.  The mass difference between BCG and second most massive group galaxy is shown in Fig. \ref{fig:BCGDetection}.  Throughout the paper, we also plot a comparison sample of massive group members with estimated stellar masses greater than 75\% of the stellar mass of the BCG identified for that group.

This yields an initial sample of 44 BCGs with no selection for star formation rate (SFR) activity. Their parent clusters span the range of log(m$_{200c}$/M$_{\odot}$) $\sim$12.9-14.3, as determined by the cluster X-ray luminosity \citep{Finoguenov:2007aa}, and calibrated via gravitational lensing measurement of the COSMOS field \citep{Leauthaud:2010aa,George:2011aa}.  We note that COSMOS is biased toward a denser region of sky than a random patch \citep{Finoguenov:2007aa}, with two large overdensities: at z $\sim$ 0.3 \citep{Masters:2011aa} and the `COSMOS Wall' at z $\sim$ 0.73 \citep{Scoville:2013aa,Iovino:2016aa}. We address the potential impact of these overdensities in Section \ref{sec:results} by running our analysis with objects in range of these overdensities in redshift space excluded.

COSMOS groups and clusters were originally detected by identifying extended sources on the scale of 32''; and 64'' using the wavelet analysis of \citet{Vikhlinin:1998aa}, in the Chandra and 1.4 Ms $XMM-Newton$ observations of the COSMOS field.  Once identified, \citet{George:2011aa} ran a red-sequence finding algorithm to identify an optical counterpart within a projected distance of 500 kpc of the X-ray center.  A galaxy is included as a member galaxy using a Bayesian analysis of galaxy properties such as location with respect to cluster or group center, redshift error, and relative number of field and cluster or group galaxies \citep[see Section 4 of][]{George:2011aa}.

Our BCG sample requires no completion correction as the COSMOS X-ray group member catalog is complete down to  K$_{s}$ = 24 mag and F184W = 24.2 mag (corresponding to an approximate stellar mass of M$_{*}$ $\geqslant$ 10$^{10.3}$ M$_{\odot}$ at z = 1; \citealt{George:2011aa}). However, group mass is limited to $M_{200} > 10^{13}$ out to $z \sim 1$, which may bias our group selection to higher masses at high redshift \citep{George:2011aa}.  Stellar mass estimates used in the COSMOS X-ray group catalog are estimated via SED fitting of galaxies with 3$\sigma$ detections in the K$_s$ band, using \citet{Bruzual:2003aa} stellar population models with a \citet{Chabrier:2003aa} IMF.  Two targets, COSMOS CLJ100028.3+024103 and COSMOS CL J095824.0+024916, had updated photometric redshifts above z = 1 in the more recent COSMOS2015 catalog \citep{Laigle:2016aa}.

\section{\textbf{COSMOS Data}}\label{sec:inst}
To accurately derive stellar masses and SFRs for each BCG, we use FUV--FIR observations of the COSMOS field \citep{Scoville:2007aa}.  COSMOS is a multi-wavelength survey that observed a 2 sq. deg. field centered at R.A.(J2000) = 10:00:28.600, decl.(J2000) = +02:12:21.00 from the X-ray to the radio, with publicly available multi-wavelength data available in the COSMOS2015 catalog \citep{Laigle:2016aa}.  A short review of each observation set is below; for further details of the data reduction in COSMOS2015, see \citet{Laigle:2016aa}.  The following photometry has been corrected for photometric and systematic offsets as detailed in Eq. 9 in \citet{Laigle:2016aa}.  For observations from \emph{GALEX} FUV-\emph{Spitzer} IRAC4, we also correct for Milky Way foreground extinction using a Galactic reddening of $R_v$ = 3.1 \citep{Morrissey:2007aa} and $E(\bv)$ values from the \citet{Schlegel:1998aa} dust maps. 

For our SED fitting, we use errors which include the observational error reported in the COSMOS2015 catalog as well as the absolute calibration uncertainty unique to each telescope. For GALEX NUV and FLUX flux errors, we include a $\pm$ 10$\%$ uncertainty\footnote{\url{https://asd.gsfc.nasa.gov/archive/galex/FAQ/counts\_background.html}}.  Subaru photometric calibration is accurate to within 0.02 mag \citep{Taniguchi:2015aa}, therefore we include a systematic error of 2\%.  For the Canada--France--Hawaii Telescope (CFHT) u*, we err on the side of caution and use the 5\% error from worst-quality data of the original $CFHT$ observations of the COSMOS field \citep{Capak:2007aa}.  According to the $Vista$/VIRCAM User Manual,\footnote{Doc. No. VIS-MAN-ESO-06000-0002} $Vista$ photometry is accurate to within 3-5\% and so we include a 5\% error in addition to the observational error for data from $Vista$/VIRCAM filters.   $Spitzer$ absolute flux calibration is considered accurate to within 3\% \citep{Van-Dyk:2013aa}, so we add an additional 3\% to our final $Spitzer$ IRAC errors.  Most $Herschel$ data used here are upper limits defined by the sensitivity of the original $Herschel$ surveys of the COSMOS fields, the PACS Evolutionary Probe \citep[PEP;][]{Lutz:2011aa} and the Herschel Multi-tiered Extragalactic Survey \citep[HerMES;][]{Oliver:2012aa}. For the handful of $Herschel$ detections in our sample, we adopt the 10\% systematic uncertainty term used in \cite{Fogarty:2017aa}.

		\subsection{GALEX}
		To constrain the degree of unobscured star formation, we use \emph{GALEX} \citep{Martin:2005aa} FUV and NUV band point-spread function (PSF)-fit photometric magnitudes presented in the COSMOS2015 catalog \citep[for details see ][]{Zamojski:2007aa}.   \emph{GALEX} observes a 1\fdg2 circular field of view through a 50 cm diameter Richey--Chretien telescope.  To measure FUV and NUV magnitudes, \citet{Zamojski:2007aa} use a PSF-fitting routine using $u^*$ band observations as a prior to minimize blending effects due to GALEX's FWHM of 5\arcsec.   
 	
		\subsection{Canada--France--Hawaii Telescope (CFHT)}
		We use COSMOS2015 CFHT/MegaPrime \citep{Aune:2003aa,Boulade:2003aa} $u^{*}$ magnitudes to constrain the blue and NUV rest-frame emission.  The COSMOS field was observed in queue mode with a consistent PSF across all observations, to a depth of m$_{u^{*}}$ $\sim$ 26.4 and seeing of 0.9\arcsec.  For further details, see \citet{Capak:2007aa}.
		
		Additionally, we use K$_{s}$ magnitudes from CFHT/WIRCam \citep{Puget:2004aa} taken during the COSMOS-WIRCam Near-Infrared Imaging Survey \citep{McCracken:2010aa}, down to a 3$\sigma$ detection limit of $m_{K_s}$=23.4 and FWHM of 1\farcs1 or less.  
		
		\subsection{Subaru}
		To constrain the optical continuum of each BCG, we retrieve Subaru/Suprime-Cam optical magnitudes from the COSMOS2015 catalog in five broad filters ($B, V, R$, $i+$, $z++$) and 11 medium filters (IA427, IA464, IA484, IA505, IA527, IA574, IA624, IA679, IA738, IA767, and IA827).  The Subaru observations with the worst resolution are from the IA464 filter with a PSF FWHM of 1\farcs89, which is still sufficient enough to resolve the BCGs in our sample.  All observations reach a $3\sigma$ depth of m$_{AB}\sim25.2$ or deeper.  For further details, see \citet{Taniguchi:2007aa,Taniguchi:2015aa}.  
		
		\subsection{Vista}
		NIR observations are an important constraint for the old stellar population which dominates the emission and stellar mass of BCGs.  We retrieve $Y, J,$ and $H$-band $Vista$ \citep{Dalton:2006aa,Emerson:2006aa} observations taken with VIRCAM \citep{Sutherland:2015aa} during the UltraVISTA-DR2 survey \citep{McCracken:2012aa}.  For UltraVISTA, the COSMOS field was observed with $Y, J,$ and $H$ filters down to limiting magnitude $m_{AB}$ of 25.3, 24.9, and 24.6 respectively with a median FWHM of 0\farcs6.

		\subsection{Spitzer}
		Additional observations of the old stellar population are available through NIR to MIR observations taken by the \emph{Spitzer Space Telescope} \citep{Werner:2004aa}.  We include archival data from \textit{Spitzer}'s Infrared Array Camera (IRAC)  3.6, 4.5, 5.7, and 7.9 \micron\; channels \citep[For more information, see][]{Fazio:2004aa}.  IRAC observes 5\farcm2 x 5\farcm2 degree fields with PSF widths 1\farcs6, 1\farcs6, 1\farcs8, and 1\farcs9  for bands IRAC1 to IRAC4 respectively.  IRAC magnitudes in COSMOS2015 are measured from observations taken as part of the SPLASH \citep{Steinhardt:2014aa} and S-COSMOS surveys \citep{Sanders:2007aa} to a 3$\sigma$ depth of $m_{AB}$ of 25.5, 25.5, 23.0, and 22.9 for IRAC1-4 respectively.  We also include Multiband Imaging Photometer \citep[MIPS, ][]{Rieke:2004aa} 24\micron\; fluxes originally presented in \citet{Le-Floch:2009aa} to a 3$\sigma$ depth of 80 $\mu$Jy.  To account for blending, photometry from $J,H,$ and $K$ observations were used as a prior during the source extraction of the 3.6 \micron\; image \citep{Laigle:2016aa}.  Each successive IRAC filter used the adjacent shorter wavelength image as a prior out to 24 $\micron$.
		
		\subsection{Herschel}
		The FIR is an important regime for observing the re-radiated energy from dust surrounding obscured star forming regions.  We use the \emph{Herschel Space Observatory} \citep{Pilbratt:2010aa} Photoconductor Array Camera and Spectrometer (PACS) \citep{Poglitsch:2010aa} in the green (100 \micron) and red (160 \micron) bands as well as the Spectral and Photometric Imaging Receiver (SPIRE) 250 \micron\; and 350 \micron\; bands.  While the large beam size at 250 and 350 \micron\; (18\farcs1, 24\farcs9 respectively) guarantees blending, all but three targets at 350\micron\; and all but six at 250\micron\; are non-detections and the fluxes that are detected have low signal-to-noise ratio (S/N) and therefore modest constraining power on the SED fits.  PACS observations were taken as part of PEP \citep{Lutz:2011aa} to a 3$\sigma$ depth of 5 and 10.2 mJy for 100 and 160 \micron\; bands respectively.  SPIRE observations originate from HerMES \citep{Oliver:2012aa} and reach a 3$\sigma$ depth of 8.1 and 10.7 mJy at 250 and 350$\micron$ respectively.  The MIPS 24 $\micron$ image of the COSMOS field was used as a prior during source extraction from the FIR images \citep{Laigle:2016aa}.
		
\section{\textbf{Methods}}\label{sec:analysis}
\subsection{SED Construction}
SEDs were composed of photometry taken from \emph{GALEX}, Subaru, \emph{Vista}, \emph{Spitzer}, and \emph{Herschel} in the COSMOS2015 public catalog,
so as to maximize coverage of UV-through-IR flux. 
Of these, 34 had \emph{GALEX} NUV
and/or FUV detections,
43 had full \emph{Spitzer}/IRAC detections with the 44th (COSMOS CLJ100013.0+023519) detected in only IRAC1-3, 18 were detected in MIPS 24 \micron, and 9 with \emph{Herschel} PACS or SPIRE detections. Therefore, every BCG in the final
catalog has rest-frame $U$-band through MIR detections,
with limiting magnitudes out to 350 \micron.

We note that our study differentiates itself from that of \citet{Gozaliasl:2016aa,Gozaliasl:2018aa}, who study a sample of 407 X-ray-selected groups, including those detected in the COSMOS field in \citet{George:2011aa} and \citet{Finoguenov:2007aa}, in terms of how we estimate stellar properties of cluster galaxies and in terms of our treatment of galaxies comparably massive to the BCG. These studies cite results from either \cite{Ilbert:2013aa} or \cite{Laigle:2016aa}, which do not incorporate FIR photometry into estimates of the SFR. Our fits take into account both the observed UV flux of the young stellar population and the flux absorbed by dust and re-emitted in the FIR, and therefore measure the total SFR of each system. Furthermore, by selecting only those COSMOS clusters with at least 30 members, and by taking into account comparably massive galaxies in clusters which lack a clearly dominant galaxy, we limit ourselves to rich systems that are comparable to higher-mass cluster analogs and account for the effects of potentially ambiguous BCG selection.

\subsection{SED Fitting}\label{sec:sedfitting}
SEDs were fit using a modified version of {\tt iSEDfit} \citep{Moustakas:2013aa,Moustakas:2017aa}. {\tt iSEDfit} is a Bayesian SED fitting tool that uses a grid of synthetic SEDs generated using a set of input priors to estimate the posterior probability distribution of parameters of the stellar population emitting an observed SED. We used the modified version of this tool described in \citet{Fogarty:2017aa} in order to take into account both the stellar and dust emission observed in the NUV-through-IR SEDs of our sample.

A detailed description of {\tt iSEDfit} is available in Appendix A of \citet{Moustakas:2013aa}. {\tt iSEDfit} takes a synthetic stellar population, IMF, and dust attenuation law, and creates a grid of synthetic SEDs that randomly sample the parameter
space of metallicity, A$_{V}$, emission line ratios, and the parameters governing the star formation history (SFH) using user-defined prior distributions. For each BCG, we assume a \citet{Salpeter:1955aa} IMF. In order to determine the extent to which our results depend on the parameterization of the SFH, we tested two SFH parameterizations, a one-component model and a two-component model. The former consists of an exponentially decaying SFH with a decay constant between 0.6 and 60 Gyr sampled logarithmically. The latter consists of an exponentially decaying SFH with a decay constant between 0.3 and 1.5 Gyr, and for half of the models in our grid we incorporated an exponentially decaying starburst at present time. For either model, the age of the BCG was allowed to vary between 6 and 9 Gyr if the BCG was at $z \leq 0.45$, or between 4 and 6 Gyr if the BCG was at $z > 0.45$. These age priors were chosen to ensure that model stellar populations do not exceed the age of the universe at the redshift of the galaxy, while still ensuring an old stellar component. This enables us to fit SEDs of both quiescent (nominally `red and dead') and star-forming BCGs. Our choices of stellar population model and parameter space are given in Table \ref{table:SED_Params}.

\begin{deluxetable*}{lrrr}
\tabletypesize{\footnotesize}
\tablecaption{SED Fitting Parameters}


\tablehead{\textit{Stellar Population Model} \label{table:SED_Params} \\
\hline
Synthetic Stellar Population & \citet{Bruzual:2003aa} \\
Initial Mass Function & \citet{Salpeter:1955aa}\\
Attenuation Law & \citet{Calzetti:2000aa} \\
Dust Emission & \citet{Draine:2007aa} \\
\hline
\hline
\textit{Model Parameter Space Constraints} \\
\hline
Parameter Name & Minimum Value & Maximum Value & Sampling Interval }
\hline
\hline
\startdata
\textit{One-component SFH} \\
Age, $t$ & 4 Gyr (6 Gyr)$^{a}$ & 6 Gyr (9 Gyr) & Linear \\
Decay Timescale, $\tau$ & 0.6 Gyr & 60.0 Gyr & Logarithmic \\
\hline
\textit{Two-component SFH} \\
Age, $t$ & 4 Gyr (6 Gyr) & 6 Gyr (9 Gyr) & Linear \\
Decay Timescale, $\tau$ & 0.3 Gyr & 1.5 Gyr & Linear \\
Burst Age, $t_{b}$ & $10^{-2}$ Gyr & 5.0 Gyr & Logarithmic$^{b}$ \\
Burst Decay Percentage & 0.01 & 0.99 & Linear \\
Burst Mass Percentage & 0.0015 & 0.85 & Logarithmic \\
\hline
Metallicity & 0.03 Z$_{\odot}$ & $1.5 Z_{\odot}$ & Linear \\
\textit{Dust Parameters} \\
Attenuation A$_{V}$ & 0 & 2 & Linear \\
PAH Abundance Index $q_{PAH}$ & 0.10 & 4.58 & Linear$^{c}$ \\
$\gamma^{d}$ & 0.0 & 1.0 & Linear \\
$U_{min}^{d}$ & 0.10 & 25.0 & Logarithmic \\
$U_{max}^{d}$ & $10^{3}$ & $10^{7}$ & Logarithmic
\enddata
\tablecomments{
$^{a}$ Minimum and maximum values for ages inside parentheses apply to BCGs at $z \leq 0.45$, while those outside the parentheses apply to BCGs at $z > 0.45$. 
$^{b}$ Burst parameters were sampled logarithmically, since their qualitative effect on the model SED of the galaxy occurs on order-of-magnitude scales. The exception to this is the burst decay percentage, which is one minus the amplitude of current star formation activity relative to the amplitude of the burst $t_{b}$ years ago. \\
$^{c}$ \citet{Draine:2007aa} model parameter sampling intervals were chosen based on the model parameter distributions of the template spectra. \\ 
$^{d}$ The \citet{Draine:2007aa} treats dust in a galaxy as consisting of two components. The first component consists of a fraction $\gamma$ of the dust is exposed to a power-law distribution of starlight intensity, ranging from $U_{min}$ to $U_{maxs}$, while the second component consists of the remainder of the dust, and is only exposed to a starlight intensity $U_{min}$. The quantities $U_{min}$ and $U_{max}$ are unitless measures of intensity relative to the ambient local radiation field. \\ 
 }
 
\end{deluxetable*}

The relative likelihood that each model SED is the observed SED is determined by calculating $e^{-\chi^2}$, where $\chi^2$ is the reduced chi square value for the model compared to the data. By randomly sampling the model grid, weighted by the relative likelihood, {\tt iSEDfit} recovers the posterior probability distribution of the model SEDs, and therefore the probability distribution of the physical stellar parameter input \citep{Moustakas:2013aa}.

Dust emission is incorporated into the synthetic SED grid, allowing us to take full advantage of the MIR and FIR data available from \emph{Spitzer}, and \emph{Herschel}. SED fits incorporating observations of the IR dust emission are preferable to those that do not since they reduce the degeneracy between A$_V$ and the SFR in fits to dusty star-forming systems. Following the prescription in \citet{Fogarty:2017aa}, we used the dust emission templates in \citet{Draine:2007aa}. The choice dust model parameter space is given in Table \ref{table:SED_Params}. The dust emission component of each synthetic SED was normalized such that the total energy re-emitted by the dust equals the total energy absorbed via attenuation of the synthetic stellar spectrum.

Our choice of dust model is the \citet{Calzetti:2000aa} attenuation law for dusty starburst galaxies. As the galaxies we study are typically dust poor, the choice of dust model has a limited impact on the SED fit. Furthermore, as was shown in \citet{Fogarty:2017aa}, even vigorously star-forming BCGs have relatively modest A$_{V}$ values, and the choice of attenuation law does not significantly affect the outcome of UV--FIR SED fitting.

Model grids consisting of $4\times 10^{4}$ models were constructed for both the one-component and two-component SFH parameterizations. These model grids were shifts to the observer frame for each galaxy studied in order to produce synthetic photometry for each fit. Parameters were sampled either linearly or logarithmically in the intervals listed in Table \ref{table:SED_Params}. 

\section{\textbf{Results}}\label{sec:results}
\subsection{BCG Stellar Population Evolution}
\begin{figure*}
\begin{center}
\epsfig{file=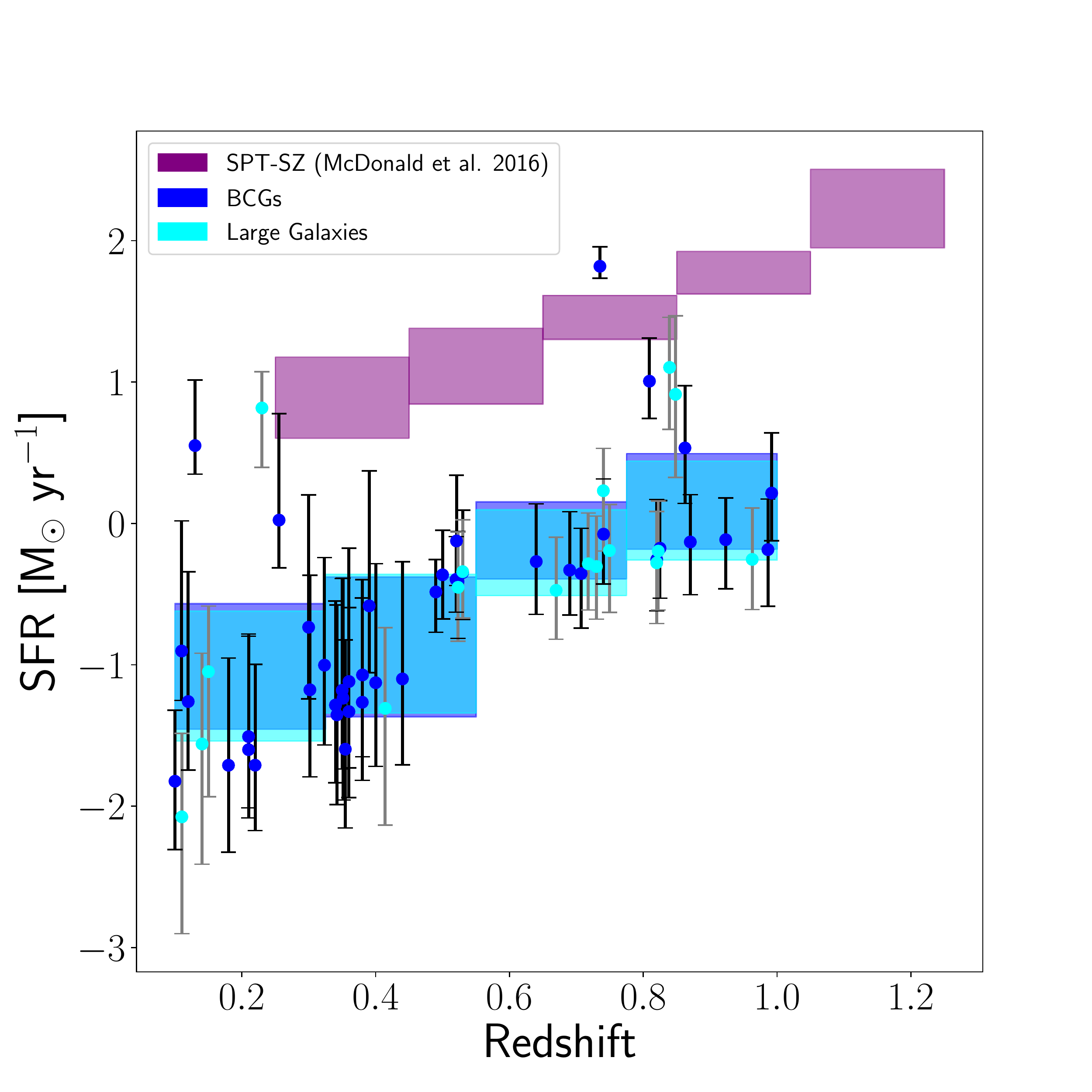,width=0.4\textwidth,angle=0}
\epsfig{file=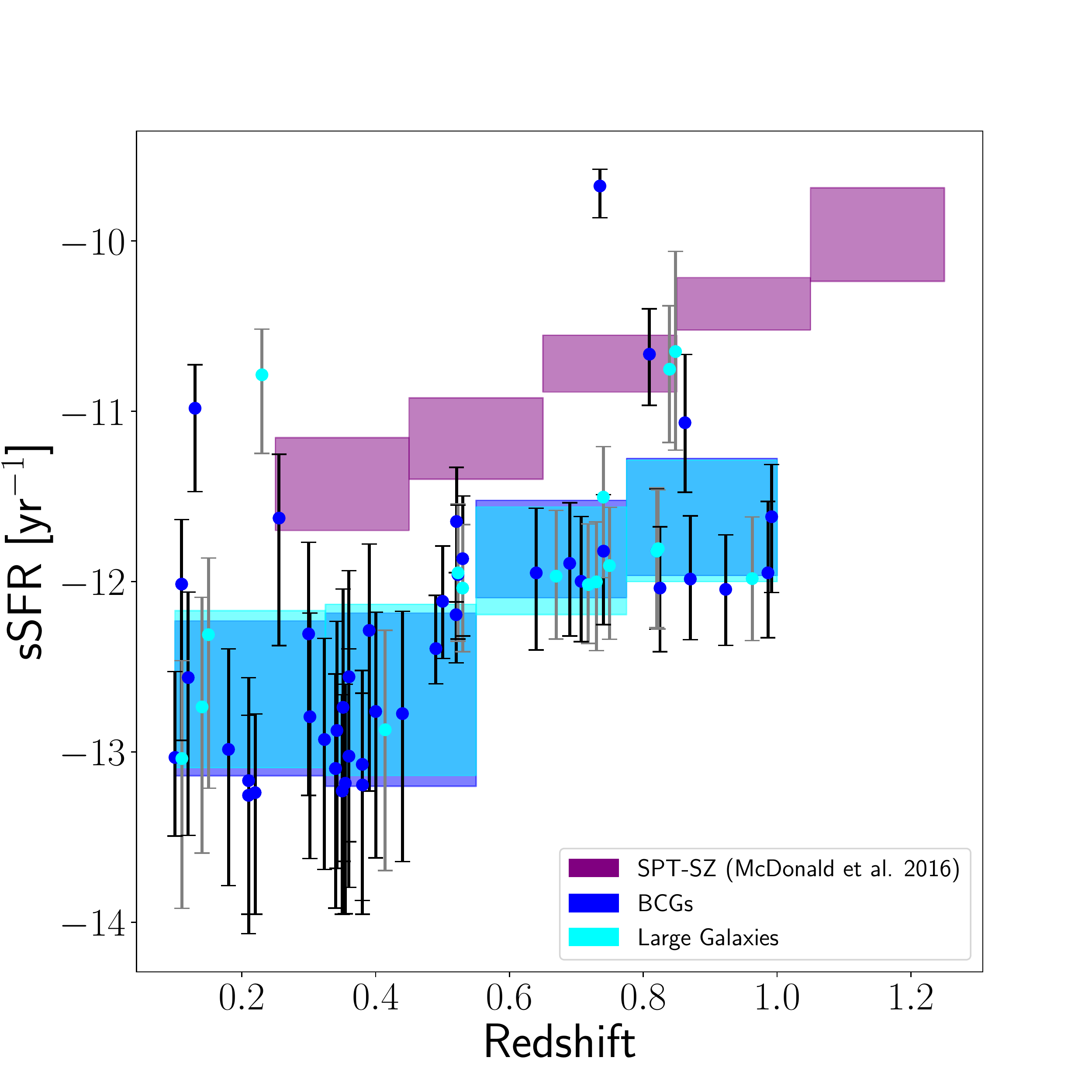,width=0.4\textwidth,angle=0}
\end{center}

\caption{\label{fig:SFR_Evolution} \emph{Left}: SFR versus redshift for BCGs (dark blue) and large group member galaxies with M$_*$/M$_{BCG}$ $>$ 0.75 (light blue). Each SFR is defined to be the median of the marginal posterior probability distribution of the SFR; the SFR weakly increases with redshift, within errors.  The error bars for SFRs denote the 68.3\% credible intervals, corresponding to 1$\sigma$ uncertainties for approximately Gaussian probability distributions. The shaded boxes show the $\chi^{2}$ weighted mean SFRs in each of four redshift bins of our sample, alongside the SZ-selected sample from \citet{McDonald:2016aa} for comparison. The horizontal width of each region depicts the width of each bin in redshift, and the vertical height depicts the 1$\sigma$ credible interval for the mean.
\emph{Right}: sSFR versus redshift for all BCGs. The definition of the best fit, color scheme, and redshift bins are the same as those in the left figure.  The trend in sSFR with redshift is shallower than for SFR and also shallower than the results from \citet{McDonald:2016aa}.}
\end{figure*}

We present the BCG SFR and sSFR as a function of redshift in Figure \ref{fig:SFR_Evolution} . Unless otherwise specified, we report results obtained with the single-component SFH parameterization throughout. SFRs and sSFRs for the individual BCGs are presented in the Appendix, along with best-fit SEDs. We demonstrate SFR is weakly dependent on redshift, within our errors, as is the sSFR. We compare our results to those obtained in \citet{McDonald:2016aa}, who estimated mean SFRs and sSFRs in redshift bins for BCGs in an SZ-selected sample of clusters with masses between $\log_{10} \frac{\textrm{M}_{500}}{\textrm{M}_{\odot}} \sim 14.5-15.2$. We also defined four redshifts bins, $0.15\leq z <  0.325$, $0.325< z\leq 0.55$, $0.55< z \leq 0.775$ and $0.775 < z$, and calculated the $\chi^{2}$-weighted mean SFRs and sSFRs for each bin. These are reported in Figure \ref{fig:SFR_Evolution} with color-coded shaded regions with vertical limits depicting the 1$\sigma$ credible interval for the mean.  For all plots we include the BCG from each group as well as any massive (M$_*$/M$_{BCG}$ $>$ 0.75) group members.  We notice no discernible difference in the trends for SFR, sSFR, or M$_*$ between these two subsamples. 

\begin{deluxetable*}{lcccc}
\tabletypesize{\footnotesize}
\tablecaption{Redshift Binned SFR and sSFR \label{table:ZBins}}


\tablehead{Parameter                            & $0.15\leq z < 0.325$ & $0.325\leq z < 0.55$ & $0.55\leq z < 0.775$ & $0.775 \leq z$ }\\
\hline
\hline
\startdata
log$_{10}$ SFR \textit{(One-component SFH)} [M$_{\odot}$ yr$^{-1}$] & $-1.1^{+0.6}_{-0.3}$ & $-0.9^{+0.6}_{-0.4}$ & $-0.1^{+0.3}_{-0.3}$ & $0.2^{+0.4}_{-0.3}$ \\
log$_{10}$ SFR \textit{(One-component SFH, All Large Gals.)}$^{a}$ & $-1.1^{+0.5}_{-0.4}$ & $-0.9^{+0.6}_{-0.4}$ & $-0.2^{+0.3}_{-0.3}$ & $0.1^{+0.4}_{-0.4}$ \\
log$_{10}$ SFR \textit{(One-component SFH, Structure Excised)}$^{b}$ & $-1.1^{+0.6}_{-0.3}$ & $-0.6^{+0.5}_{-0.4}$ & $-0.3^{+0.3}_{-0.3}$ & $0.2^{+0.4}_{-0.3}$ \\
log$_{10}$ SFR \textit{(Two-component SFH)} & $-1.6^{+1.0}_{-0.7}$ & $-1.5^{+1.2}_{-0.8}$ & $-0.6^{+1.1}_{-0.5}$ & $-0.2^{+1.2}_{-0.5}$  \\
log$_{10}$ sSFR \textit{(One-component SFH)} [yr$^{-1}$] & $-12.6^{+0.4}_{-0.6}$ & $-12.6^{+0.4}_{-0.6}$ & $-11.8^{+0.3}_{-0.3}$ & $-11.6^{+0.3}_{-0.4}$ \\
log$_{10}$ sSFR \textit{(One-component SFH), All Large Gals.)} & $-12.5^{+0.4}_{-0.6}$ & $-12.6^{+0.4}_{-0.6}$ & $-11.9^{+0.3}_{-0.3}$ & $-11.6^{+0.3}_{-0.4}$ \\
log$_{10}$ sSFR \textit{(One-component SFH, Structure Excised)} & $-12.6^{+0.4}_{-0.6}$ & $-12.3^{+0.4}_{-0.5}$ & $-11.9^{+0.3}_{-0.3}$ & $-11.6^{+0.3}_{-0.4}$ \\
log$_{10}$ sSFR \textit{(Two-component SFH)} & $-13.0^{+0.7}_{-1.1}$ & $-13.2^{+0.8}_{-1.2}$ & $-12.3^{+0.5}_{-1.1}$ & $-11.9^{+0.5}_{-1.2}$ \\
\enddata 
\tablecomments{\\
$^{a}$ Includes both BCGs and galaxies within $0.75\times$ the stellar mass of their respective BCG.\\
$^{b}$ Binned results obtained when excluding BCGs in the overdense structures at $z \sim 0.35$ and $z \sim 0.73$}
  
\end{deluxetable*}

Redshift binned results are reported in Table \ref{table:ZBins}. SFR declines by approximately one order of magnitude from z $\sim$ 1 to the present day. Across all redshift bins, we find a typical BCG SFR of $\sim$0.1--1 M$_{\odot}$ yr$^{-1}$. The SFR trends are similar whether we consider the one-component or two-component SFH, although SFRs and sSFRs are about 0.5 dex lower when measured using the two-component SFH (given the uncertainties, however; this difference is marginally significant).

Finally, we considered the impact of excising clusters in the overdense redshift regions discussed in Section \ref{sec:sample}. These results are given in Table \ref{table:ZBins} as well. The overdensities at z$\sim 0.35$ and $z \sim 0.73$ extend in redshift space between $0.325 \lesssim z \lesssim 0.38$ and $0.69 \lesssim z \lesssim 0.79$ respectively \citep{Masters:2011aa, Iovino:2016aa}, resulting in our excluding 12 clusters from bins 1-3. The overall shift in our results is well within our uncertainties.

\begin{figure}
\begin{center}
\begin{tabular}{c}
\includegraphics[height=7cm]{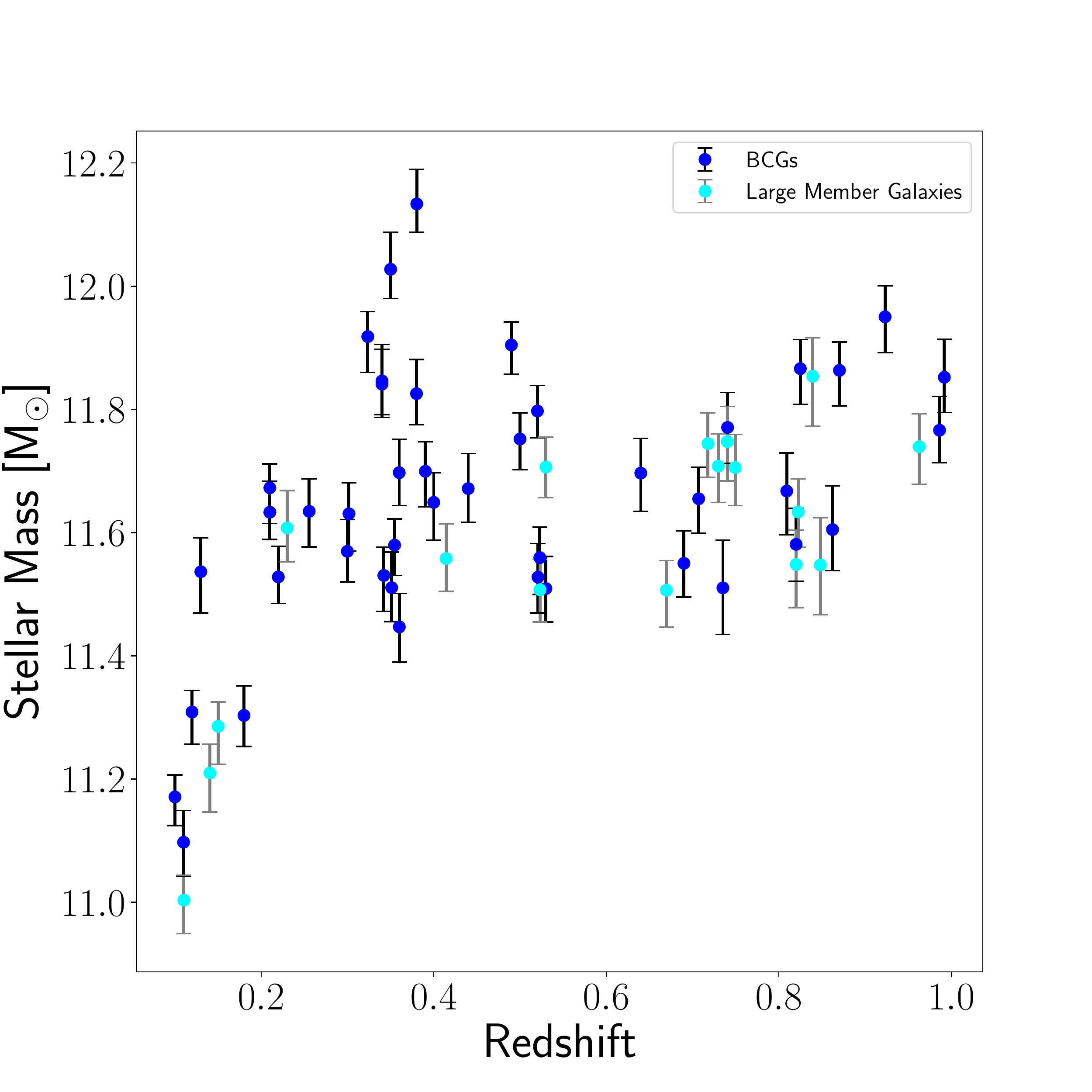}
\end{tabular}
\end{center}
\caption
{ \label{fig:SED_Masses} Stellar mass versus redshift for BCGs (dark blue) and large group member galaxies with M$_*$/M$_{BCG}$ $>$ 0.75 (light blue). The median estimation process of each M$_*$ is the same as for the SFRs in Figure \ref{fig:SFR_Evolution}.}
\end{figure}

The mean sSFR decreases by about 1 dex across the redshift range in our study. In Figure \ref{fig:SFR_Evolution} we show that, as redshift increases to 1, the typical sSFR for a BCG in our sample grows from $\sim 0.03\%$ per Gyr to $\sim 0.3\%$ per Gyr. Since the catalog we use probes masses uniformly across the range of redshifts we study, our results in Figure \ref{fig:SFR_Evolution} suggest some growth in the weighted mean sSFR may occur at $z \sim 0.6$, which would be consistent with \cite{McDonald:2016aa} even after accounting for the order of magnitude offset between our sample and their SZ-selected sample. We find that across our redshift range, the sample we study probes the mass range 11.2 M$_{\odot}$ - 12.5 M$_{\odot}$ relatively uniformly, as seen in Figure \ref{fig:SED_Masses}.

\begin{figure}
\begin{center}
\begin{tabular}{c}
\includegraphics[height=9cm, angle=0]{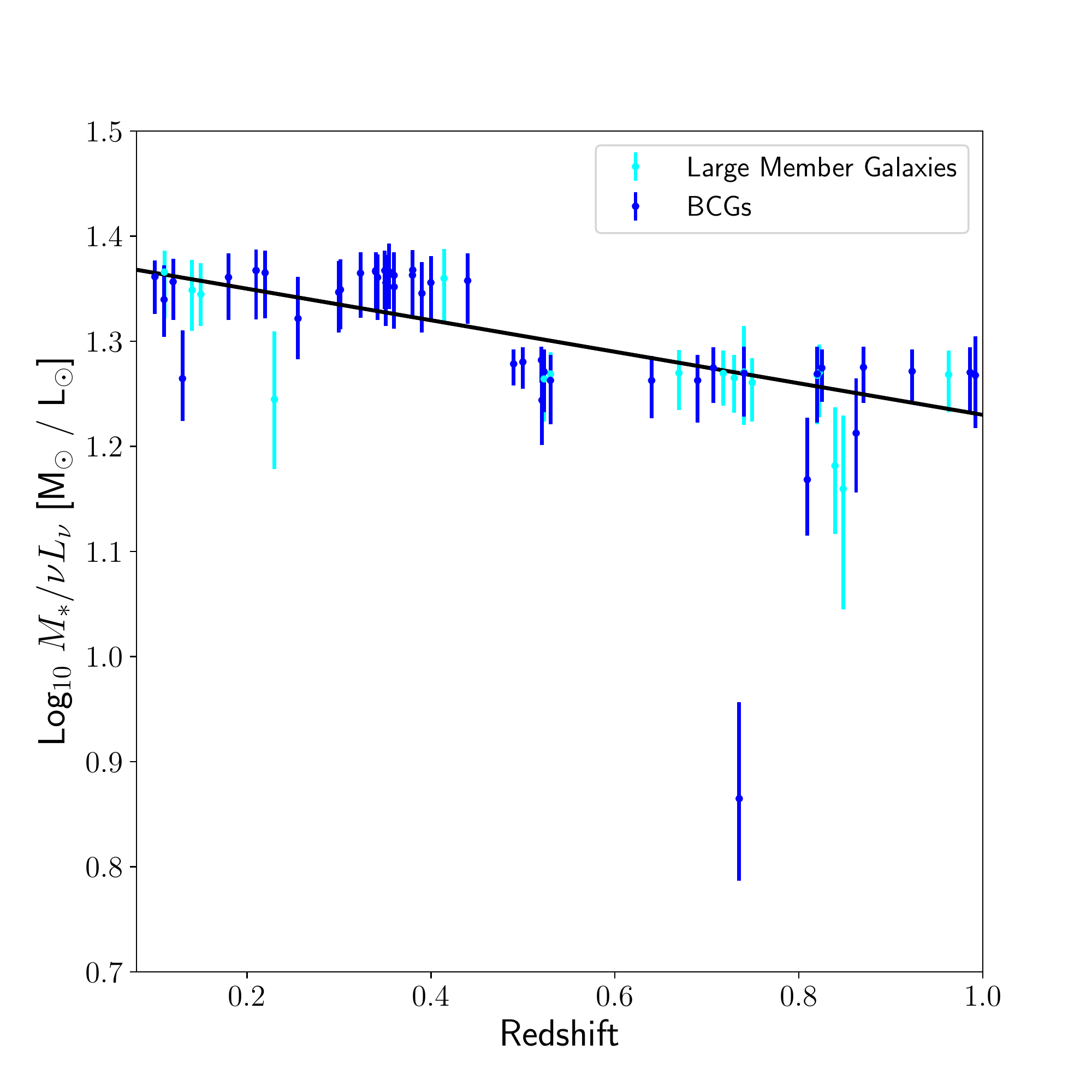}
\end{tabular}
\end{center}
\caption
{ \label{fig:Mass_Bias} $M_{*}$/$\nu L_{3.4 \micron}$ values are plotted against redshift to assess the effect of passive evolution on the BCG sample, in which the population ages and becomes redder over time for both BCGs (blue) and massive group members with M$_*$/M$_{BCG}$ $>$ 0.75 (light blue).  Luminosities are K-corrected and corrected for distance modulus. The solid line is a linear fit to the BCG points from \cite{2010Hogg_Methods}.} 
\end{figure}

Finally, we examined the redshift evolution of the BCG mass-to-light ratio. These results are presented in Figure \ref{fig:Mass_Bias}.  Studies such as \citet{Fraser-McKelvie:2014ab} and \citet{Wen:2013aa} use the $M_{*}$--$WISE$ W1 luminosity relationship to estimate stellar masses of BCGs at $z < 0.1$ and $z <$ 0.35 respectively.   However, further estimation of BCG stellar masses at higher redshifts requires estimates of the evolution of $M_{*}/L_{3.4\micron}$ with redshift.  Therefore, we computed $WISE$ W1 rest frame luminosities by estimating the best-fit model W1 photometry in {\tt iSEDfit}.  This band is sensitive to the population of old stars which compose the majority of the BCG's stellar mass, in addition to being less sensitive to recent star formation and dust than other bands. We fit a linear model to the $M_{*}/L_{3.4\mu m}$ ratio versus redshift using the least-squares method in \cite{2010Hogg_Methods}. The resulting redshift evolution of the stellar mass (M$_{*}$) to light ($\nu$L$_{\nu}$ (3.4$\micron$)) ratio is:

\begin{equation}\label{eq:w1}
\begin{aligned}
log_{10} \Bigg( \frac{M_{Stellar} (M_{\odot})}{\nu L(3.4 \micron)_\nu (L_{\odot})} \Bigg) = 1.38^{+0.01}_{-0.01} \\ 
-0.15^{+0.01}_{-0.02} \times z. 
\end{aligned}
\end{equation}
Across this range of redshift space, the $M_*/L_{3.4\micron}$ ratio changes by a factor of $\sim$ 1.3.

\citet{Wen:2013aa} find $log_{10} M_*/\nu L_{3.4\micron}$ ratios of $\sim$1.5-2 for some massive early-type galaxies with $M_{*} > 11.5 M_{\odot}$, consistent with an extrapolation to low redshift from Eq. \ref{eq:w1} (Fig. \ref{fig:Mass_Bias}).  The nearest contaminant in the NIR which could effect our result is the 3.3 $\micron$\; polycyclic aromatic hydrocarbon (PAH) emission feature, which will not be detected since the rest-frame W1 central wavelengths span 1.8--2.5 \micron\; across the redshifts observed in our sample.

\subsection{An Active Galactic Nucleus Outlier in COSMOS CL J100035.2+020346}
The fit with the highest $\chi^2$ corresponds to COSMOS CL J100035.2+020346's BCG (R.A. = 150.148794; decl. = 2.060569), with a $\chi^{2}$ of 7.6.  Upon examination of its SED (see Appendix), its positive NIR slope indicates the possible presence of an active galactic nucleus (AGN) component.  We test this hypothesis by comparing the IRAC band fluxes of this target with the criteria set by \citet{Donley:2012aa} as well as the obscured AGN models of \citet{Lacy:2004aa,Lacy:2007aa}.  This BCG does not satisfy the \citet{Donley:2012aa} criterion of an IR AGN; however the ratio of IRAC band flux ratios do match well with the model from \citet{Polletta:2008aa} for an elliptical galaxy hosting an obscured AGN contributing 0-20$\%$ of the total galaxy NIR output.  No radio counterpart is detected in the COSMOS catalogs, nor in the soft X-ray Chandra observations.  However, this target is detected in the hard X-ray band (2--7 keV) by \citet{Elvis:2009aa} at a S/N of 4.49, and a corresponding L$_X(0.5-7 \textrm{keV})$ of 6.29 $\times$ 10$^{42}$ ergs s$^{-1}$.  We believe this target is an active X-ray AGN with minimal dust content surrounding the black hole.

\section{\textbf{Discussion}}\label{sec:disc}

\subsection{Contribution of star formation to BCG mass growth}\label{sec:comparison}
Our typical sSFR values are $<$ 10$^{-11.0} \; yr^{-1}$ across all redshift bins, indicative of a doubling time of $>10^{11}$ yr, and thus a stellar growth rate due to star formation on the order of $<$ 1\% Gyr$^{-1}$ across the redshift range studied. After accounting for stellar mass loss and recycling, this rate is corrected to $<$ 0.4\% Gyr$^{-1}$ \citep[e.g.][]{Kennicutt:1994aa,Brinchmann:2004aa}. 

Our results are closest to the lower bound of \citet{McIntosh:2008aa}'s 1.4-6.4\%/Gyr growth rate and consistent with \citet{Oliva-altamirano:2014aa}'s lack of significant change at lower redshift ($0.09 < z < 0.27$). However, our results indicate an order of magnitude less growth than from \citet{Bai:2014aa} and  \citet{Inagaki:2015aa}, who used red sequence and X-ray luminosity selected BCGs.  \citet{Inagaki:2015aa} fit SZ-effect selected BCGs using only SDSS $ugriz$ magnitudes via {\tt kcorrect} and {\tt NewHyperZ}, using \citet{Bruzual:2003aa} stellar population models and a Chabrier IMF.  Our result is consistent with their lower limit of 2\%, but not their upper limit of 14\%.  They also noted that {\tt NewHyperZ} yielded higher masses than their {\tt kcorrect} models and that their selection of early-type galaxy models may have influenced the result.  \citet{Bai:2014aa} used the GALFIT luminosity of their targets combined with the M/L ratio given by the \citet{Maraston:2009aa} luminous red galaxy models.  This difference in stellar and M/L ratio assumptions may be contributors to our different results.

We also find that our sample is significantly more quiescent than the sample of groups presented in \cite{Gozaliasl:2018aa}, since they find that the mean SFR in their sample can account for as much as 45\% of the growth rate of brightest group galaxies. We suspect that this result is driven by star-forming BGGs in either the lower-mass or galaxy-poorer halos in their sample, which is consistent with the median SFR in their sample being consistent with our mean across redshift space. Taken together, these results suggest distinct evolutionary histories between BCGs and BGGs that can be distinguished in halos with total masses of M$_{200}$ $\lesssim 10^{14}$ M$_{\odot}$. Alternatively, the degeneracy between SFR and  $A_{V}$ in SEDs without MIR and FIR constraints may result in the overestimation of the SFR in a minority of cases.

This work's mean sSFR is higher than that of 9.42 $\times 10^{-12} \: yr^{-1}$ in \citet{Cooke:2016aa}, who investigated the Sloan Giant Arcs Survey (SGAS) and Cluster Lensing and Supernova Survey with $Hubble$ (CLASH) samples.  This is expected, as their estimators only measured unobscured rates while our inclusion of IR upper limits constrains the star formation obscured by dust as well as the un-obscured component.  

As noted above, this is an order of magnitude less growth than in \citet{McDonald:2016aa}, which diverges further from our results at higher redshift as wet merger-driven stellar mass growth drives higher SFRs.  Using detections from the SPT, \citet{McDonald:2016aa}'s sample probes the most massive BCGs across the southern hemisphere, each more massive than our BCG sample.  We believe the discrepancy in stellar mass growth rate is due to their mass selection requiring a more aggressive merger-driven mass growth in their past in order to reach their observed masses, an aggressive growth rate not required by our lower mass sample.

\subsubsection{Comparison with X-Ray and SZ-selected Cluster Samples}

\begin{figure}
\begin{center}
\begin{tabular}{c}
\includegraphics[height=9cm]{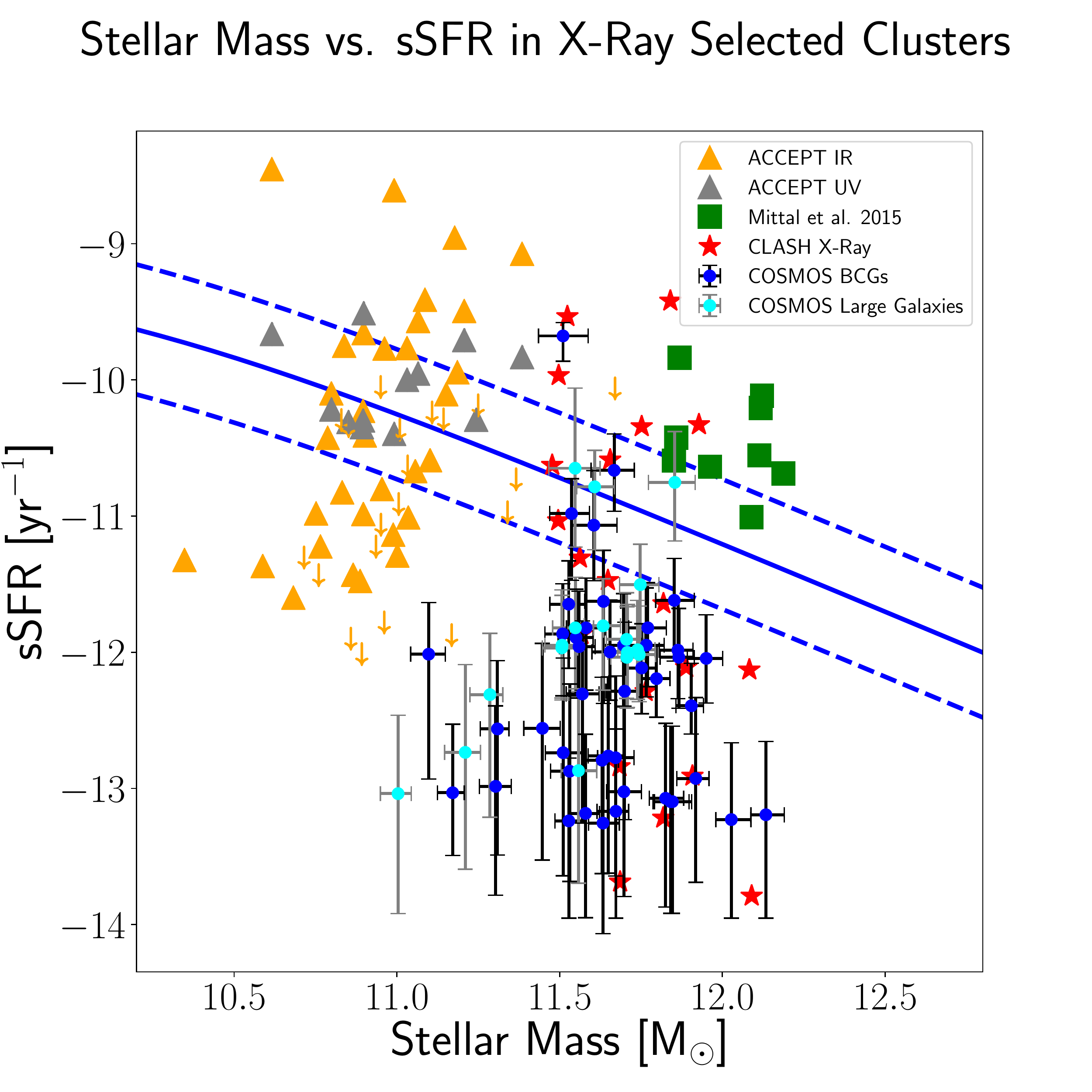}
\end{tabular}
\end{center}
\caption
{ \label{fig:SFMS} sSFR vs. stellar mass for COSMOS BCGs (dark blue) and massive group members with M$_*$/M$_{BCG}$ $>$ 0.75 (light blue), with \cite{2011Hoffer_SFREntropy}, \cite{Mittal:2015aa}, and CLASH X-ray-selected clusters shown for comparison. The blue line represents the star-forming main sequence as found by \citet{Lee:2015aa} at the mean redshift of all BCGs plotted. The blue dashed lines trace the region that is within a factor of 3 of the main-sequence trend; galaxies above the dashed region are active star formers, while those below the dashed region are quiescent. Our sample is predominantly quiescent with around five outliers still on or above the main sequence. Dark blue points correspond to BCGs in the COSMOS sample, light blue to COSMOS group members with M$_*$/M$_{BCG}$ $>$ 0.75, yellow triangles to BCGs with 70 $\mu$m detections in the ACCEPT catalog, with yellow downward arrows corresponding to upper limits, gray triangles depict ACCEPT BCGs with UV observations, red stars show BCGs in the CLASH X-ray-selected sample, and green boxes show the \citet{Mittal:2015aa} BCGs. }
\end{figure}

We use available archival data from three other X-ray-selected BCG studies in order to provide a larger sample to test whether BCG star formation is typical of other massive galaxies: the ACCEPT survey \citep{2009Cavagnolo_ACCEPT}, the BCGs studied by \cite{Mittal:2015aa},  and the CLASH \citep{Postman:2012aa}. Of these samples, the CLASH BCGs occupy a stellar mass and redshift range most like the present sample, having masses above 10$^{11} M_{\odot}$ and a redshift range of $0.2 < z < 0.7$. However, the CLASH survey selected massive ($kT > 5$ keV), morphologically symmetrical (ellipticity $\leq 0.3$) clusters, and so contains more massive clusters than COSMOS and a large fraction ($> 50\%$) of cool cores \citep{Postman:2012aa}. Meanwhile, the ACCEPT clusters overlap with both the COSMOS and the CLASH cluster masses. Stellar mass and SFRs were calibrated to a common Salpeter IMF for comparison.

The ACCEPT survey selected 239 X-ray clusters in the temperature range $T_{x} \sim 1-20$ keV and the bolometric luminosity range $L_{bol} \sim 10^{42-46}$ erg s$^{-1}$, spanning redshifts $0.05-0.89$ \citep{2009Cavagnolo_ACCEPT}. SFRs and stellar masses of ACCEPT cluster BCGs were measured in \cite{2011Hoffer_SFREntropy}. We include sSFR values for BCGs with SFRs estimated using 70 $\micron$ \textit{Spitzer} MIPS observations or NUV \textit{GALEX} observations from \cite{2011Hoffer_SFREntropy}. 

  The CLASH survey \citep{Postman:2012aa} observed 25 galaxy clusters, of which 20 were selected by X-ray temperature and morphology.  Five more strongly lensing clusters with Einstein radii $>$ 35$\arcsec$; were also included.  \citet{Fogarty:2015aa} and \citet{Donahue:2015aa} independently investigate the star formation characteristics of the CLASH sample. 

	As seen in Fig. \ref{fig:SFMS}, the sSFR of BCGs decreases as a function of stellar mass for the overall comparison sample. The behavior of actively star-forming BCGs is consistent on average with the star-forming main sequence at the mean redshift of all of the samples\citep{Noeske:2007ab,Lee:2015aa}. This behavior is not evident in the individual samples we plot, however, since each sample occupies a relatively narrow range of stellar masses. Furthermore, while the ACCEPT and CLASH samples overlap the main-sequence star forming range, our COSMOS sample is systematically more quiescent than what would be predicted by the star-forming main sequence, while the \cite{Mittal:2015aa} sample is systematically more active. 
 
We suspect individual BCGs have evolved off the star-forming main sequence, so the apparent trend between stellar mass and sSFR evident when comparing the BCGs from these different X-ray-selected samples is consistent with BCG star formation being driven by processes in the halo external to the BCG. One noteworthy aspect of Fig. \ref{fig:SFMS} is that the BCGs in the X-ray samples appear to straddle the star forming main sequence, which suggests a link between the halo-driven fueling process in BCGs and field massive ellipticals. It is also likely that whether or not the mean star formation behavior of a sample of BCGs is consistent with the star forming main sequence depends on how the sample selects for cool cores, as would be implied by the X-ray luminosity dependence on cool-core and star-forming BCG selection observed by \citet{Green:2016aa}. The morphology selection of CLASH increases the incidence of cool-core clusters in the sample, thereby increasing the incidence of cooling-induced BCG star formation, which likely contributes to why the COSMOS sample is systematically more quiescent than the CLASH sample despite occupying the same stellar mass bin.

\begin{figure}
\begin{center}
\begin{tabular}{c}
\includegraphics[height=9cm]{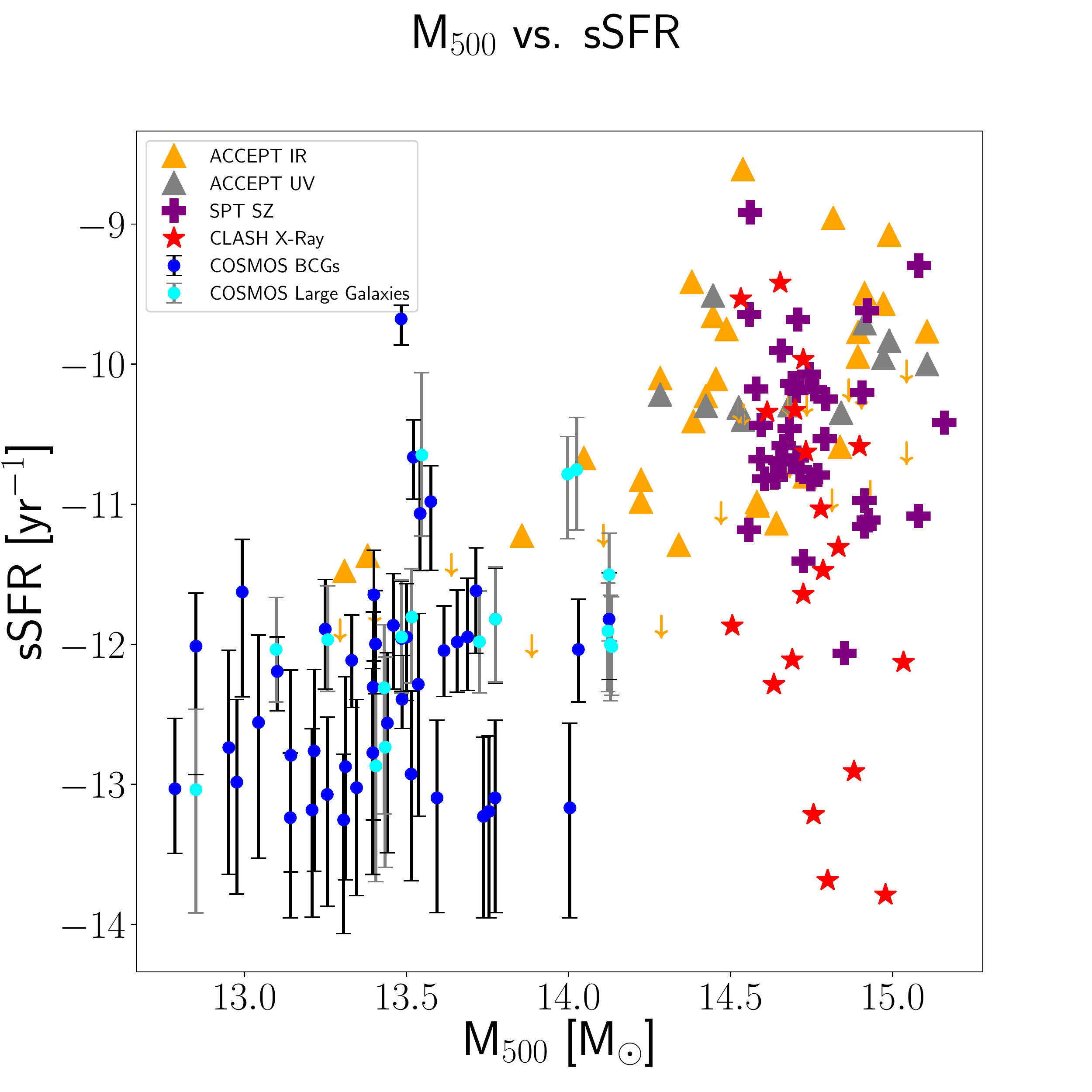}
\end{tabular}
\end{center}
\caption
{ \label{fig:SSFR_Mass}  sSFR vs. cluster M$_{500}$. Purple crosses correspond to BCGs in the \cite{McDonald:2016aa} SPT SZ-selected clusters; otherwise the colors and symbols used are identical to Figure \ref{fig:SFMS}. Cluster masses from the \cite{Mittal:2015aa} sources were not listed in the paper. Our sample populates the same low sSFR and M$_{500}$ parameter space as the low M$_{500}$ clusters from \cite{2011Hoffer_SFREntropy}, indicating a possible correlation between cluster M$_{500}$ and BCG sSFR.}
\end{figure}

The tendency of COSMOS BCGs toward quiescence may be a function of the cluster's mass. Figure \ref{fig:SSFR_Mass} displays sSFR vs. cluster M$_{500}$ for the COSMOS, ACCEPT, and CLASH samples, as well as the SZ-selected sample in \cite{McDonald:2016aa}. Cluster masses were estimated for COSMOS and ACCEPT by converting X-ray luminosities to $M_{500}$ using the scaling relation in \cite{2009Pratt_Scaling}. Masses for CLASH were estimated through a combination of strong and weak lensing \citep{Merten:2015aa}.

These results suggest BCG evolution may be affected by cluster mass, although it is possible that different effects might dominate at different redshifts. First, we consider the COSMOS and SPT samples. The COSMOS sample is both significantly more quiescent and lower mass than the SPT sample. The majority of the high-sSFR BCGs in the SPT sample occur at high redshift, while the sSFR characteristics of the COSMOS sample are still systematically lower at low redshift (log$_{10}$ sSFR $\sim -12.6$ yr$^{-1}$ as opposed to $-11.7$ to $-11.2$ yr$^{-1}$).  \citet{McDonald:2016aa} cite merger driven star-formation in young cluster environments at high redshift as driving the evolution of star formation in their sample.  The discrepancy between our findings and those of \cite{McDonald:2016aa} may be explained by the hypothesis that evolution observed in \cite{McDonald:2016aa} is merger driven at $z\gtrsim 0.6$. In the high redshift bins, more massive clusters will have undergone more cluster mergers, which at high redshift may serve as a gas delivery mechanism to drive star formation. As a result, a high-cluster mass sample like the SPT sample would have BCGs with higher sSFRs than a low-cluster mass sample like COSMOS at high redshifts. 

The CLASH and ACCEPT samples, meanwhile, occupy the full range of cluster masses and BCG sSFRs bracketed by the COSMOS and SPT samples. These X-ray-selected samples are lower redshift ($\langle z\rangle = 0.39$ for CLASH and $0.14$ for the ACCEPT clusters used in this paper, as opposed to $0.75$ for the SPT sample), and therefore we expect star formation to be induced by cooling. Taken together, they show that the BCG sSFR depends on mass in X-ray-selected clusters at low to moderate redshifts in the sense that star formation can be more vigorous as cluster mass increases, but need not be (while the ACCEPT BCGs show a trend between sSFR and M$_{500}$, we suspect the bolometric luminosity-derived masses in ACCEPT may be biased by cool cores, and implying this trend may actually be reflective of the correlation between BCG star formation and the presence of cool cores). A plausible physical explanation at lower redshifts is that the larger sSFRs of BCGs in more massive clusters have the potential to be supported by proportionally larger reservoirs if these BCGs are in the cores of cool-core clusters.

\subsection{Evolution of $M/L_{3.4 \micron}$ with Redshift}\label{sec:mlratio}

Previous studies at low redshift \citep[e.g.][]{Wen:2013aa,Fraser-McKelvie:2014ab} found a correlation between stellar mass and rest-frame W1 luminosity density at $z < 0.1$.  This correlation will evolve with redshift as  the W1 band observations include emission from a younger and brighter stellar population.   Therefore, we measured the evolution of the $M_{*}/L_{3.4 \micron}$ ratio across a redshift range between 0.2 and 1.0 using our COSMOS dataset.  Our estimated $M_{*}/L_{3.4 \micron}$ ratio of $10^{1.38}$ at $z = 0$ is consistent with the high-mass end of the $M_{*}/L_{3.4 \micron}$ relationship in \citet{Wen:2013aa}. For a 10$^{12} M_{\odot}$ galaxy, \citet{Wen:2013aa} predict a ratio of 10$^{1.29}$.  Since the masses used in their results were estimated from colors calibrated assuming a \citet{Chabrier:2003aa} IMF, it matches our results given the $\sim$0.2 dex offset between the \citet{Chabrier:2003aa} and \citet{Salpeter:1955aa} IMFs.

The slope in our overall relationship is consistent with the passive evolution of a stellar population becoming redder over time, following a \citet{Bruzual:2003aa} evolutionary track from $\sim$ 3 to $\sim$ 10 Gyr old.  Our measurement of the $M_{*}/L_{3.4 \micron}$ ratio is consistent with the body of results supporting dry-merger-driven stellar mass growth in BCGs \citep[e.g.,][]{De-Lucia:2007aa,Whiley:2008aa,Vulcani:2016aa}.  In particular, our results from the COSMOS sample imply that the addition of new mass to a BCG does not change its mass-to-light ratio, so the stellar population of cannibalized galaxies must be a similar age to the existing population in the galaxy.  Since observations indicate that BCG masses grow by a factor of $\sim$ 2 between $z$ = 0.9 and 0.2, stars from early-type stellar populations must accrete onto the BCGs without triggering star formation \citep{Lidman:2012aa,Lidman:2013aa,Rodriguez-Gomez:2017aa}. Our findings also align with observations of the evolution of the $M_{*}/L_{B}$ ratio out to $z = 0.5$, which also imply passive BCG evolution \citep{van-der-Marel:2007aa}.

\section{\textbf{Conclusions}}\label{sec:conc}
We examined the role of star formation in the stellar mass growth of BCGs in low-mass clusters at intermediate redshifts by fitting the SEDs of 44 BCGs below $ z < 1$ in the COSMOS survey.  By using publicly available archival data from the FUV to FIR, we estimated SFR and M$_{*}$ across four redshift bins ($0.15 < z < 0.325$, $0.325 < z < 0.55$, $0.55 < z < 0.775$ and $0.775 < z$). By comparing our estimates with similar work examining more massive clusters in the literature, we conclude the following:

\begin{enumerate}
\item{BCG SFR weakly declines with redshift from $z \sim 1$ to the present day. We find evidence that the sSFRs of BCGs in low-mass clusters evolve at least down to $z\sim 0.5$, but that at all redshifts these galaxies are systematically more quiescent than their higher-mass cluster counterparts.}
\item{An evolution of the baryonic M/L$_{3.4 \mu m}$ ratio with redshift is observed and fit.  This redshift-dependent correlation provides an extension of that found by \citet{Wen:2013aa}, previously limited to $z < 0.1$.}
\item{Star formation plays very little role in BCG mass growth in the COSMOS sample. Our estimates for the contribution of star formation to BCG stellar mass at $z<1$ ($< 1\%$ per Gyr) is consistent with or below the low end of similar estimates in the literature.}
\end{enumerate}

While we find evidence for evolving SFRs in COSMOS BCGs, when compared to the massive SZ-limited sample of \cite{McDonald:2016aa}, our sample is systematically more quiescent across redshift bins. We suspect that the processes governing the evolution of star formation are the same in these homogeneously selected samples of clusters, but the magnitude of star formation is a selection effect. By comparing with both this sample and X-ray selected samples of clusters at lower redshifts (which have a greater tendency than either COSMOS or SZ-selected samples to select clusters that have formed cool cores), we are led to suspect that this effect is as function of how samples select cluster mass and ICM dynamical state.

\acknowledgements{
\linespread{1}

We thank the referee for their constructive feedback which strengthened this study. Based on the COSMOS2015 catalog reduced and compiled by Clotilde Laigle using data taken as part of the COSMOS Collaboration.  Based on observations made with the NASA Galaxy Evolution Explorer. GALEX is operated for NASA by the California Institute of Technology under NASA contract NAS5-98034.  Based on observations made with the NASA/ESA Hubble Space Telescope, obtained from the Data Archive at the Space Telescope Science Institute, which is operated by the Association of Universities for Research in Astronomy, Inc., under NASA contract NAS 5-26555.
Based on observations obtained with Vista/VIRCAM.
Based on observations obtained with MegaPrime/MegaCam, a joint project of CFHT and CEA/DAPNIA, at the Canada-France-Hawaii Telescope (CFHT) which is operated by the National Research Council (NRC) of Canada, the Institut National des Sciences de l'Univers of the Centre National de la Recherche Scientifique of France, and the University of Hawaii. Based on observations obtained with WIRCam, a joint project of CFHT, Taiwan, Korea, Canada, France, and the Canada-France-Hawaii Telescope (CFHT).
Based (in part) on data collected at Subaru Telescope, which is operated by the National Astronomical Observatory of Japan.
This work is based (in part) on observations made with the Spitzer Space Telescope, which is operated by the Jet Propulsion Laboratory, California Institute of Technology under a contract with NASA.  Herschel is an ESA space observatory with science instruments provided by European-led Principal Investigator consortia and with important participation from NASA.  HCSS / HSpot / HIPE is a joint development (are joint developments) by the Herschel Science Ground Segment Consortium, consisting of ESA, the NASA Herschel Science Center, and the HIFI, PACS and SPIRE consortia.
}


\newcommand\invisiblesection[1]{%
  \refstepcounter{section}%
  \addcontentsline{toc}{section}{\protect\numberline{\thesection}#1}%
  \sectionmark{#1}}

\invisiblesection{Bibliography}
\bibliography{CookeFogarty2017bib}

\appendix

Fit parameters  for individual COSMOS BCGs are presented in Table \ref{table:BCG_Params}. We report the median SFR and sSFR and 1$\sigma$ uncertainties obtained for each galaxy. Values for $\chi^{2}$ are reported for the best-fitting model in the Monte Carlo grid for each BCG. Best-fit model spectra for each BCG are shown in Figure \ref{figure:Best_SEDs}.

\invisiblesection{BCG Best-Fit Parameters}\label{sec:BCGparams}
\LongTables
\begin{deluxetable*}{lccccc} 
\tabletypesize{\footnotesize}  

\tablecaption{BCG Best-fit Parameters}
 
  
\tablehead{
BCG & Redshift & $\chi^{2}$ & SFR$^{a}$ & sSFR$^{a,b}$  \\  
 & & & log$_{10}$(M$_{\odot}$ yr$^{-1})$ & yr$^{-1}$  } 
\label{table:BCG_Params} 
  
\startdata
COSMOS CL J100045.6+013926 & 0.21 & 1.98 & $-1.51^{+0.73}_{-0.58}$ & $-13.17^{+0.6}_{-0.78}$ \\ 
COSMOS CL J100201.2+021330 & 0.825 & 2.68 & $-0.18^{+0.34}_{-0.35}$ & $-12.04^{+0.36}_{-0.37}$ \\ 
COSMOS CL J100013.6+021230 & 0.18 & 1.83 & $-1.71^{+0.76}_{-0.61}$ & $-12.98^{+0.59}_{-0.8}$ \\ 
COSMOS CL J100005.7+021211 & 0.923 & 1.22 & $-0.12^{+0.3}_{-0.35}$ & $-12.04^{+0.32}_{-0.33}$ \\ 
COSMOS CL J100056.8+021225 & 0.36 & 1.1 & $-1.12^{+0.94}_{-0.61}$ & $-12.56^{+0.62}_{-0.97}$ \\ 
COSMOS CL J100109.1+021637 & 0.11 & 1.69 & $-0.9^{+0.92}_{-0.35}$ & $-12.01^{+0.38}_{-0.92}$ \\ 
COSMOS CL J100051.5+021648 & 0.862 & 1.2 & $0.53^{+0.44}_{-0.39}$ & $-11.07^{+0.4}_{-0.41}$ \\ 
COSMOS CL J100139.8+022548 & 0.13 & 2.68 & $0.55^{+0.46}_{-0.2}$ & $-10.98^{+0.26}_{-0.49}$ \\ 
COSMOS CL J095951.4+014049 & 0.38 & 2.66 & $-1.07^{+0.67}_{-0.58}$ & $-13.19^{+0.54}_{-0.76}$ \\ 
COSMOS CL J100013.9+022249 & 0.4 & 1.74 & $-1.13^{+0.84}_{-0.59}$ & $-12.76^{+0.58}_{-0.86}$ \\ 
COSMOS CL J095833.6+022056 & 0.992 & 0.93 & $0.21^{+0.43}_{-0.34}$ & $-11.62^{+0.31}_{-0.45}$ \\ 
COSMOS CL J095907.2+022358 & 0.351 & 1.41 & $-1.24^{+0.85}_{-0.72}$ & $-12.74^{+0.69}_{-0.9}$ \\ 
COSMOS CL J100027.4+022123 & 0.22 & 1.84 & $-1.71^{+0.71}_{-0.46}$ & $-13.24^{+0.46}_{-0.71}$ \\ 
COSMOS CL J100021.8+022328 & 0.21 & 2.95 & $-1.6^{+0.8}_{-0.41}$ & $-13.25^{+0.47}_{-0.81}$ \\ 
COSMOS CL J095847.9+022410 & 0.355 & 2.84 & $-1.6^{+0.77}_{-0.56}$ & $-13.18^{+0.58}_{-0.77}$ \\ 
COSMOS CL J095931.8+022654 & 0.36 & 2.25 & $-1.33^{+0.74}_{-0.61}$ & $-13.02^{+0.63}_{-0.77}$ \\ 
COSMOS CL J100016.0+023850 & 0.707 & 2.04 & $-0.35^{+0.32}_{-0.39}$ & $-12.0^{+0.38}_{-0.35}$ \\ 
COSMOS CL J095941.6+023129 & 0.741 & 1.14 & $-0.08^{+0.39}_{-0.35}$ & $-11.82^{+0.33}_{-0.43}$ \\ 
COSMOS CL J095940.6+023603 & 0.256 & 2.41 & $0.02^{+0.75}_{-0.34}$ & $-11.63^{+0.38}_{-0.75}$ \\ 
COSMOS CL J100056.0+022834 & 0.38 & 1.51 & $-1.26^{+0.74}_{-0.55}$ & $-13.07^{+0.55}_{-0.8}$ \\ 
COSMOS CL J100138.5+023514 & 0.1 & 1.94 & $-1.82^{+0.5}_{-0.48}$ & $-13.03^{+0.5}_{-0.46}$ \\ 
COSMOS CL J095957.1+023506 & 0.69 & 1.29 & $-0.33^{+0.41}_{-0.32}$ & $-11.89^{+0.36}_{-0.43}$ \\ 
COSMOS CL J100013.0+023519 & 0.64 & 2.85 & $-0.27^{+0.41}_{-0.37}$ & $-11.95^{+0.38}_{-0.45}$ \\ 
COSMOS CL J100028.3+024103 & 0.35 & 2.81 & $-1.18^{+0.79}_{-0.56}$ & $-13.23^{+0.56}_{-0.72}$ \\ 
COSMOS CL J100111.9+014037 & 0.523 & 0.91 & $-0.41^{+0.35}_{-0.4}$ & $-11.96^{+0.41}_{-0.38}$ \\ 
COSMOS CL J095924.4+014623 & 0.12 & 1.85 & $-1.26^{+0.92}_{-0.48}$ & $-12.56^{+0.5}_{-0.93}$ \\ 
COSMOS CL J095901.5+024740 & 0.49 & 3.21 & $-0.48^{+0.23}_{-0.28}$ & $-12.39^{+0.31}_{-0.21}$ \\ 
COSMOS CL J095824.0+024916 & 0.34 & 1.56 & $-1.29^{+0.73}_{-0.55}$ & $-13.1^{+0.55}_{-0.82}$ \\ 
COSMOS CL J100020.7+023153 & 0.87 & 1.22 & $-0.13^{+0.34}_{-0.37}$ & $-11.98^{+0.37}_{-0.36}$ \\ 
COSMOS CL J100027.0+023321 & 0.5 & 2.02 & $-0.36^{+0.32}_{-0.31}$ & $-12.12^{+0.32}_{-0.34}$ \\ 
COSMOS CL J100043.2+014607 & 0.34 & 1.48 & $-1.28^{+0.74}_{-0.55}$ & $-13.1^{+0.55}_{-0.82}$ \\ 
COSMOS CL J100049.6+014923 & 0.302 & 3.6 & $-1.18^{+0.81}_{-0.62}$ & $-12.79^{+0.61}_{-0.83}$ \\ 
COSMOS CL J100139.3+015051 & 0.39 & 1.8 & $-0.58^{+0.96}_{-0.47}$ & $-12.29^{+0.51}_{-0.94}$ \\ 
COSMOS CL J095805.4+015256 & 0.342 & 1.03 & $-1.35^{+0.78}_{-0.63}$ & $-12.87^{+0.64}_{-0.81}$ \\ 
COSMOS CL J100217.7+015601 & 0.52 & 2.68 & $-0.4^{+0.3}_{-0.23}$ & $-12.19^{+0.25}_{-0.28}$ \\ 
COSMOS CL J100128.6+015958 & 0.82 & 1.23 & $-0.26^{+0.43}_{-0.36}$ & $-11.82^{+0.37}_{-0.46}$ \\ 
COSMOS CL J100147.3+020314 & 0.53 & 1.89 & $-0.35^{+0.44}_{-0.33}$ & $-11.86^{+0.37}_{-0.45}$ \\ 
COSMOS CL J100035.2+020346 & 0.986 & 0.94 & $-0.19^{+0.36}_{-0.4}$ & $-11.95^{+0.42}_{-0.38}$ \\ 
COSMOS CL J100200.6+020405 & 0.521 & 1.82 & $-0.12^{+0.46}_{-0.32}$ & $-11.65^{+0.32}_{-0.47}$ \\ 
COSMOS CL J100141.0+015904 & 0.3 & 1.81 & $-0.73^{+0.94}_{-0.51}$ & $-12.31^{+0.54}_{-0.95}$ \\ 
COSMOS CL J100139.2+022435 & 0.809 & 1.0 & $1.01^{+0.31}_{-0.26}$ & $-10.66^{+0.27}_{-0.3}$ \\ 
COSMOS CL J095945.1+023622 & 0.324 & 3.09 & $-1.0^{+0.76}_{-0.56}$ & $-12.93^{+0.59}_{-0.76}$ \\ 
COSMOS CL J100031.5+015108 & 0.735 & 7.6 & $1.82^{+0.14}_{-0.08}$ & $-9.68^{+0.1}_{-0.19}$ \\ 
COSMOS CL J100135.3+024617 & 0.44 & 1.29 & $-1.1^{+0.83}_{-0.61}$ & $-12.77^{+0.6}_{-0.87}$ \\ 
\enddata
\tablecomments{
$^{a}$ Uncertainties denote the 1$\sigma$ credible intervals for each value. \\
$^{b}$ Best-fit sSFRs based on {\tt iSEDfit}. \\
}

\end{deluxetable*}
\invisiblesection{Best-fit SEDs from {\tt iSEDfit}}\label{sec:bestSEDs}

\begin{figure*}
\epsfig{file=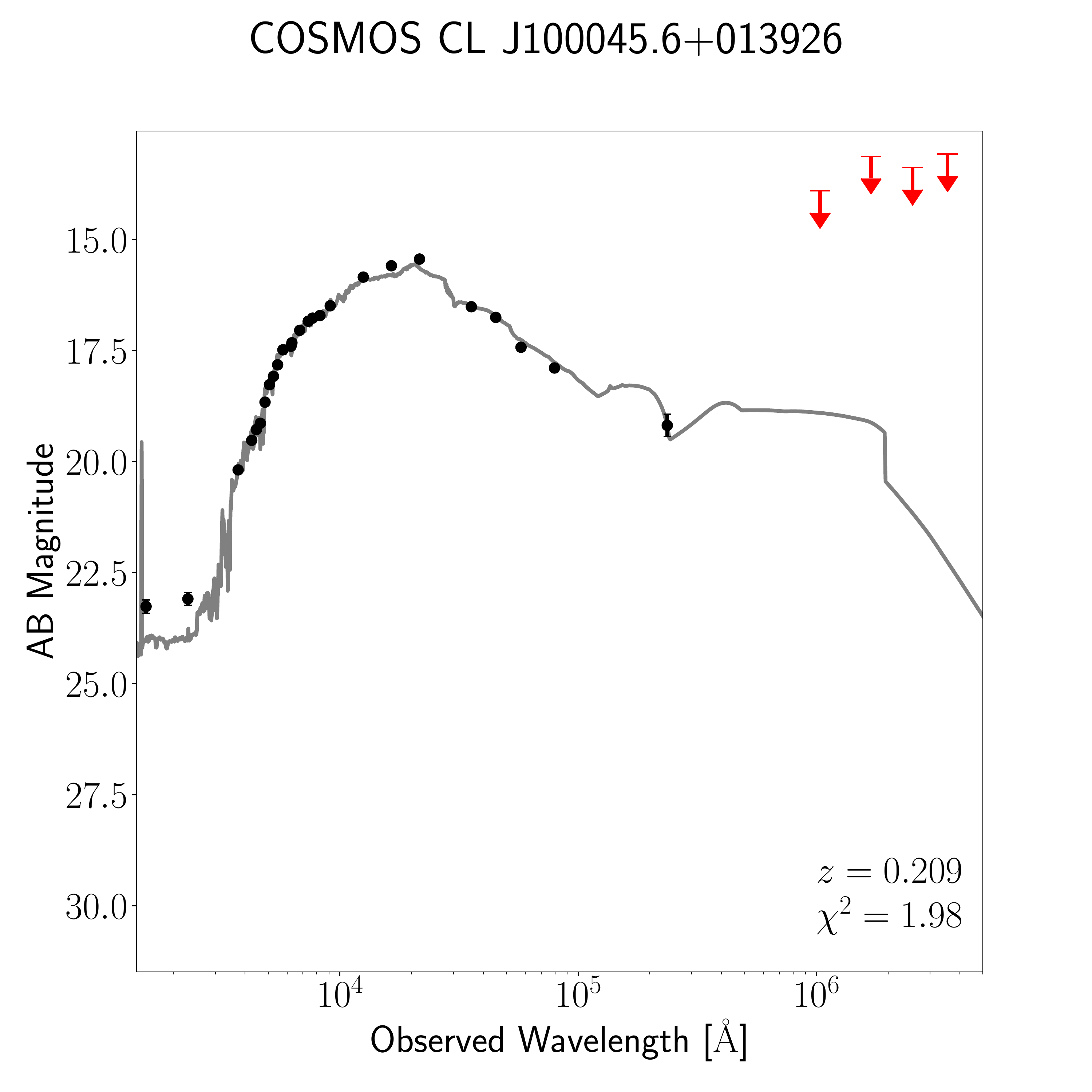,width=4cm,angle=0}
\epsfig{file=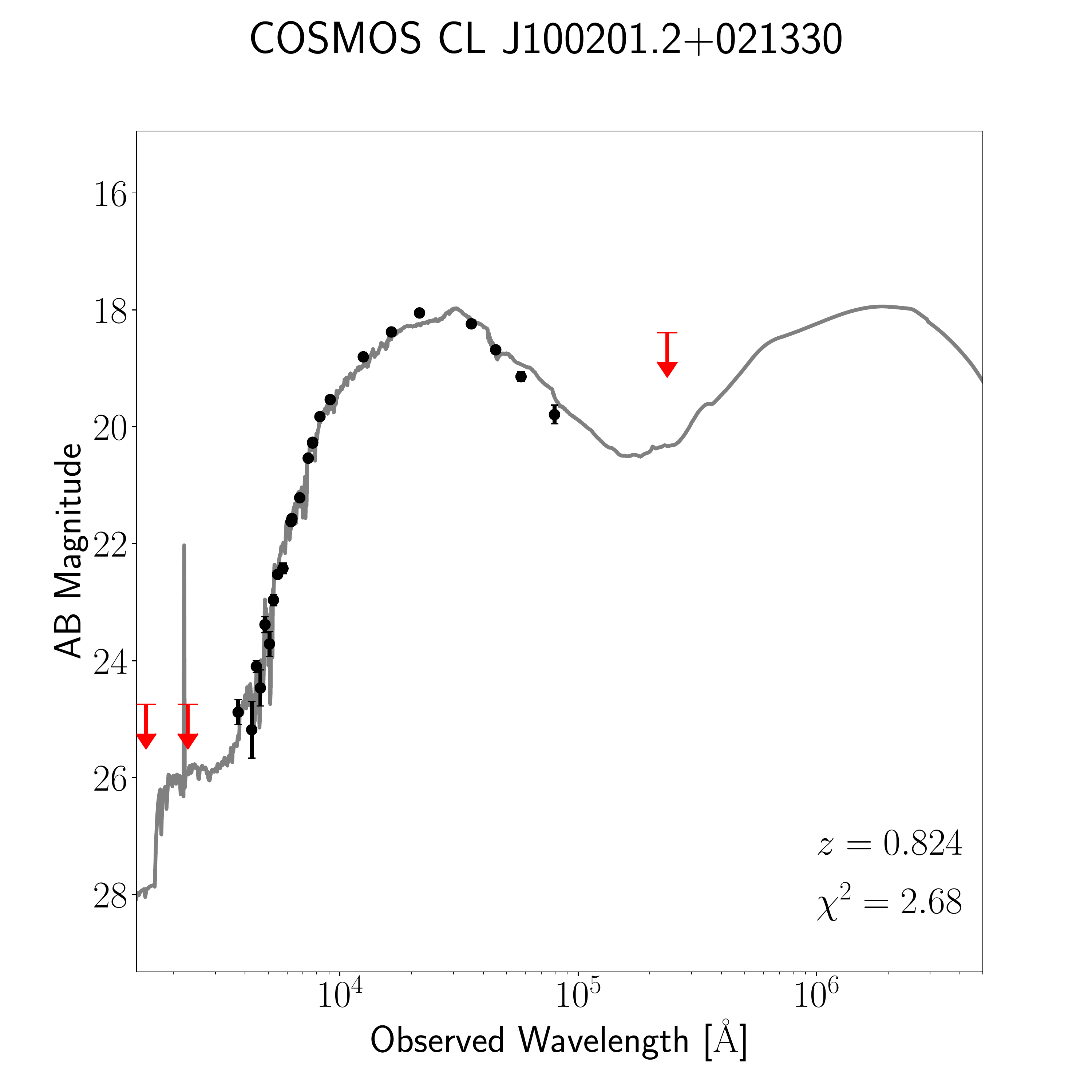,width=4cm,angle=0}
\epsfig{file=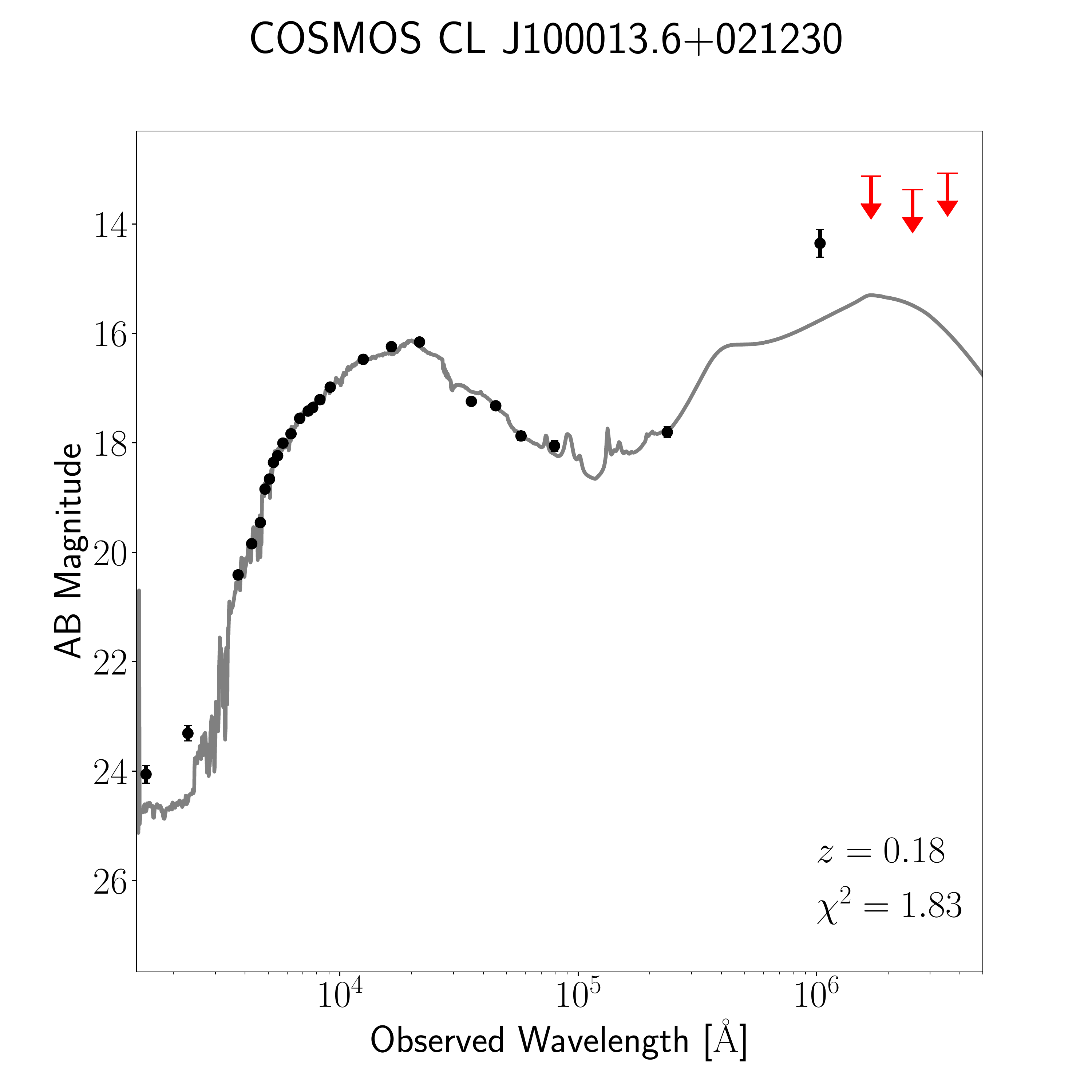,width=4cm,angle=0}
\epsfig{file=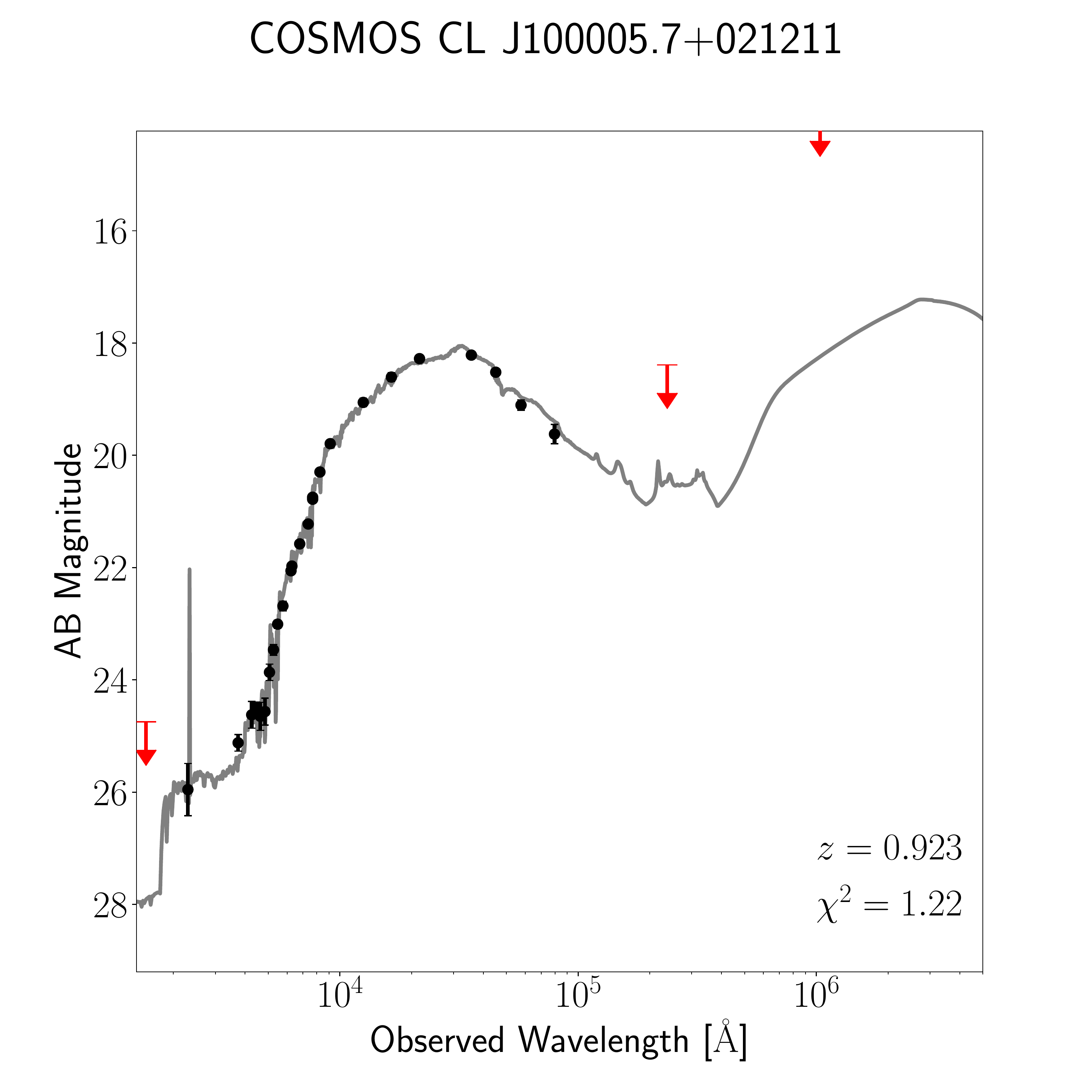,width=4cm,angle=0}
\epsfig{file=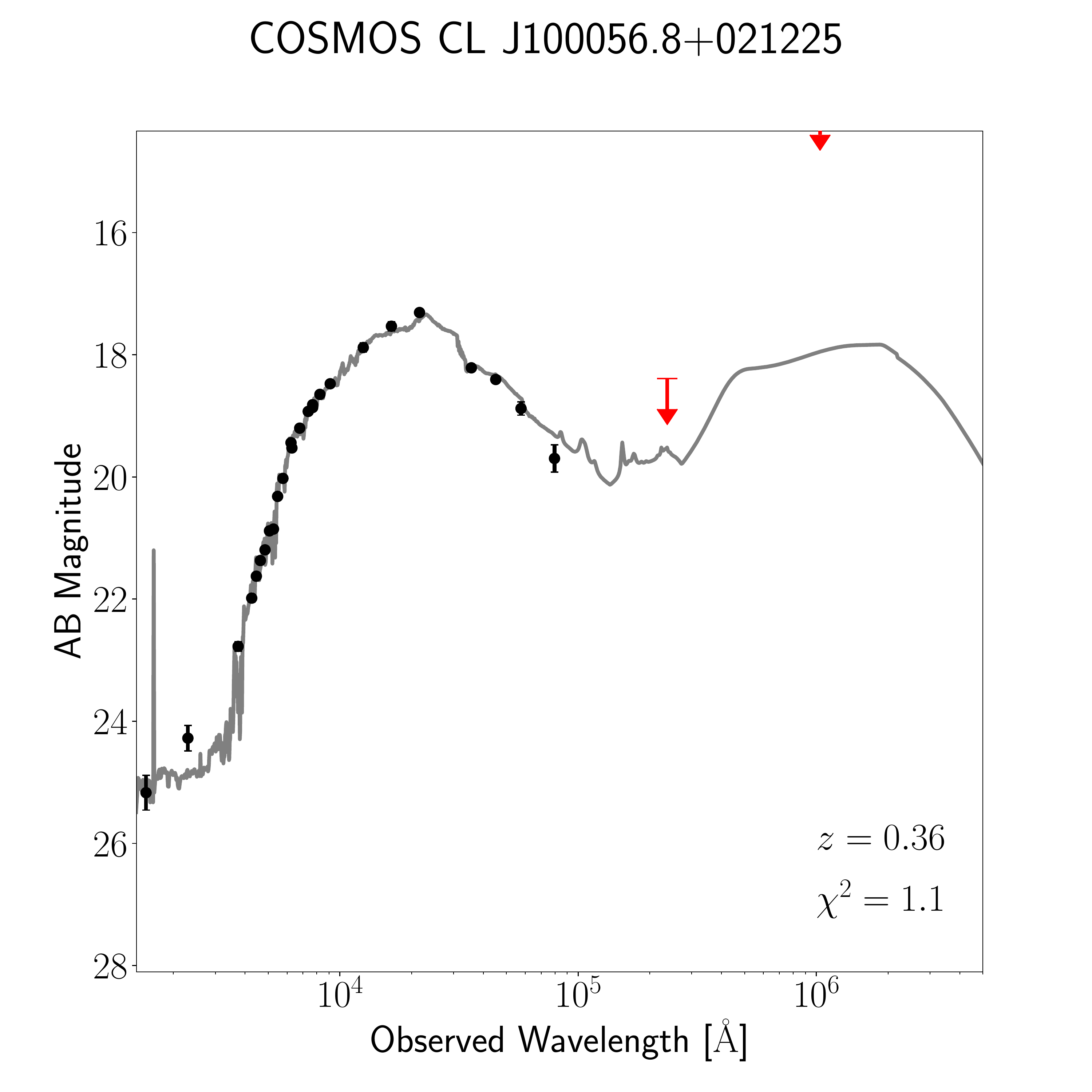,width=4cm,angle=0}
\epsfig{file=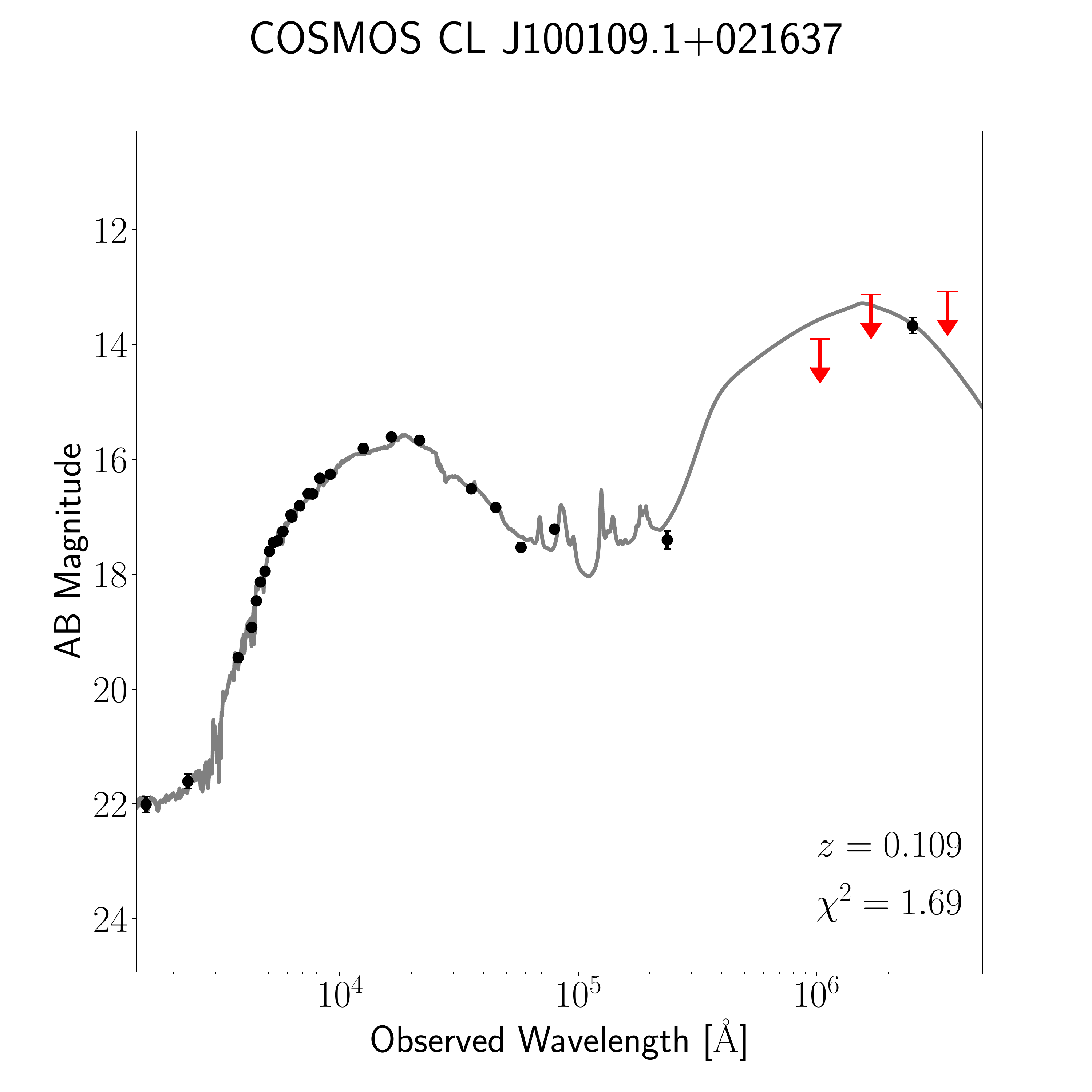,width=4cm,angle=0}
\epsfig{file=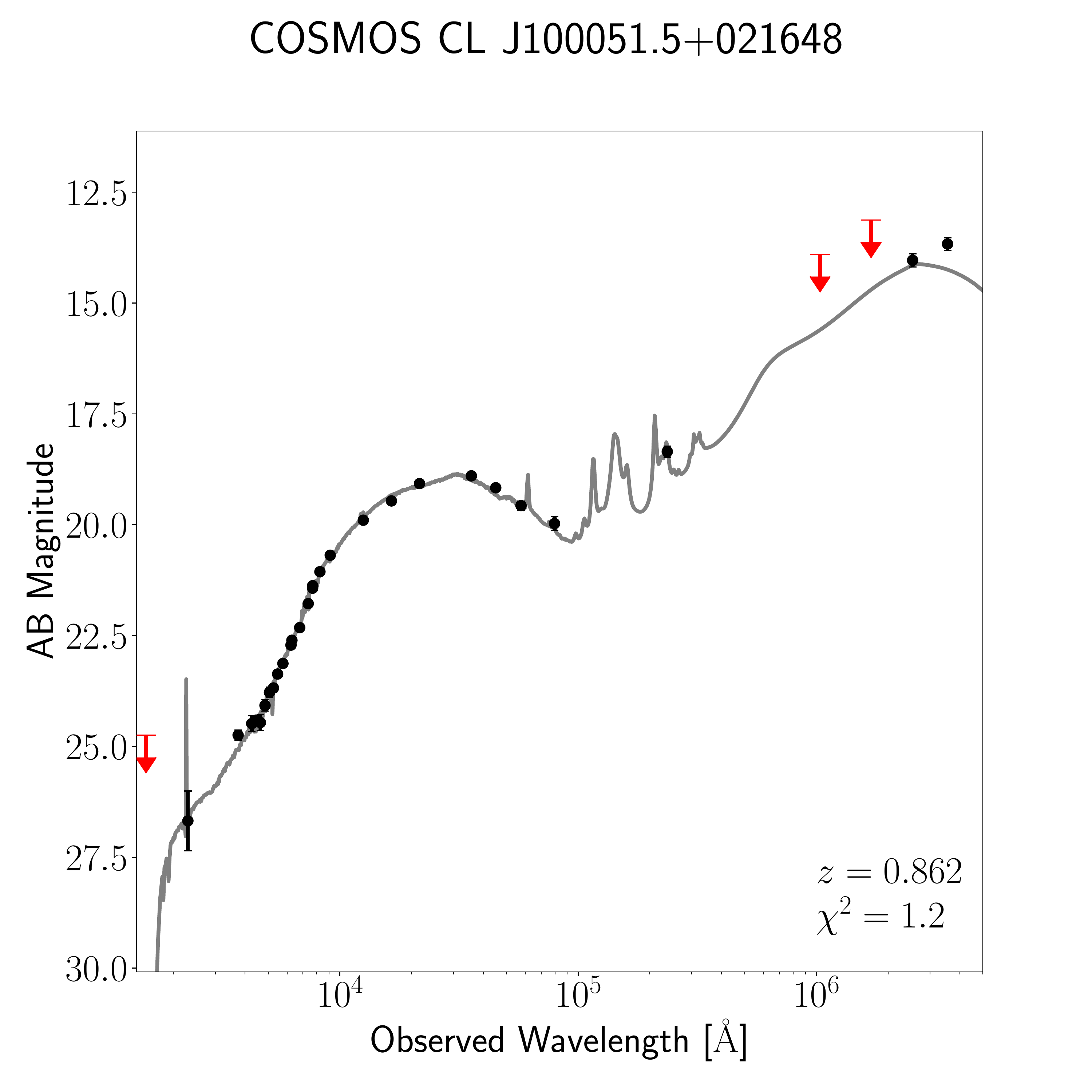,width=4cm,angle=0}
\epsfig{file=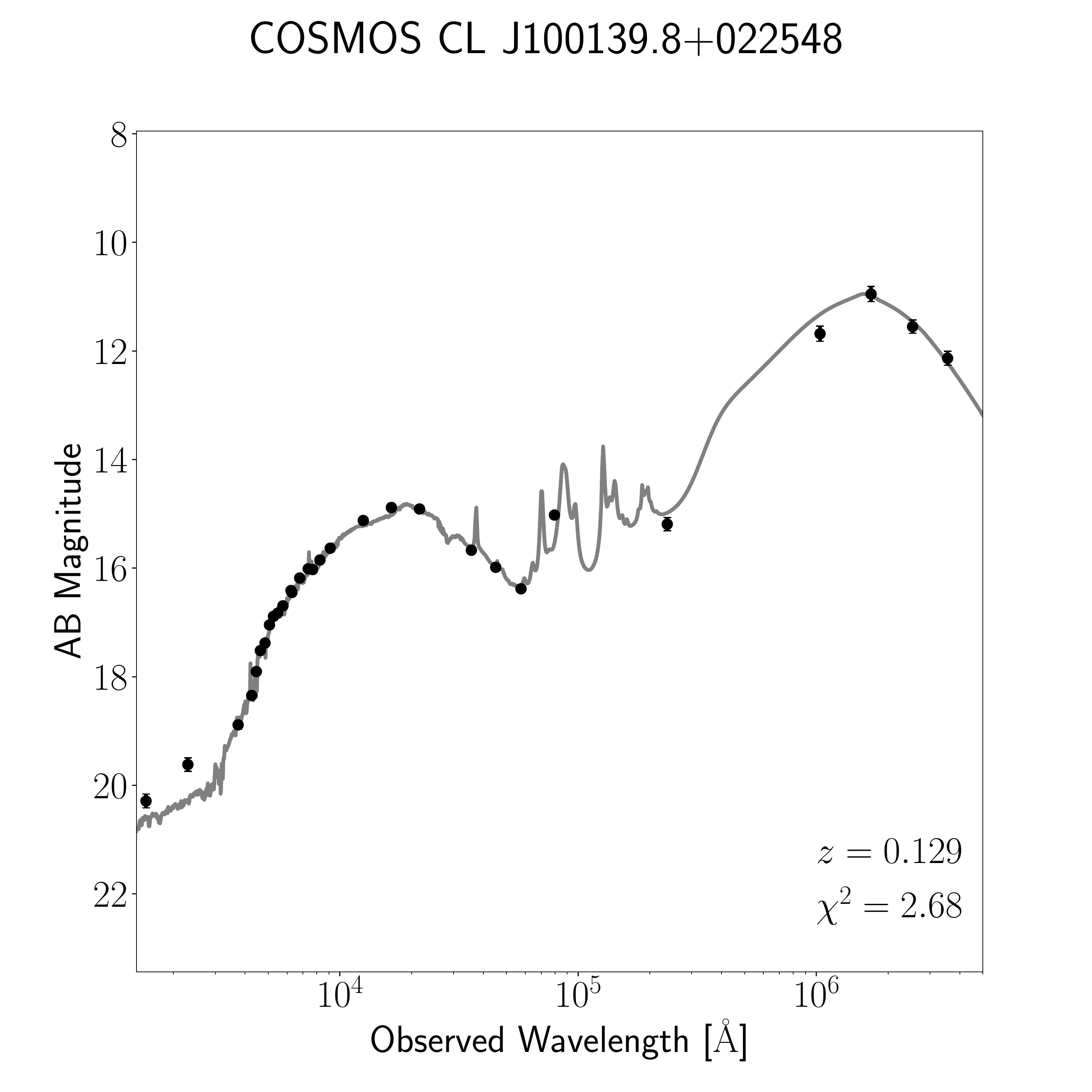,width=4cm,angle=0}
\epsfig{file=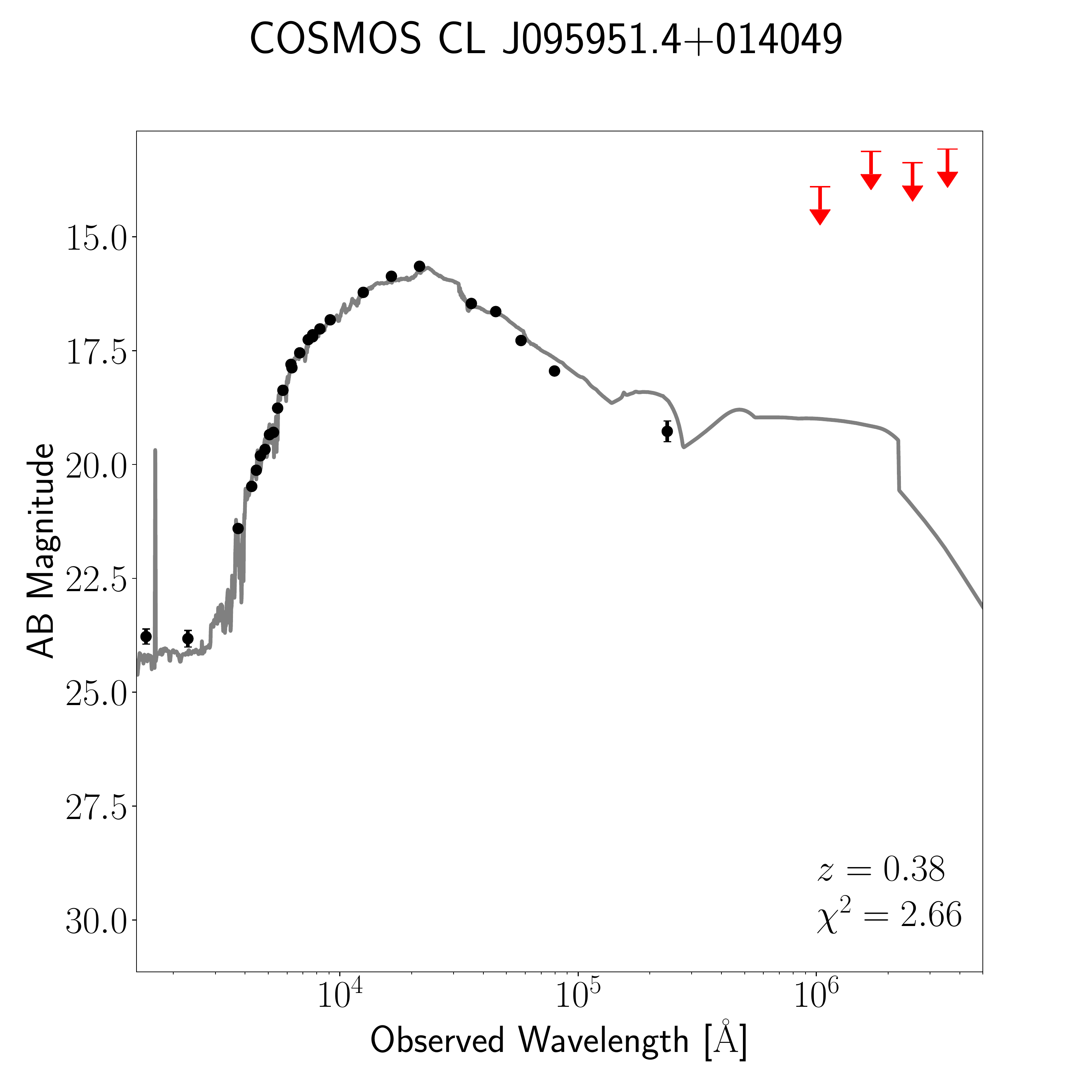,width=4cm,angle=0}
\epsfig{file=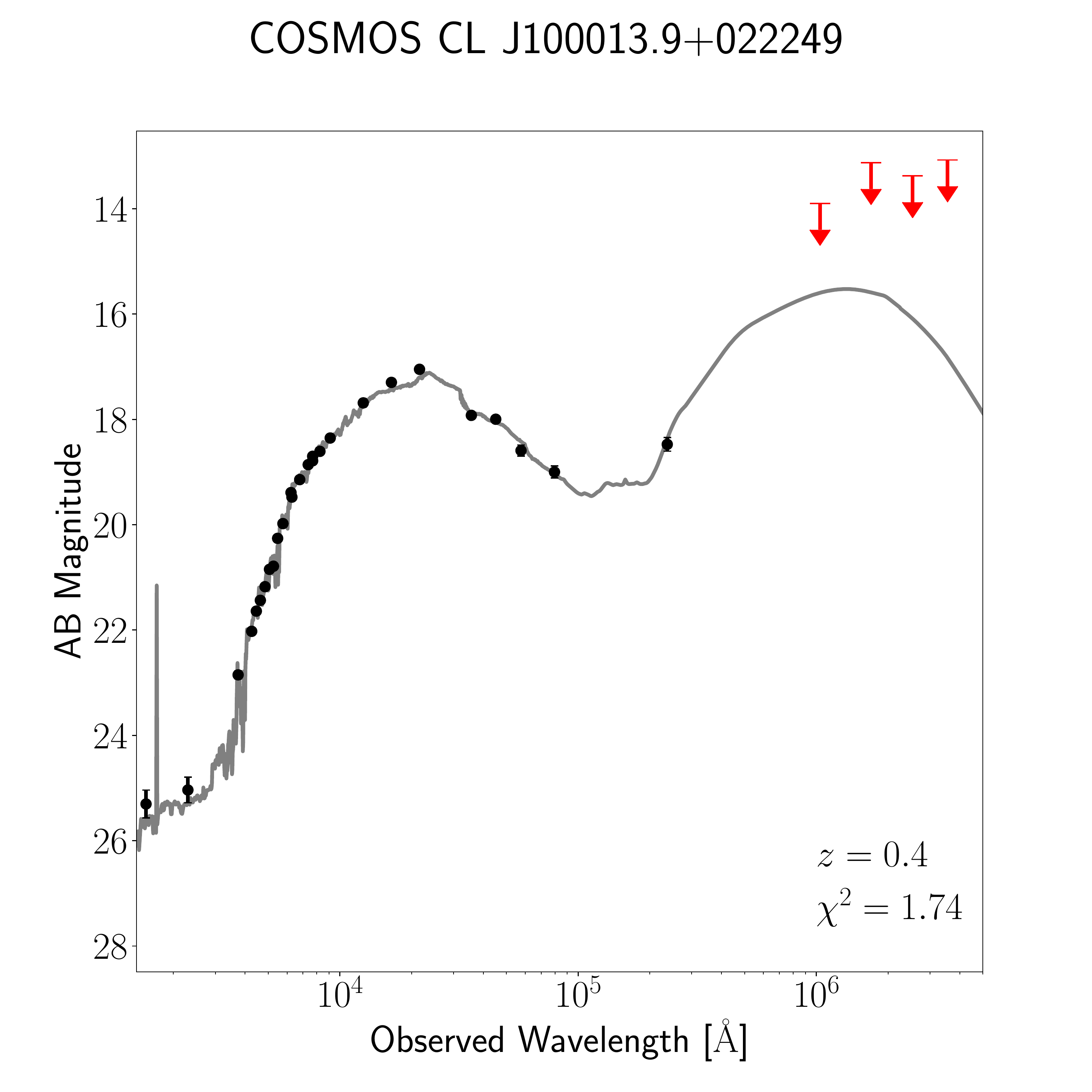,width=4cm,angle=0}
\epsfig{file=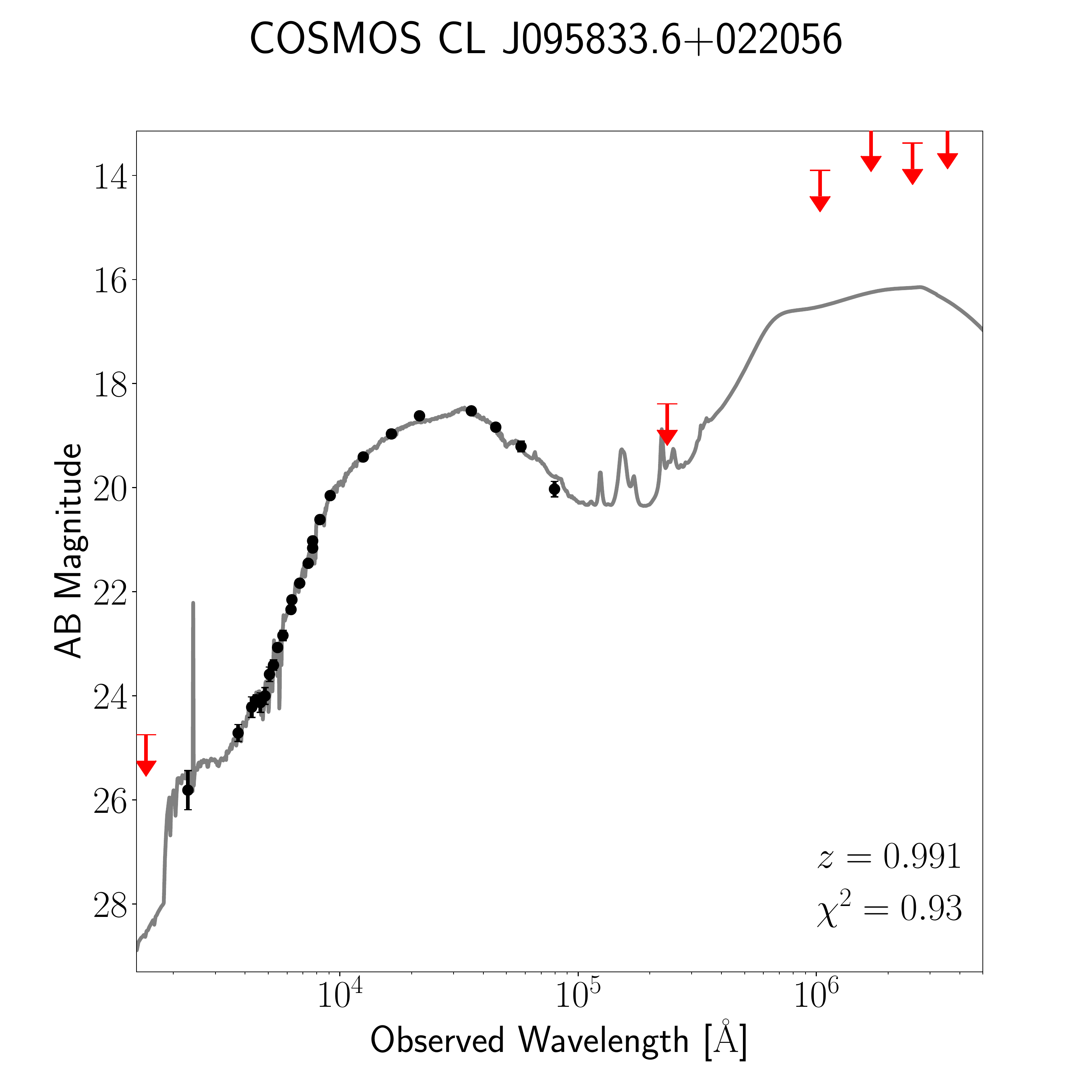,width=4cm,angle=0}
\epsfig{file=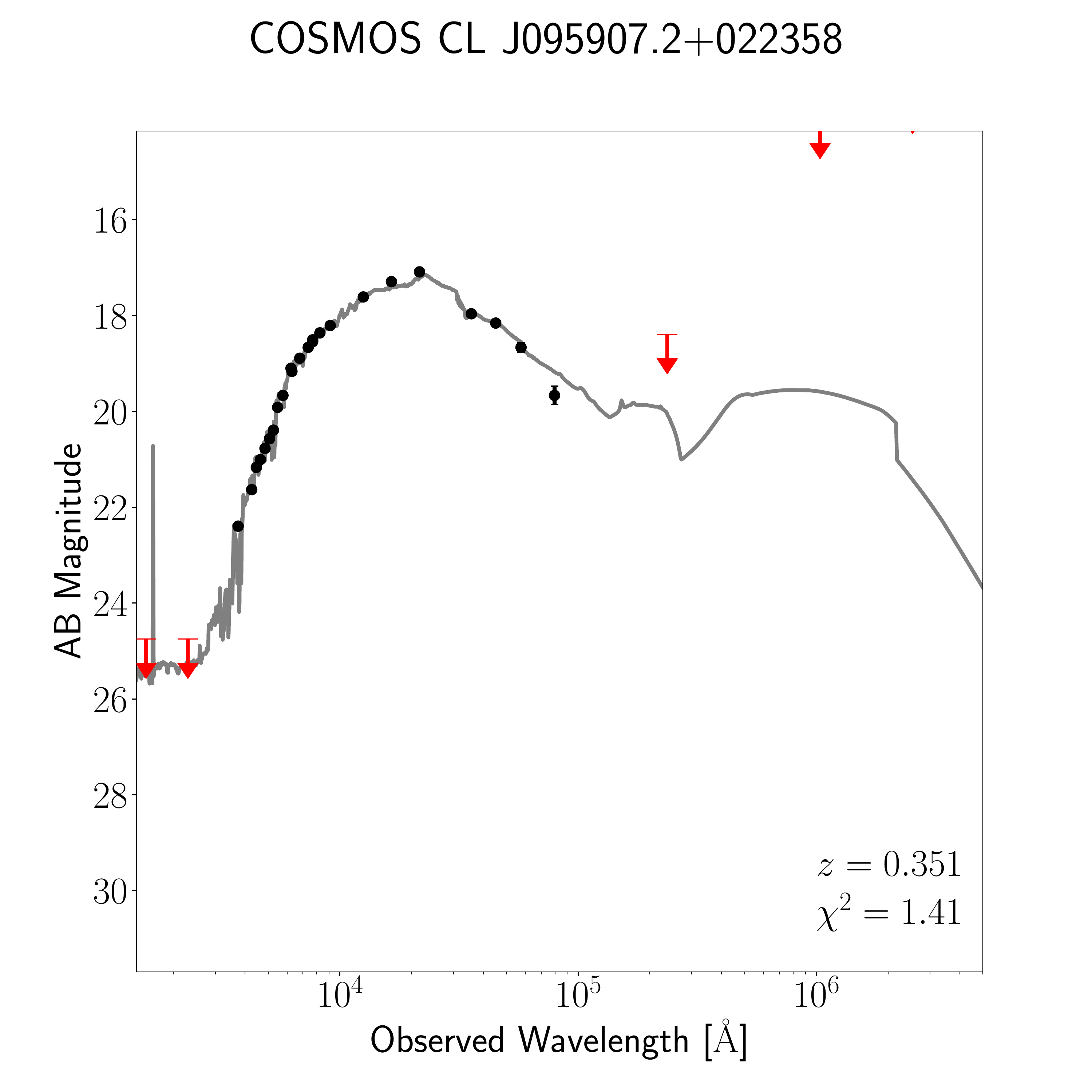,width=4cm,angle=0}
\epsfig{file=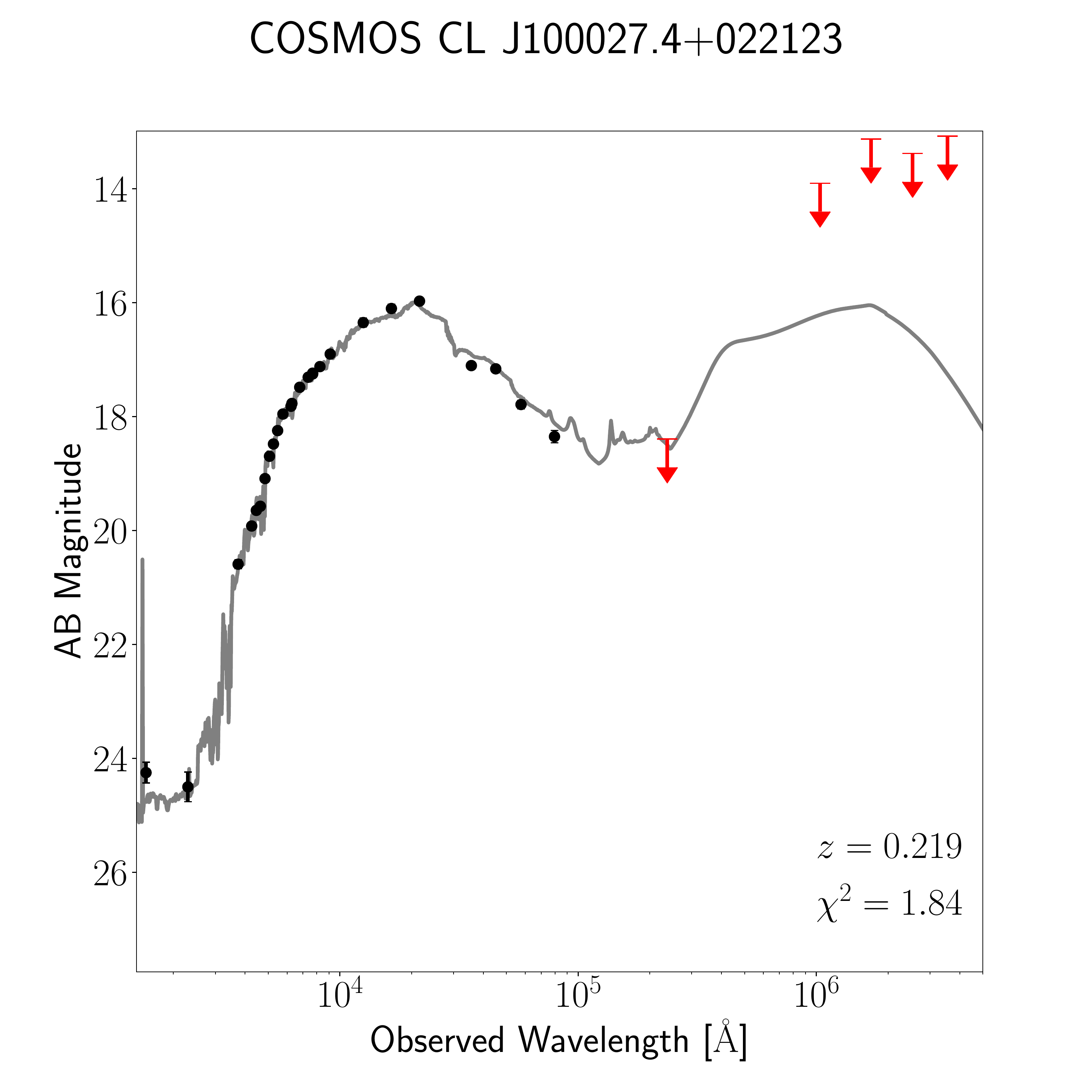,width=4cm,angle=0}
\epsfig{file=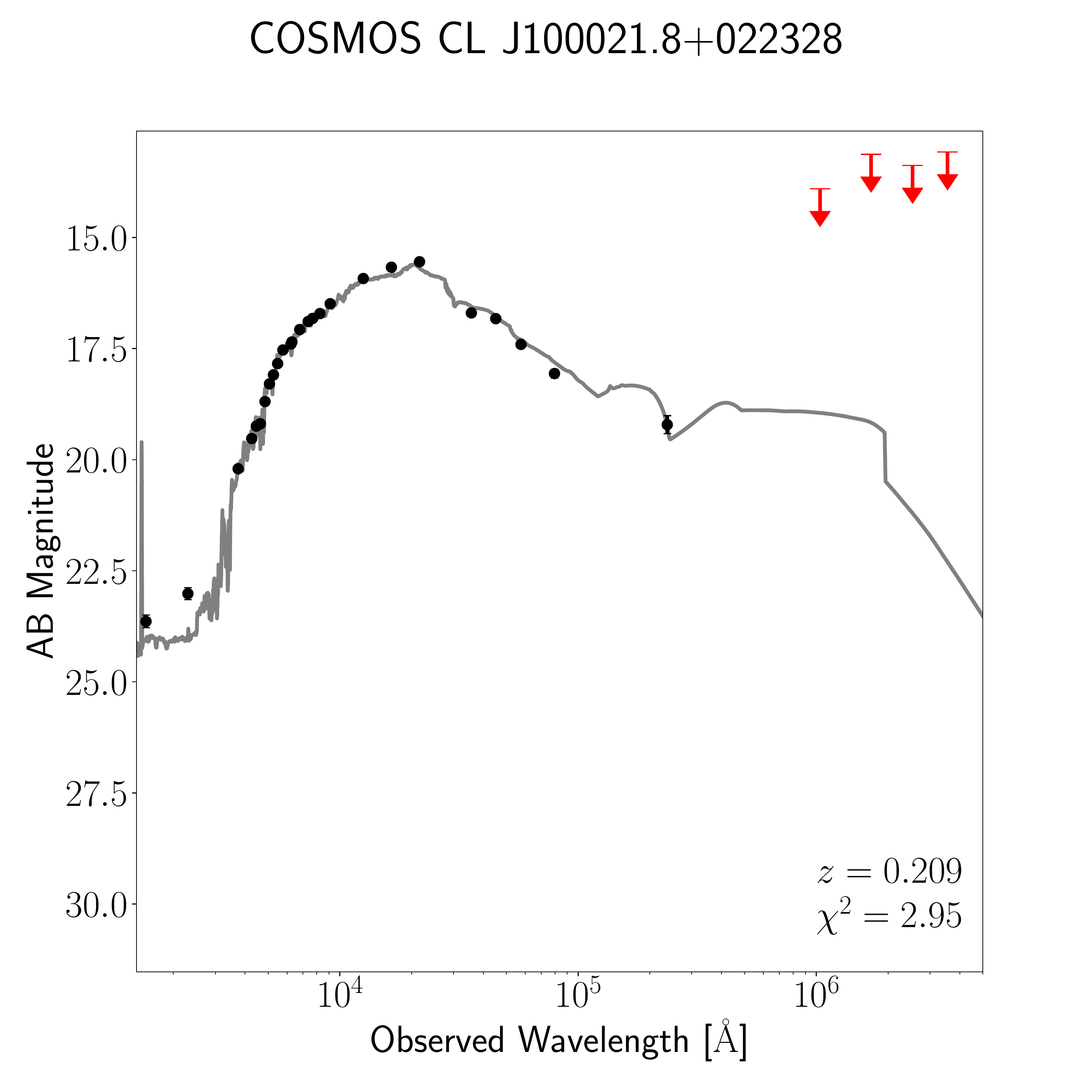,width=4cm,angle=0}
\epsfig{file=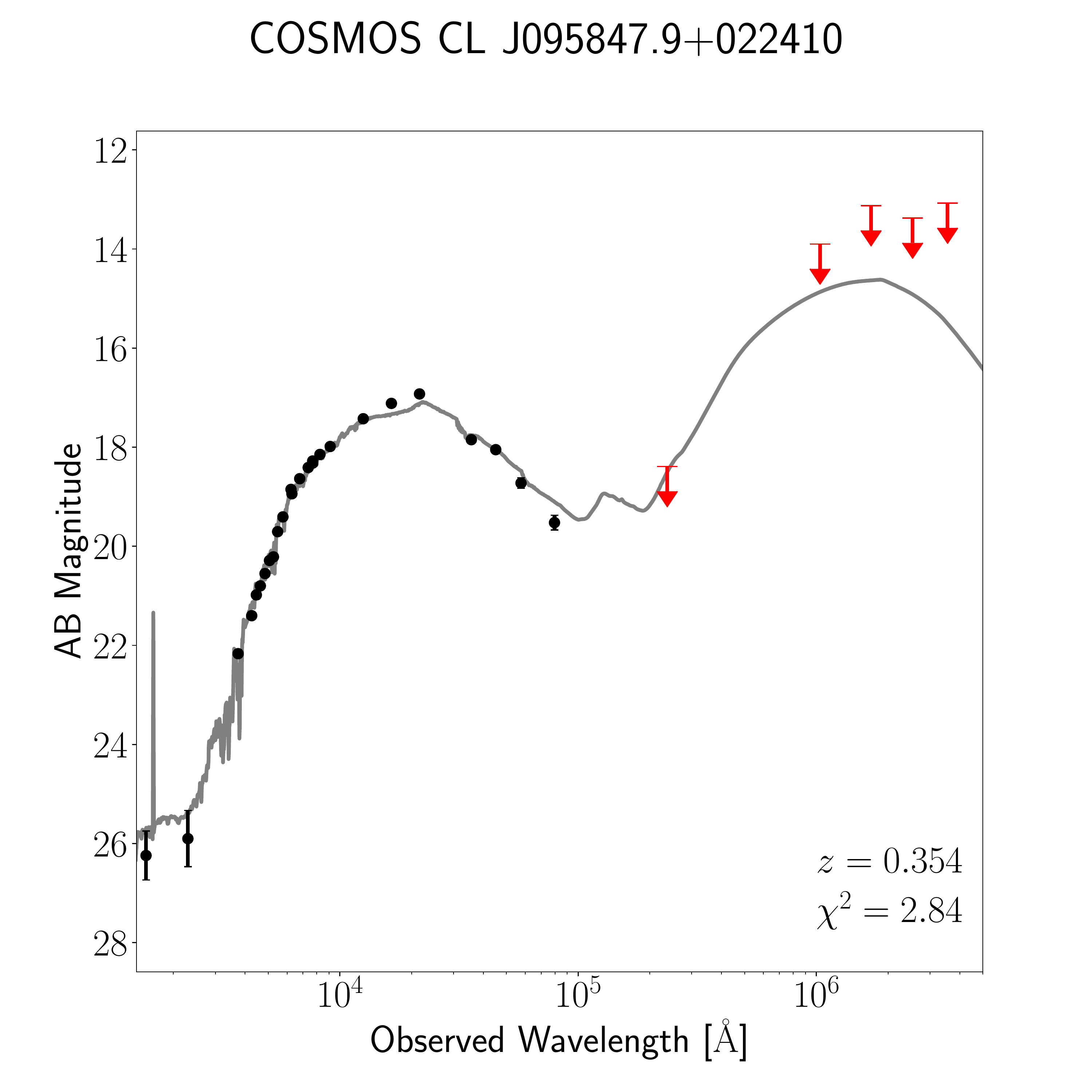,width=4cm,angle=0}
\epsfig{file=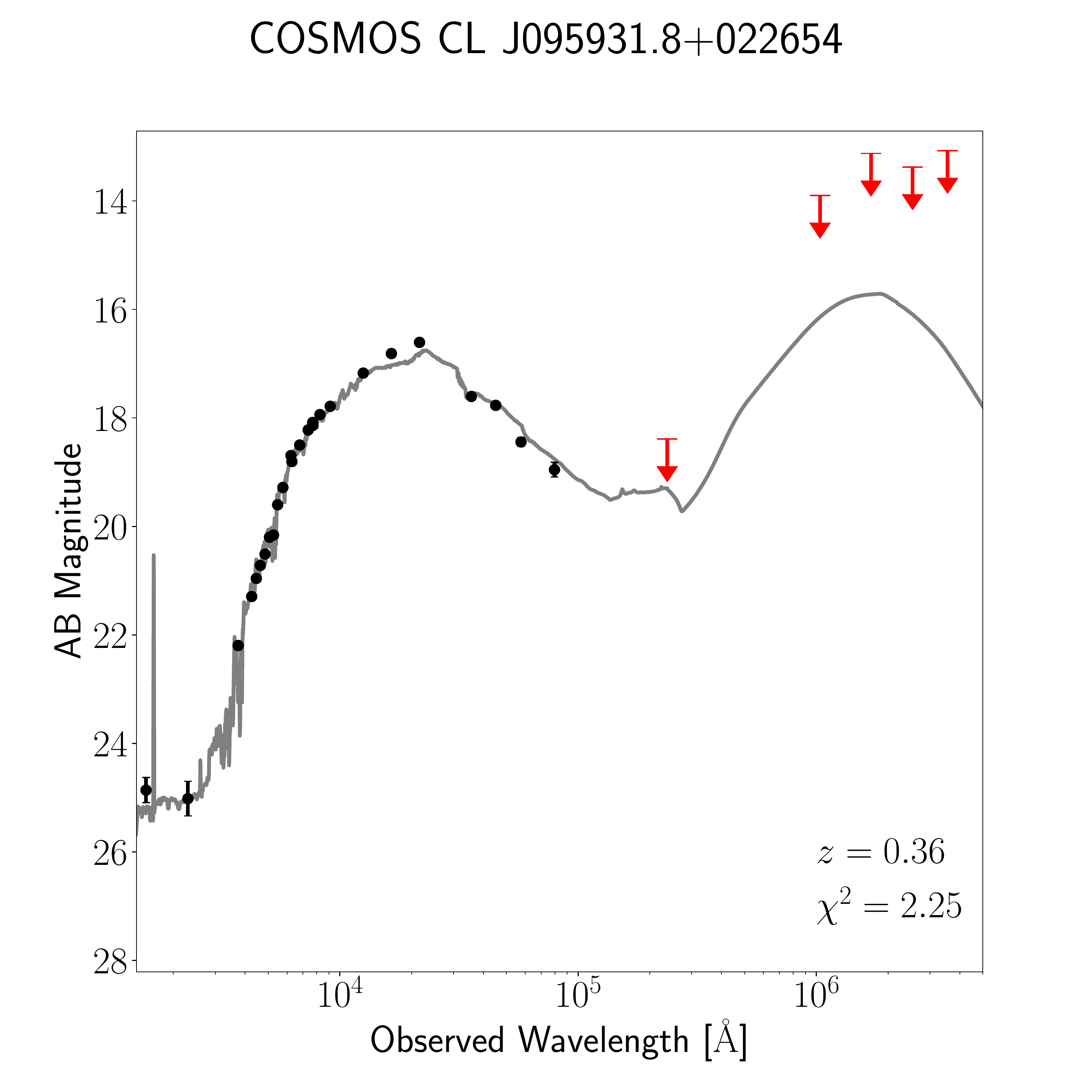,width=4cm,angle=0}
\epsfig{file=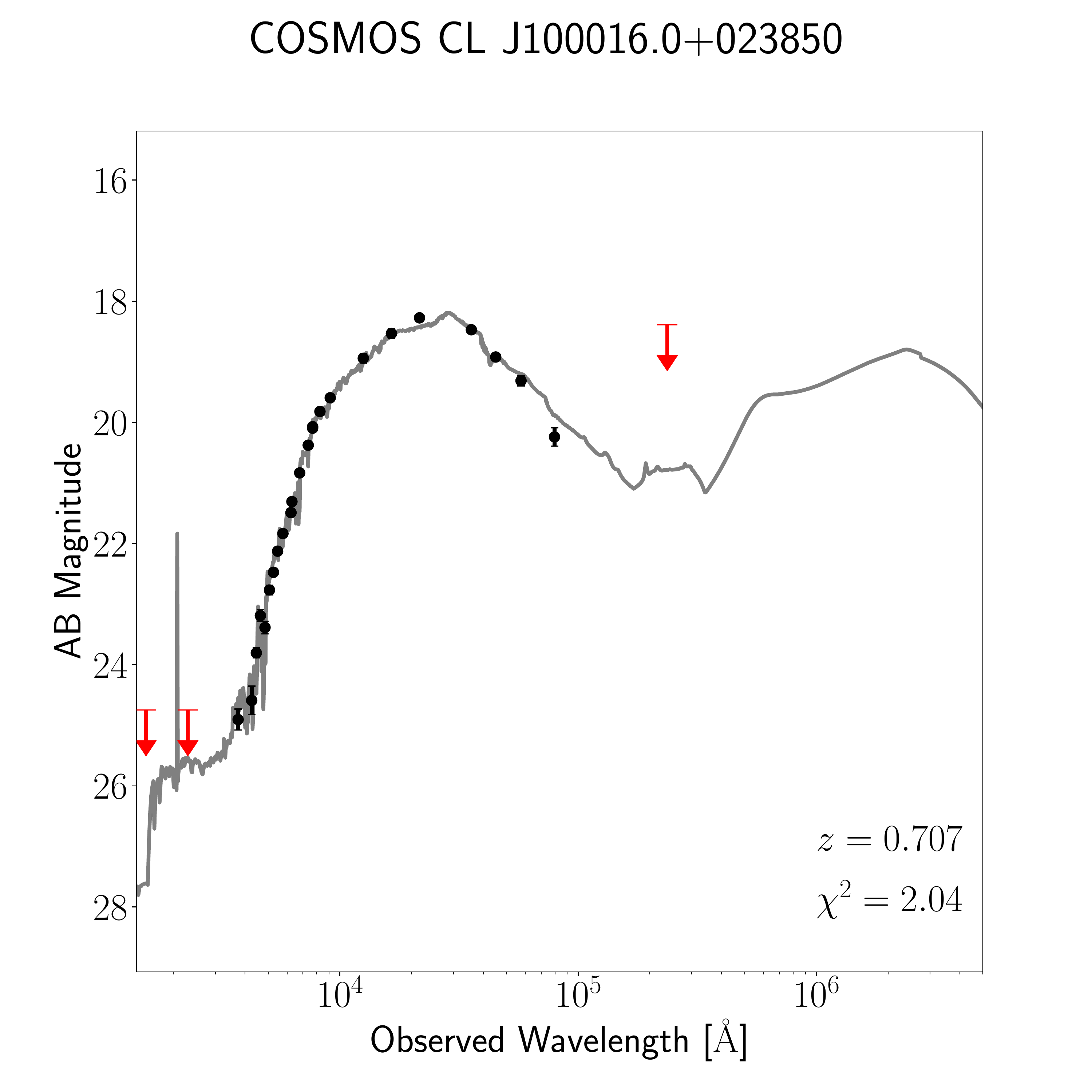,width=4cm,angle=0}
\epsfig{file=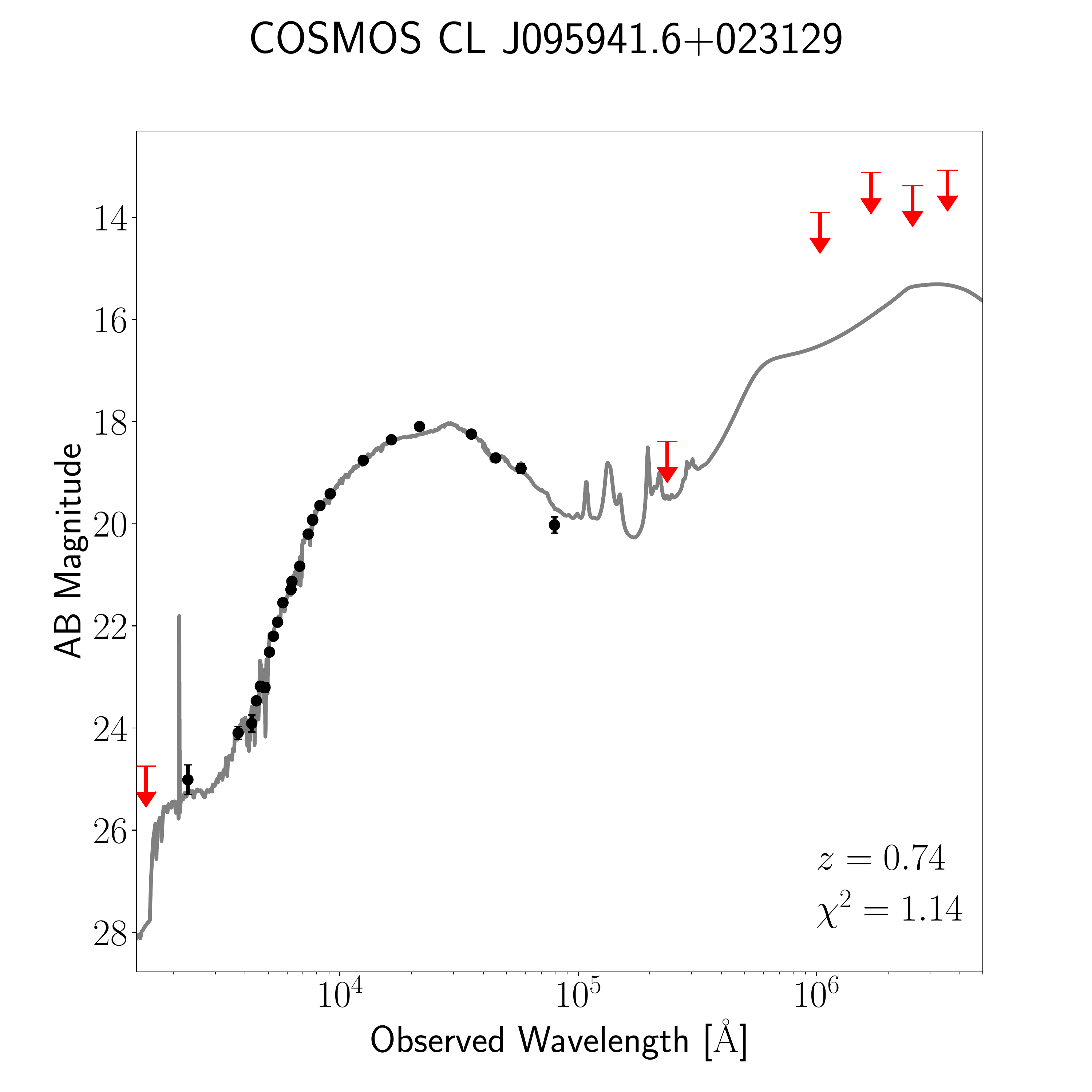,width=4cm,angle=0}
\epsfig{file=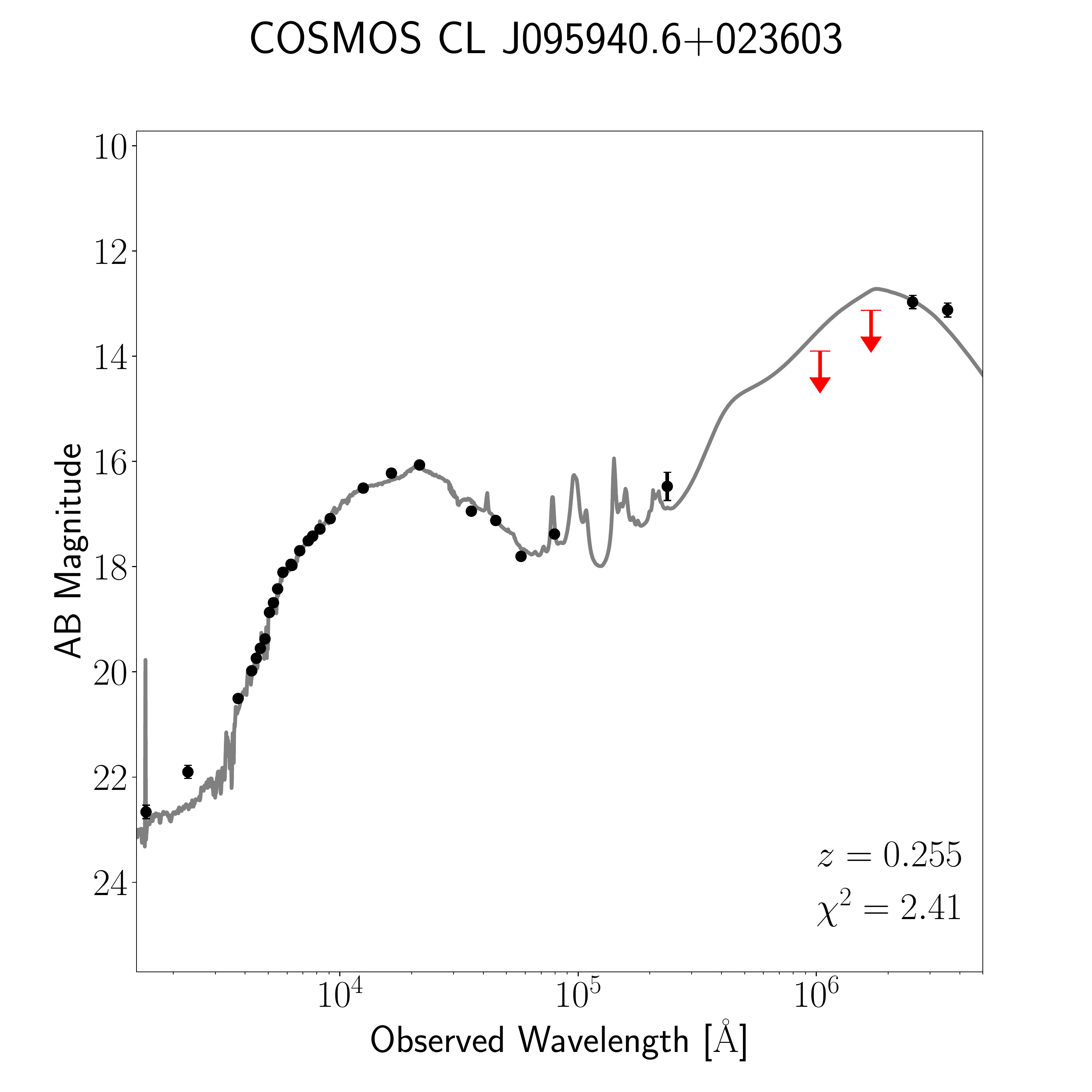,width=4cm,angle=0}
\epsfig{file=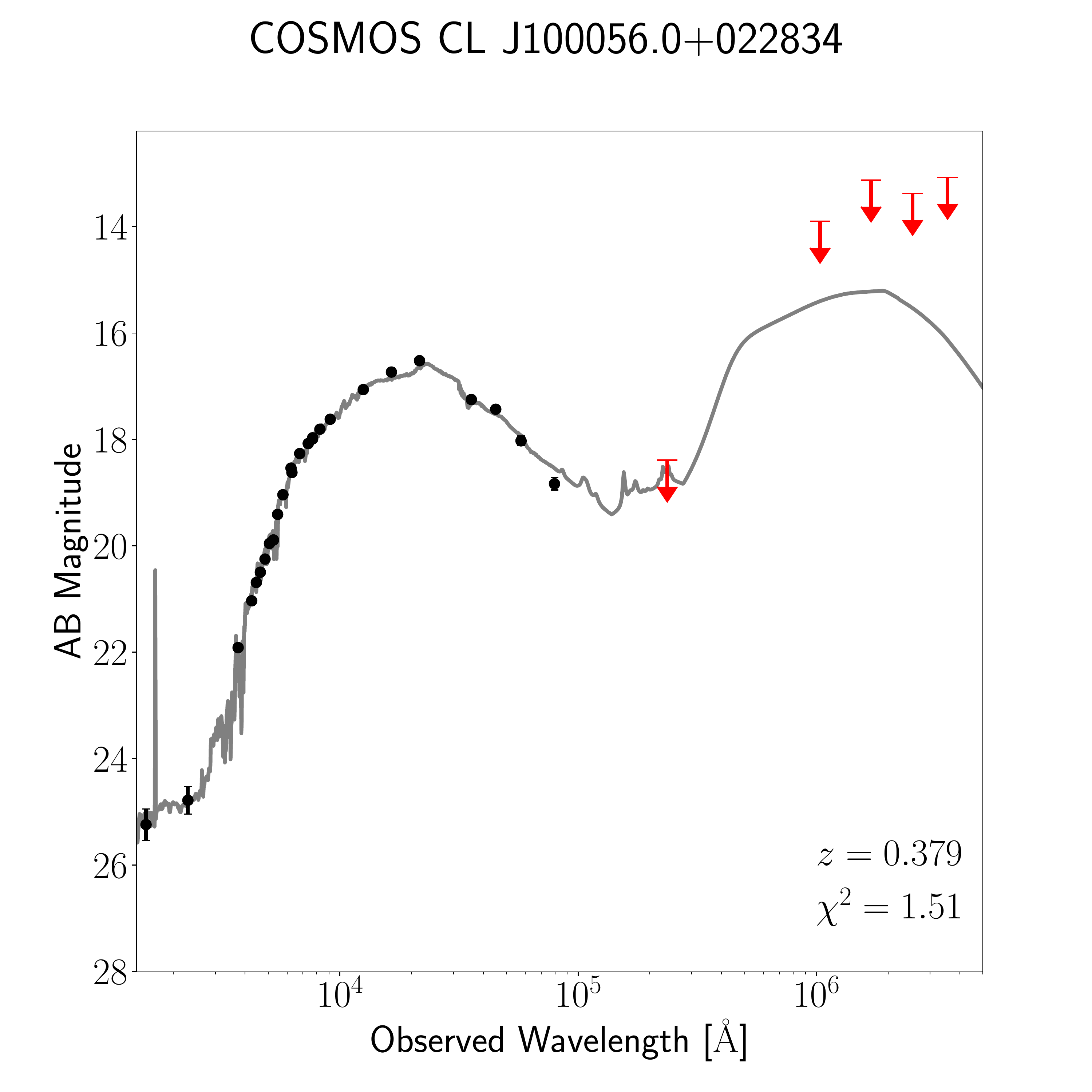,width=4cm,angle=0}

\caption{\label{figure:Best_SEDs} Best-fit models for the COSMOS BCGs. COSMOS SEDs are plotted in each figure as black points (for detections) and red arrows (for 3$\sigma$ upper limits). Error bars depicted on the SEDs are 1$\sigma$ error bars. The gray line in each plot depicts the best-fit spectrum, corresponding to the model producing the smallest reduced $\chi^{2}$ in the {\tt iSEDfit} Monte Carlo grid.}
\end{figure*}

\renewcommand{\thefigure}{A\arabic{figure}}

\setcounter{figure}{0}

\begin{figure*}
\epsfig{file=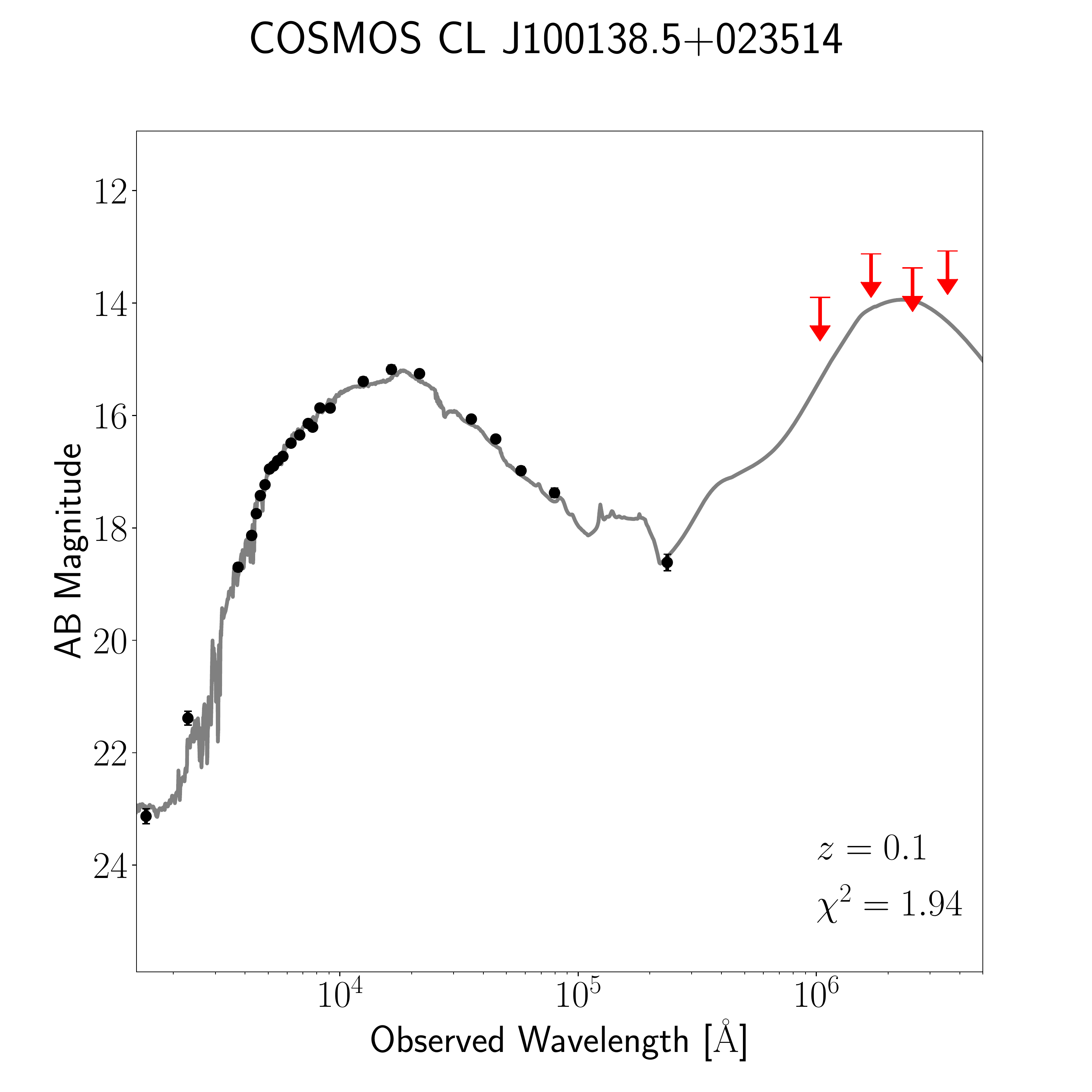,width=4cm,angle=0}
\epsfig{file=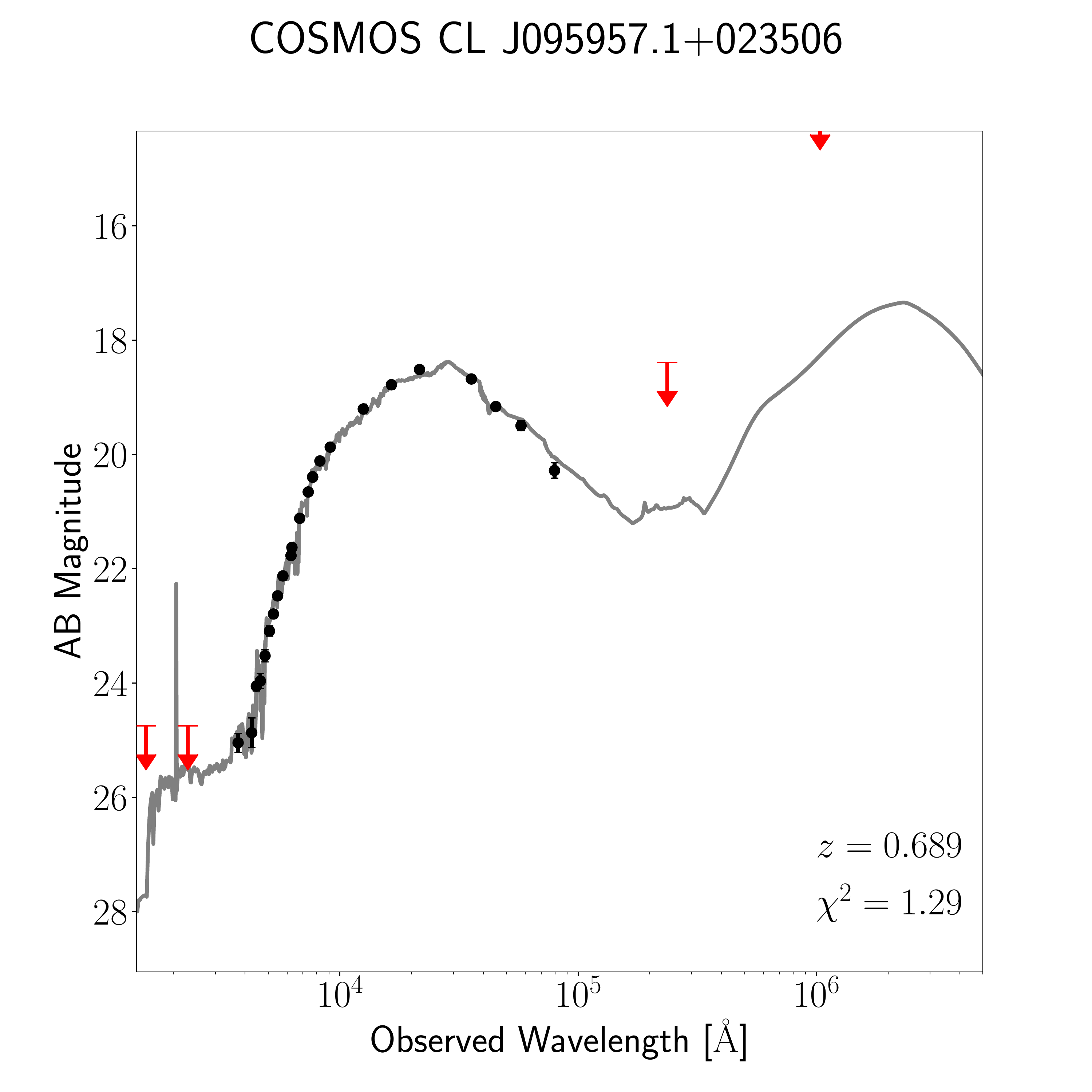,width=4cm,angle=0}
\epsfig{file=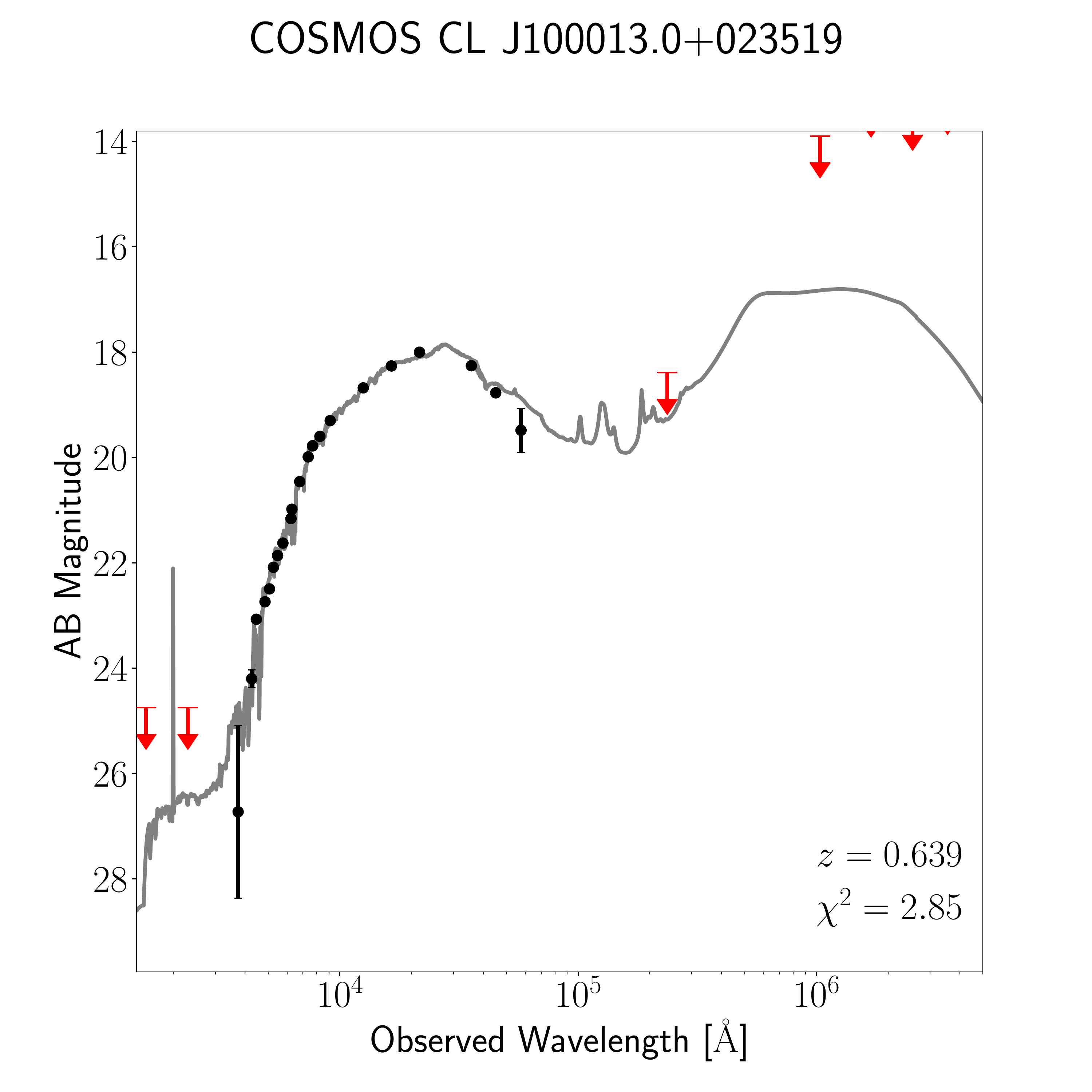,width=4cm,angle=0}
\epsfig{file=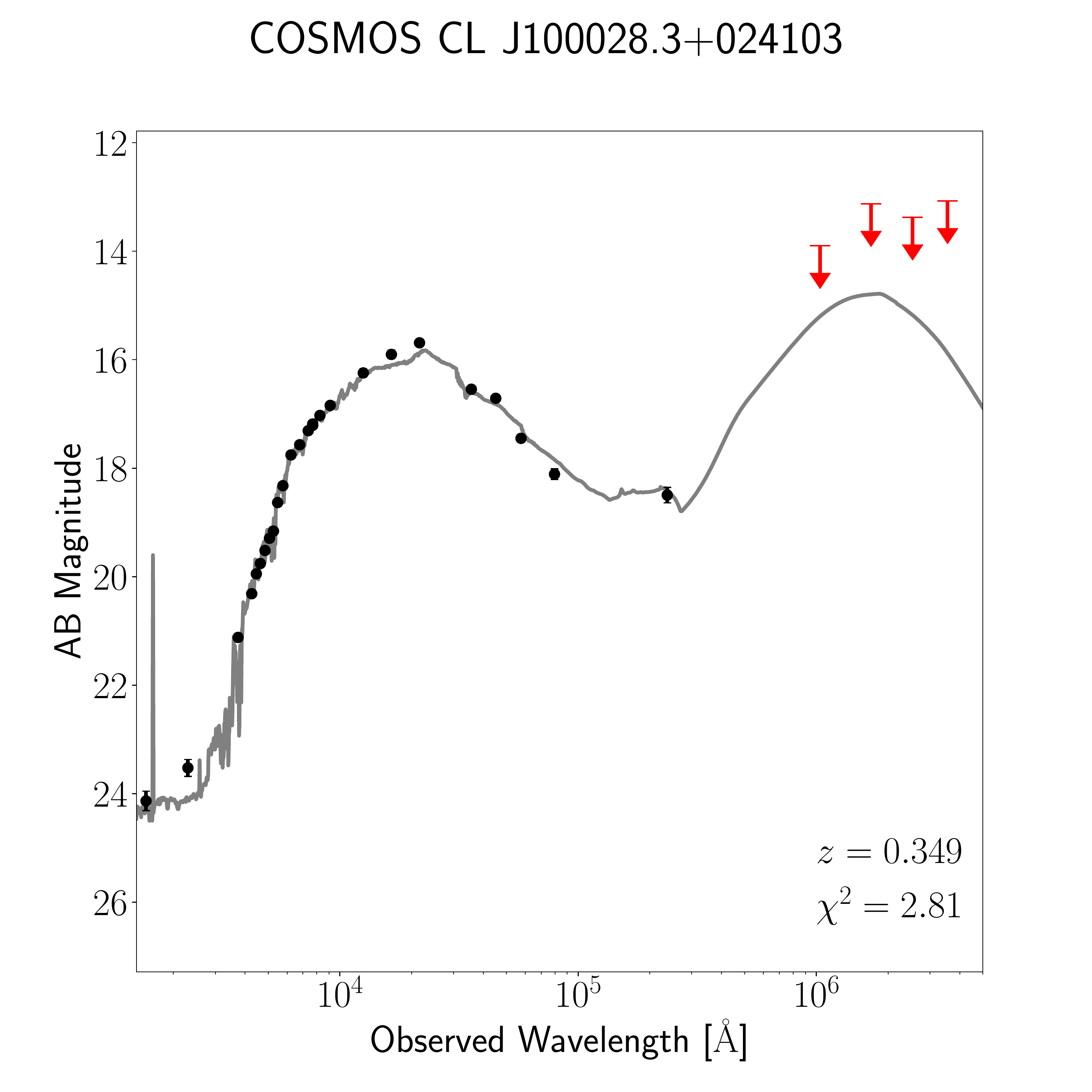,width=4cm,angle=0}
\epsfig{file=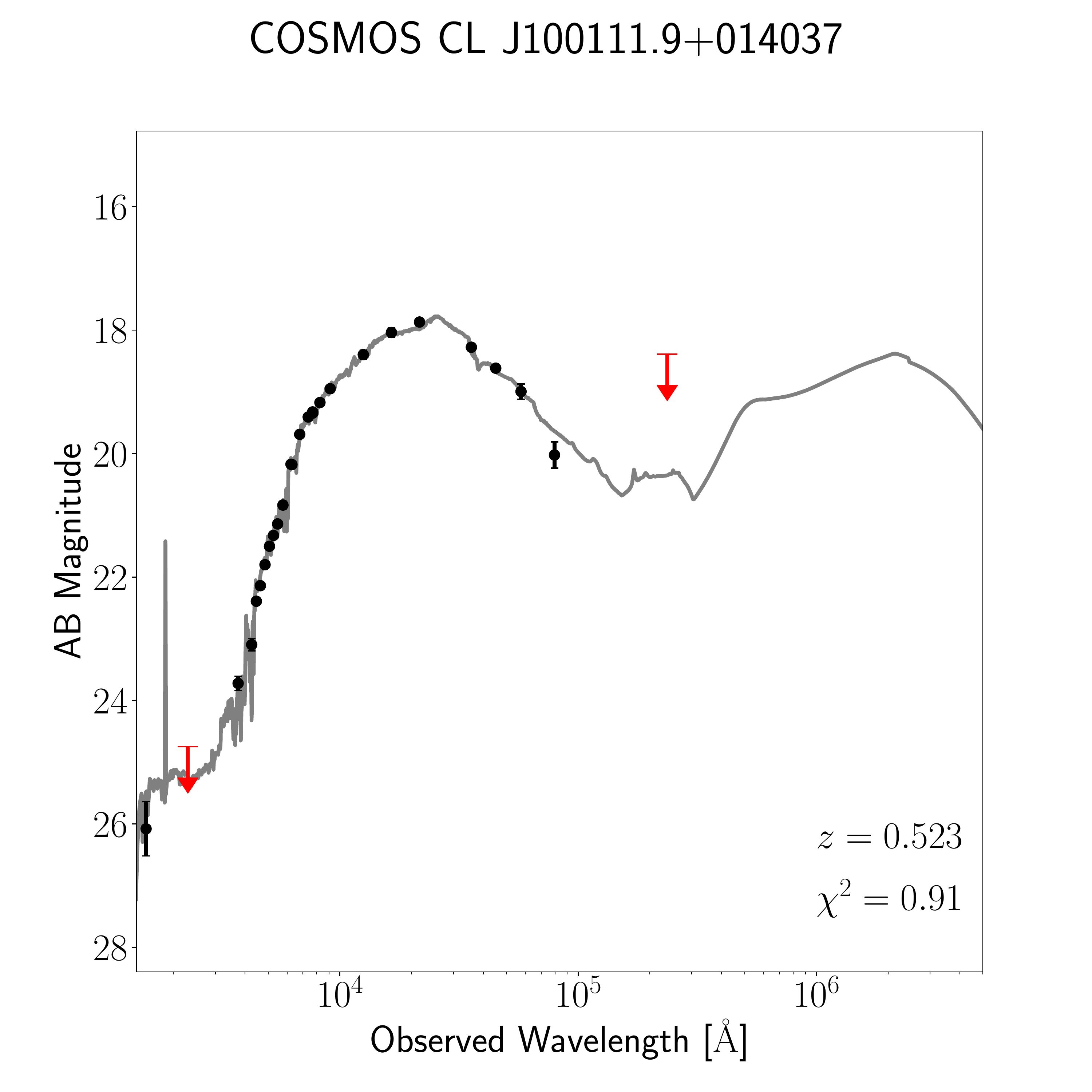,width=4cm,angle=0}
\epsfig{file=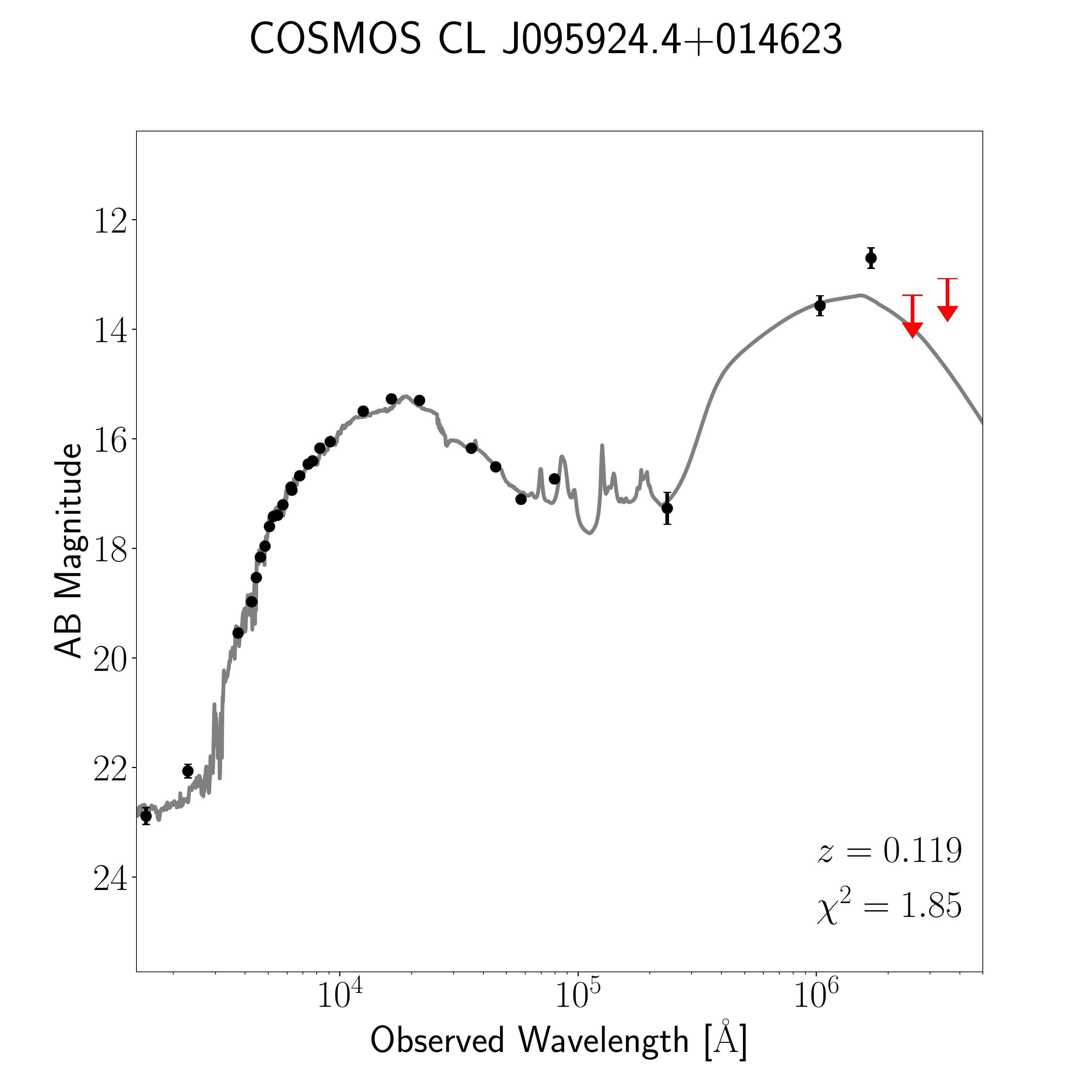,width=4cm,angle=0}
\epsfig{file=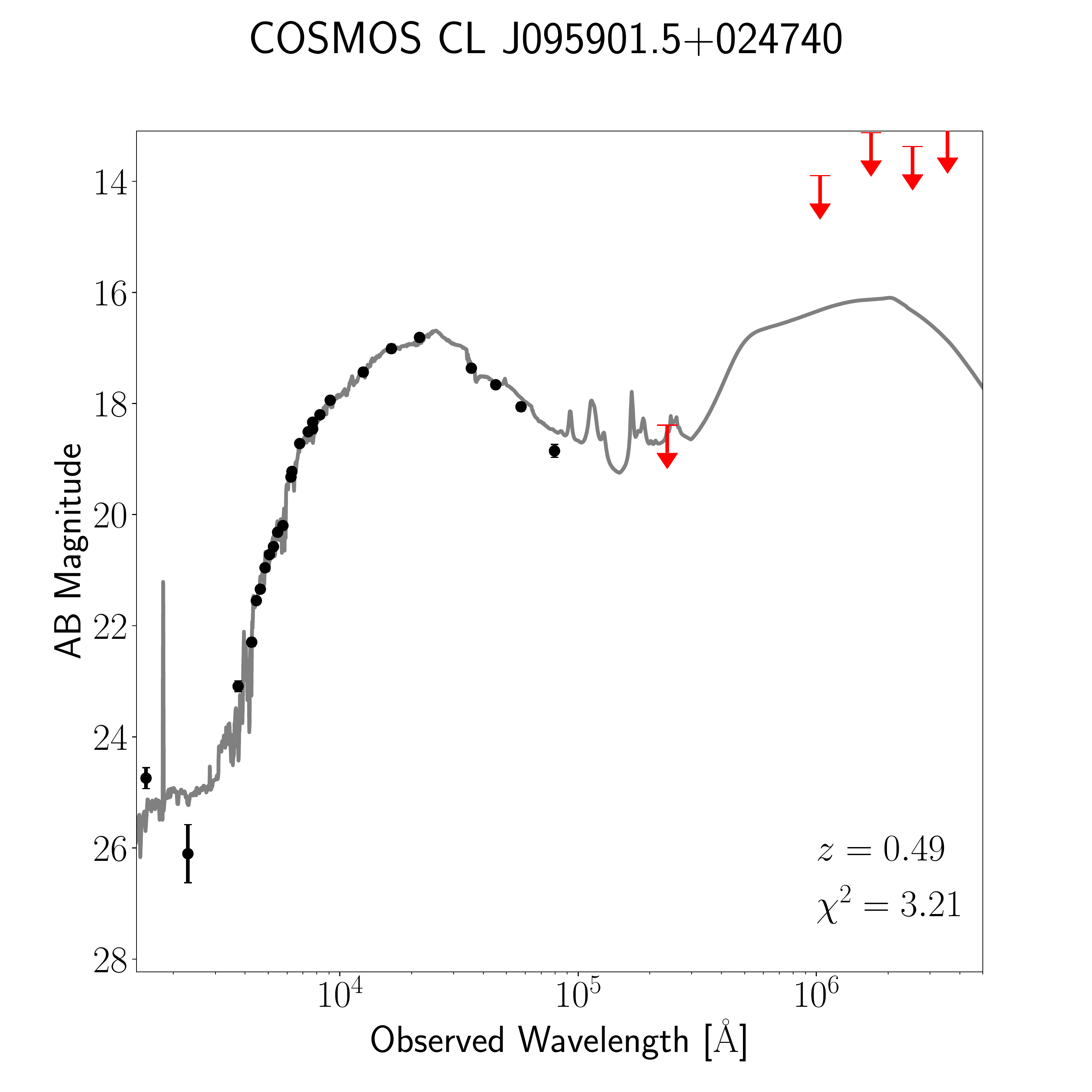,width=4cm,angle=0}
\epsfig{file=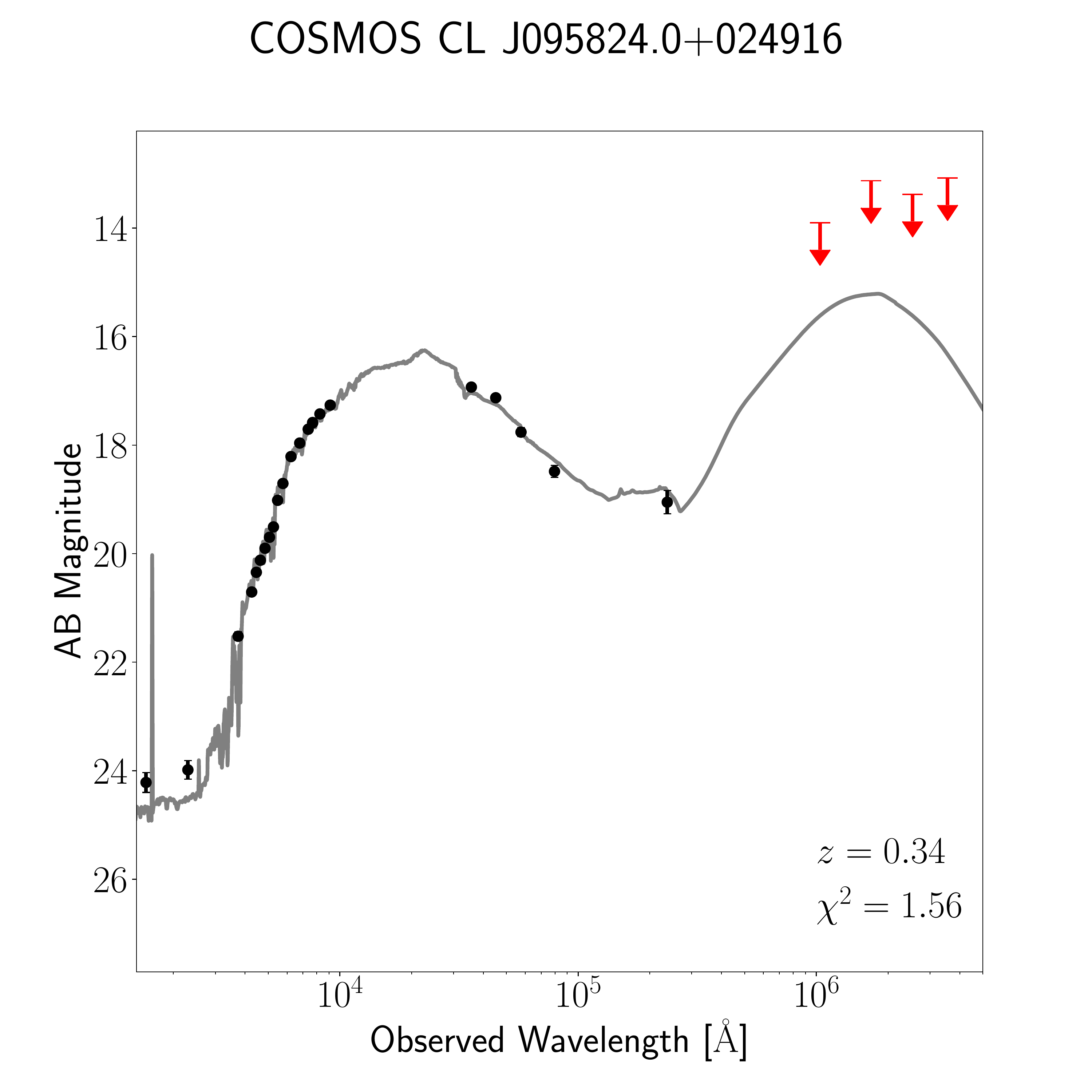,width=4cm,angle=0}
\epsfig{file=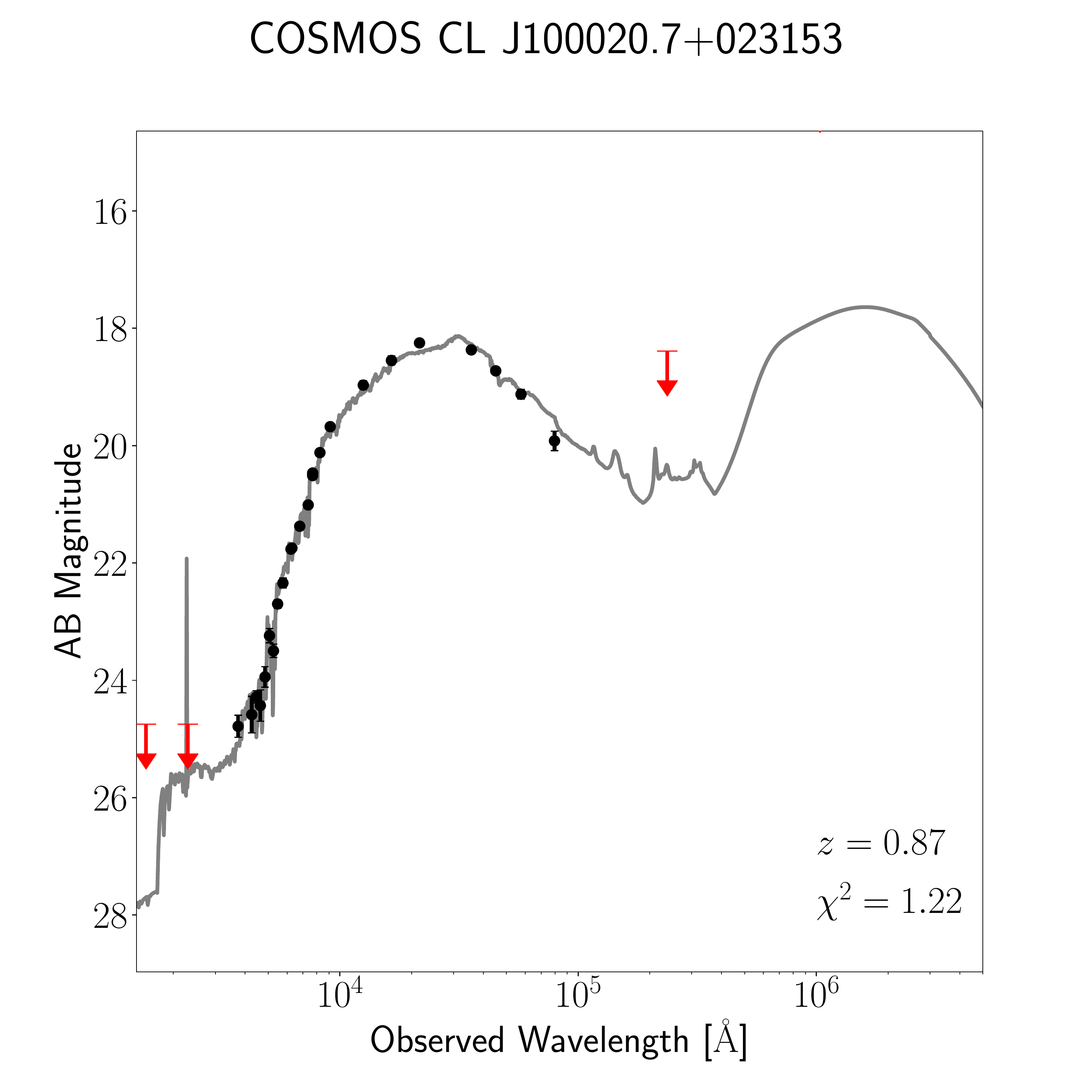,width=4cm,angle=0}
\epsfig{file=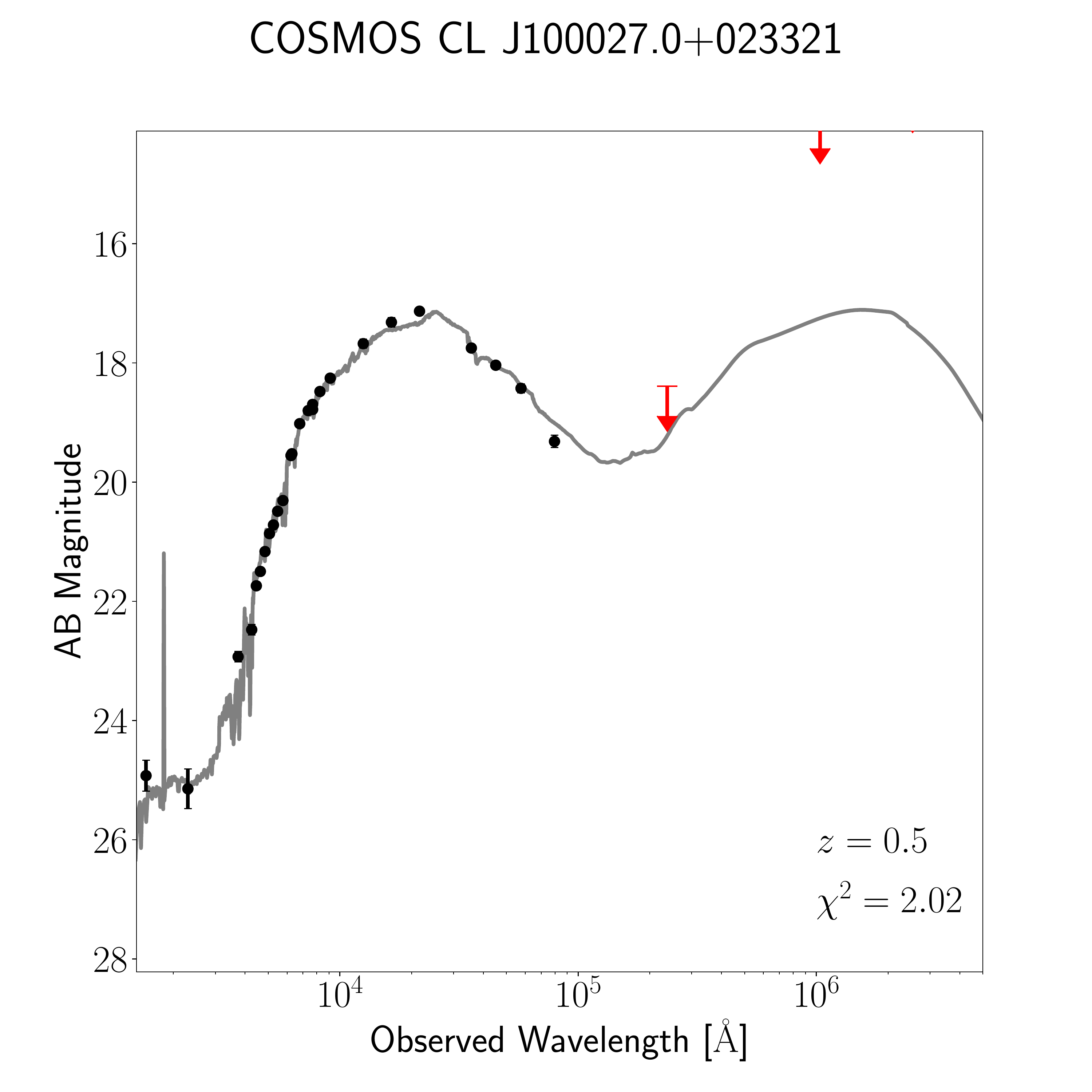,width=4cm,angle=0}
\epsfig{file=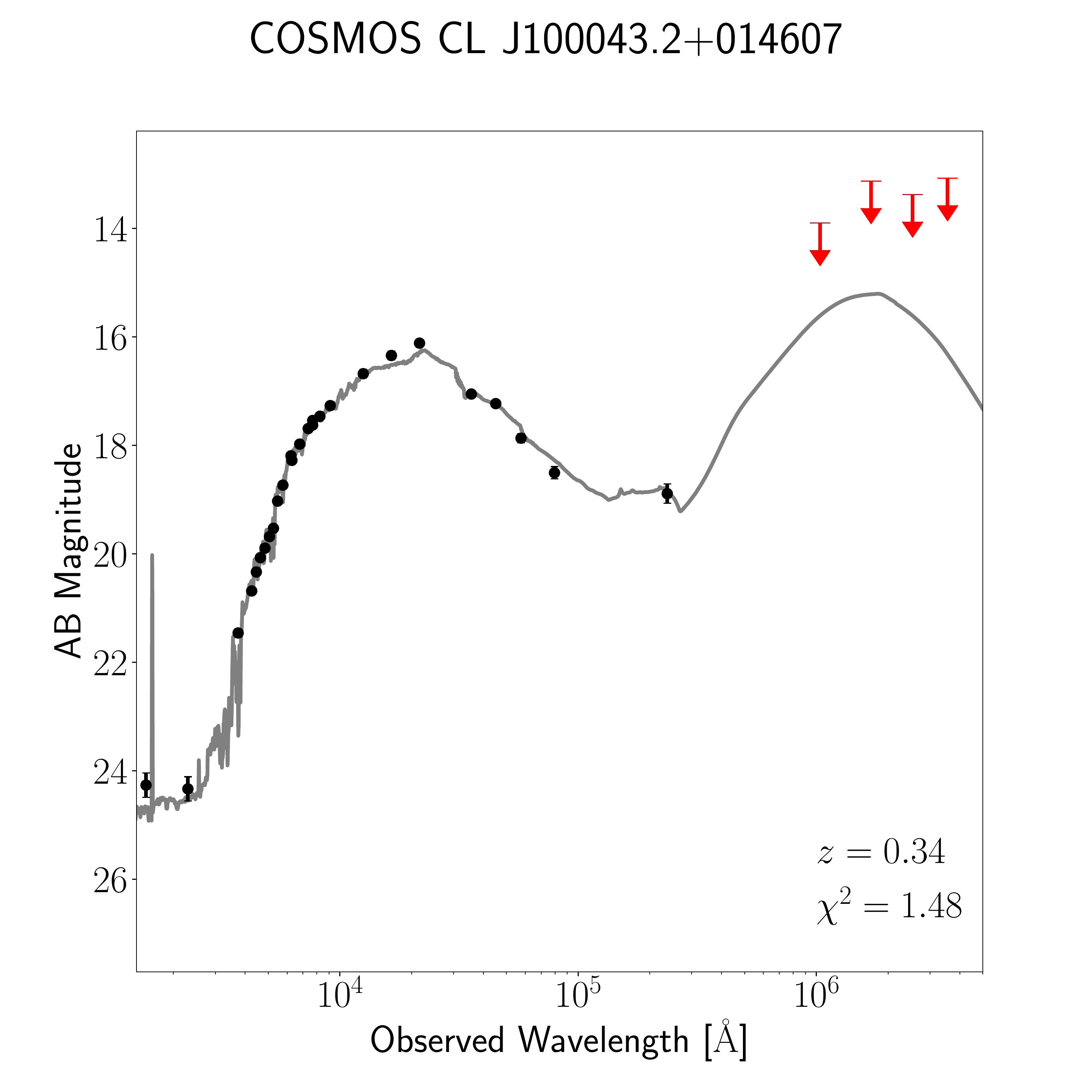,width=4cm,angle=0}
\epsfig{file=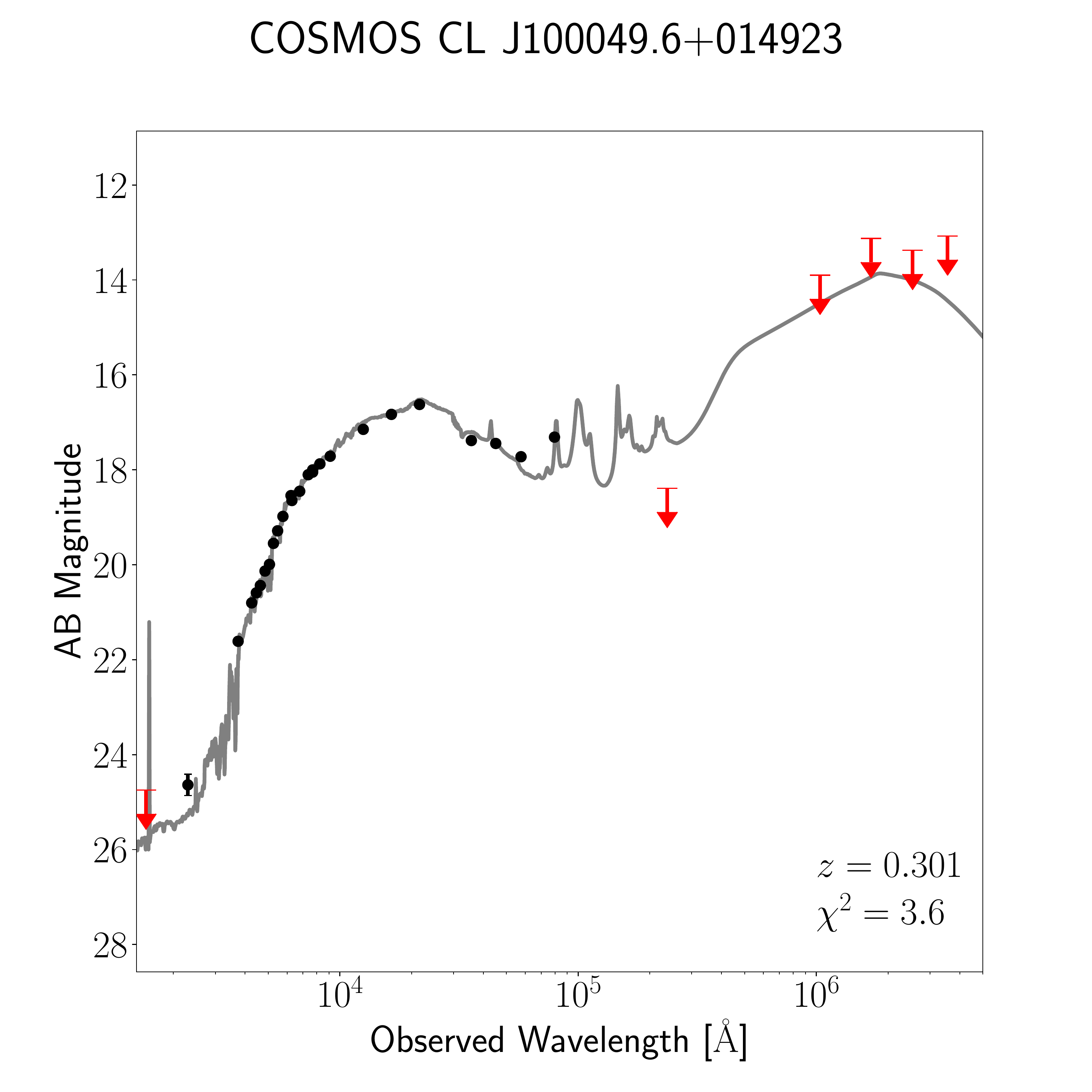,width=4cm,angle=0}
\epsfig{file=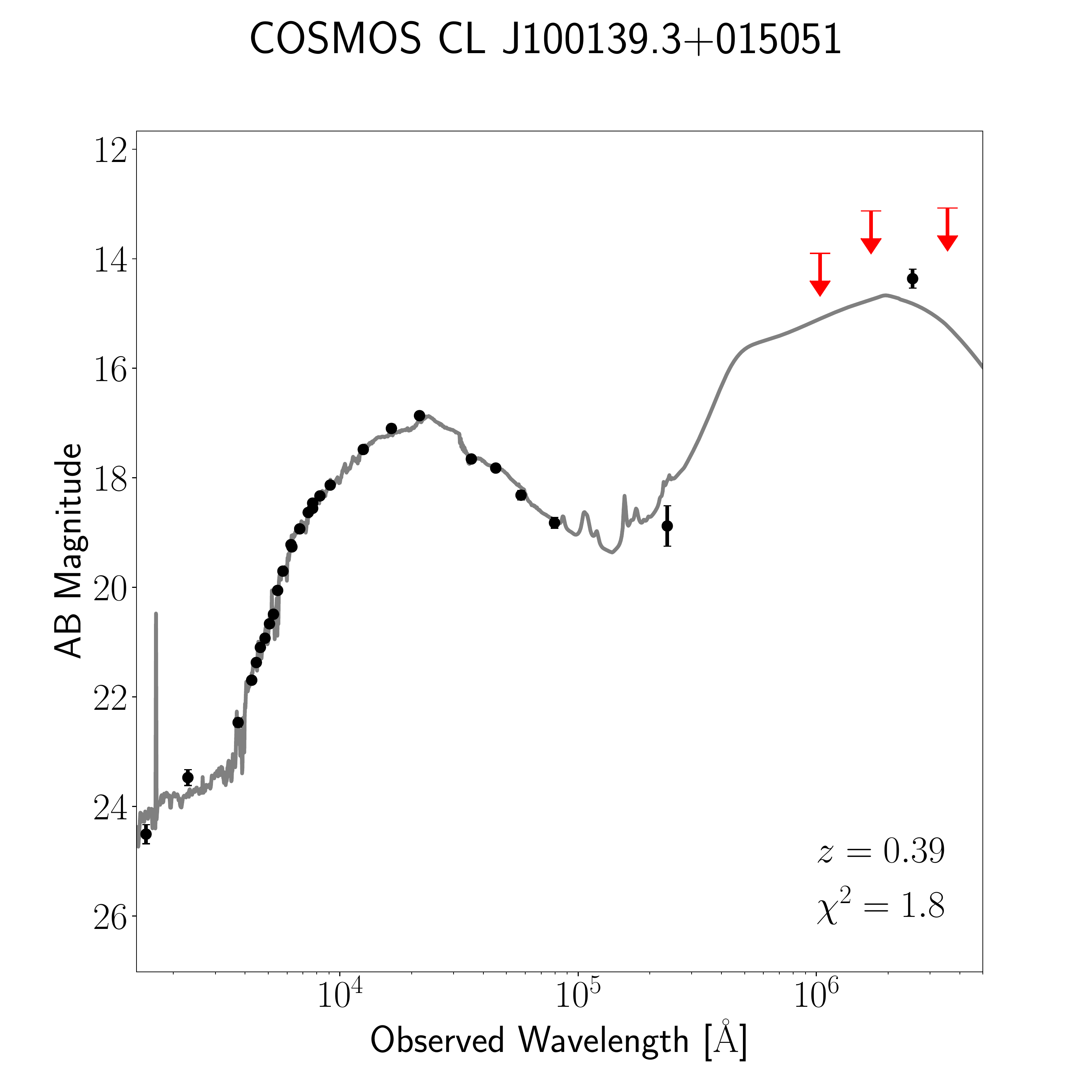,width=4cm,angle=0}
\epsfig{file=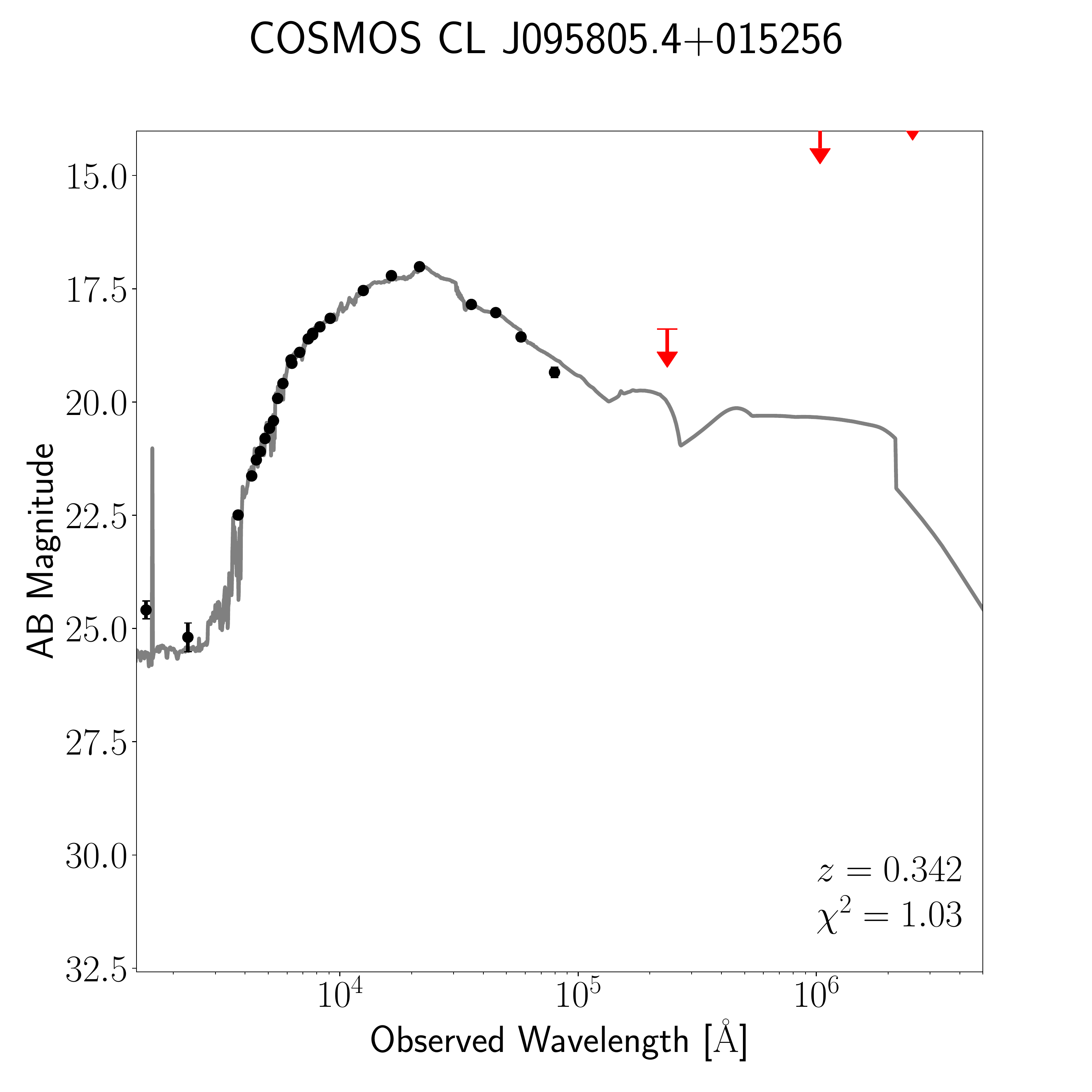,width=4cm,angle=0}
\epsfig{file=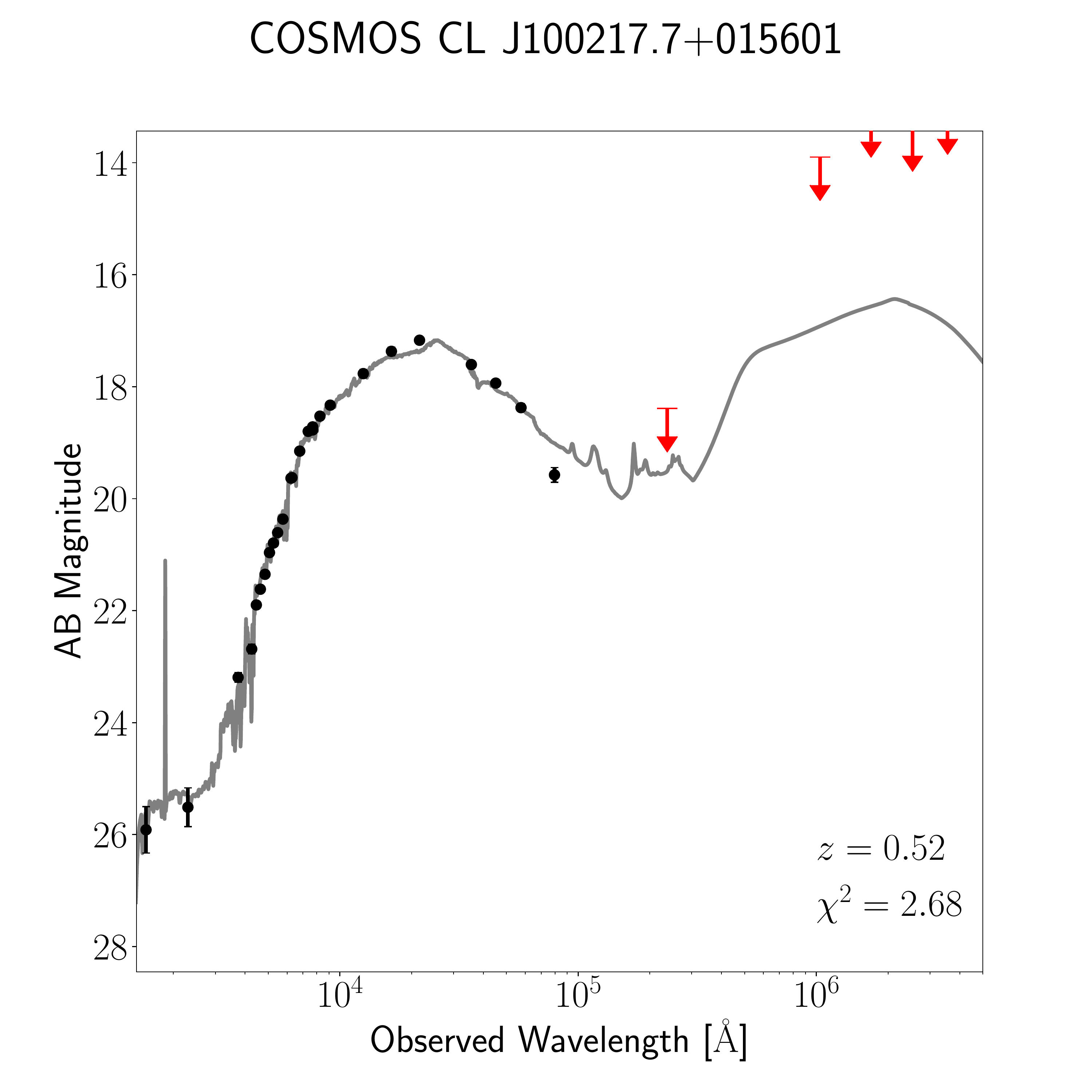,width=4cm,angle=0}
\epsfig{file=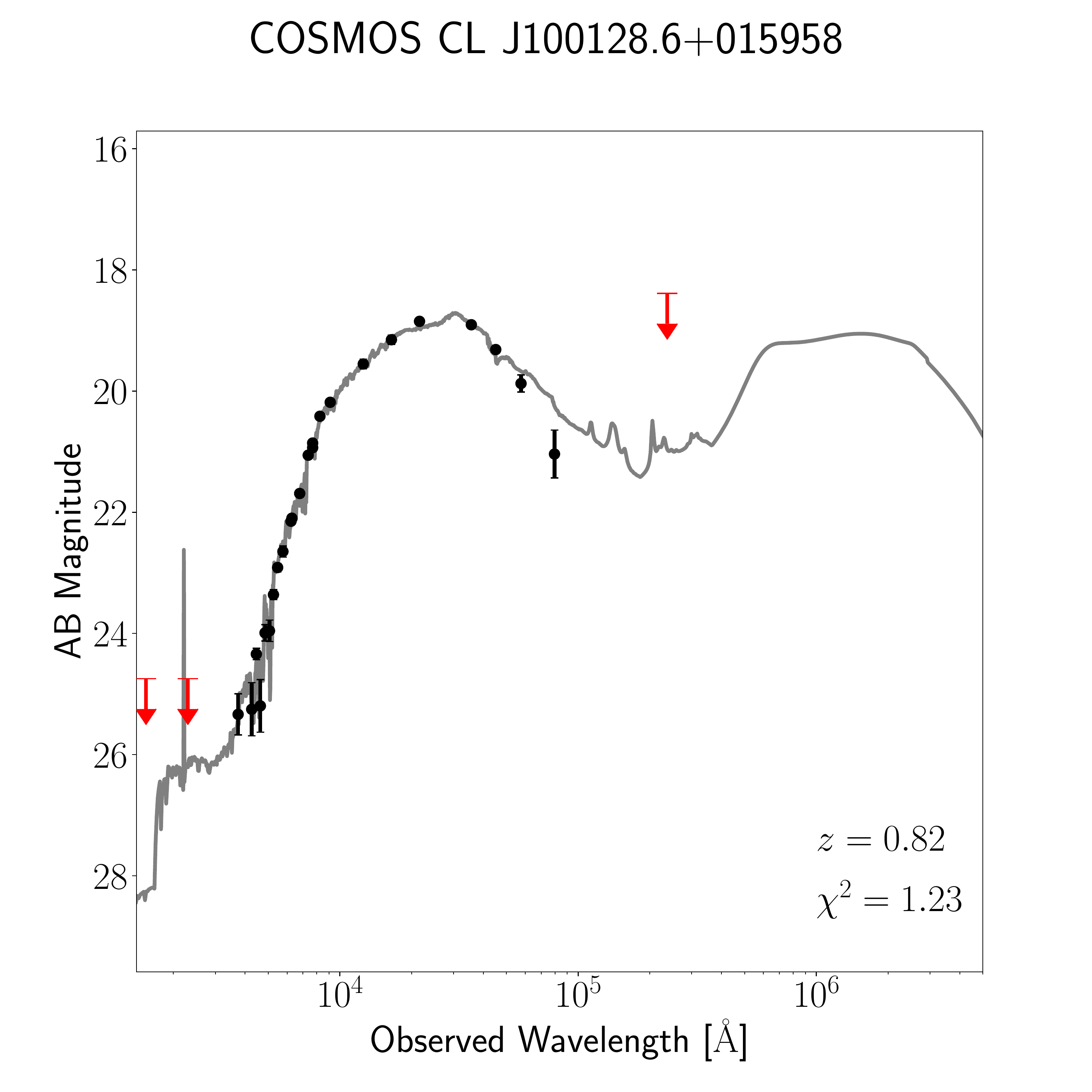,width=4cm,angle=0}
\epsfig{file=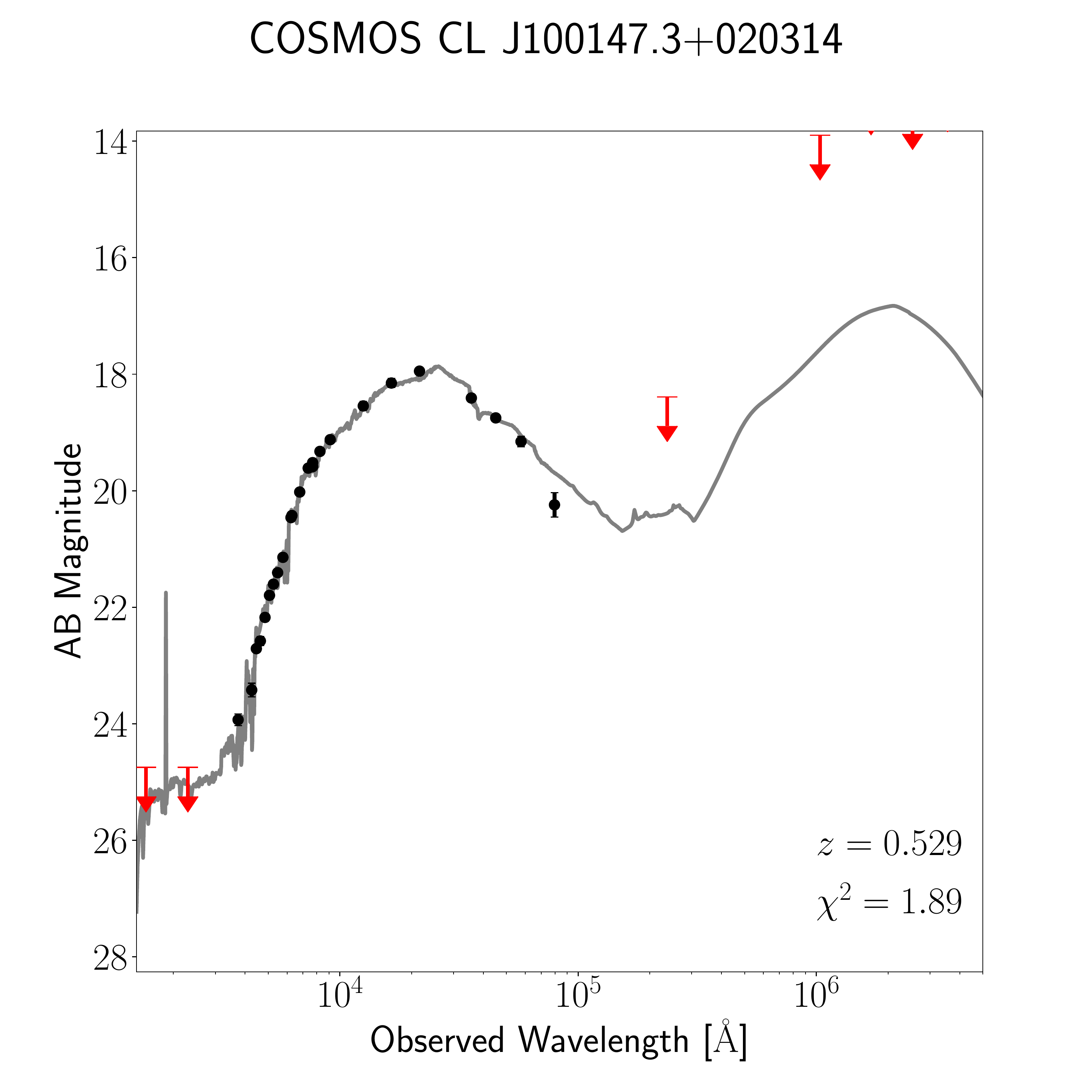,width=4cm,angle=0}
\epsfig{file=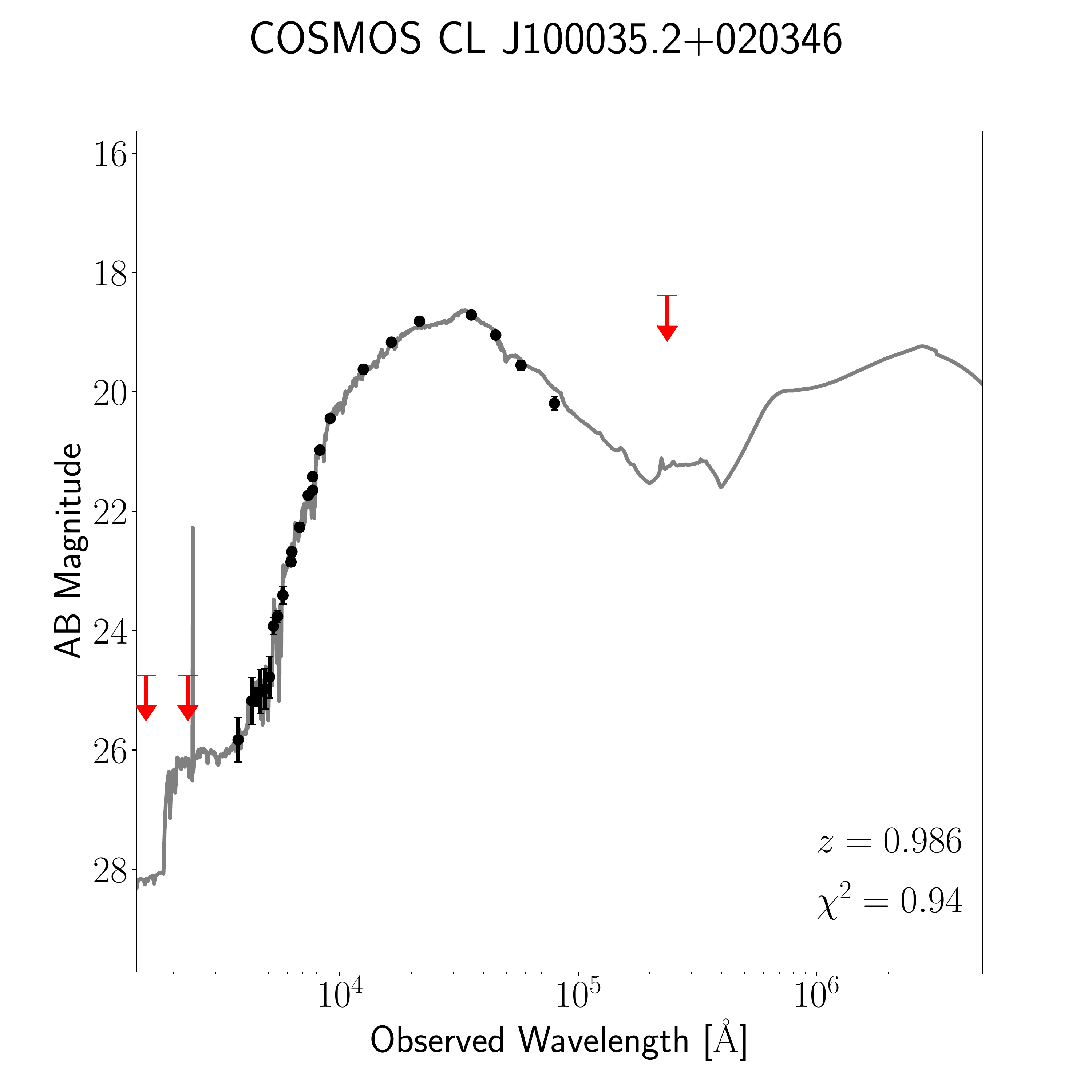,width=4cm,angle=0}
\epsfig{file=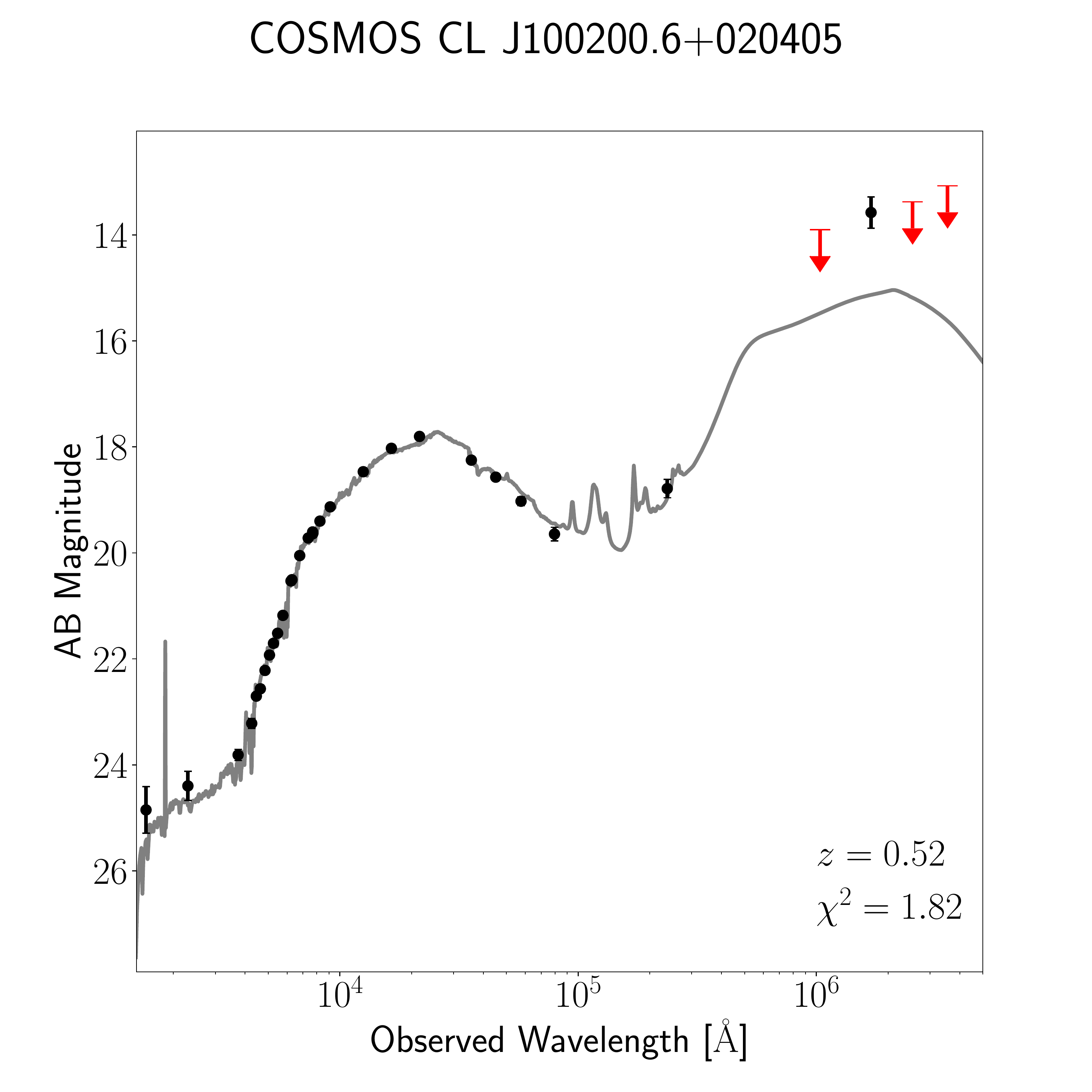,width=4cm,angle=0}
\epsfig{file=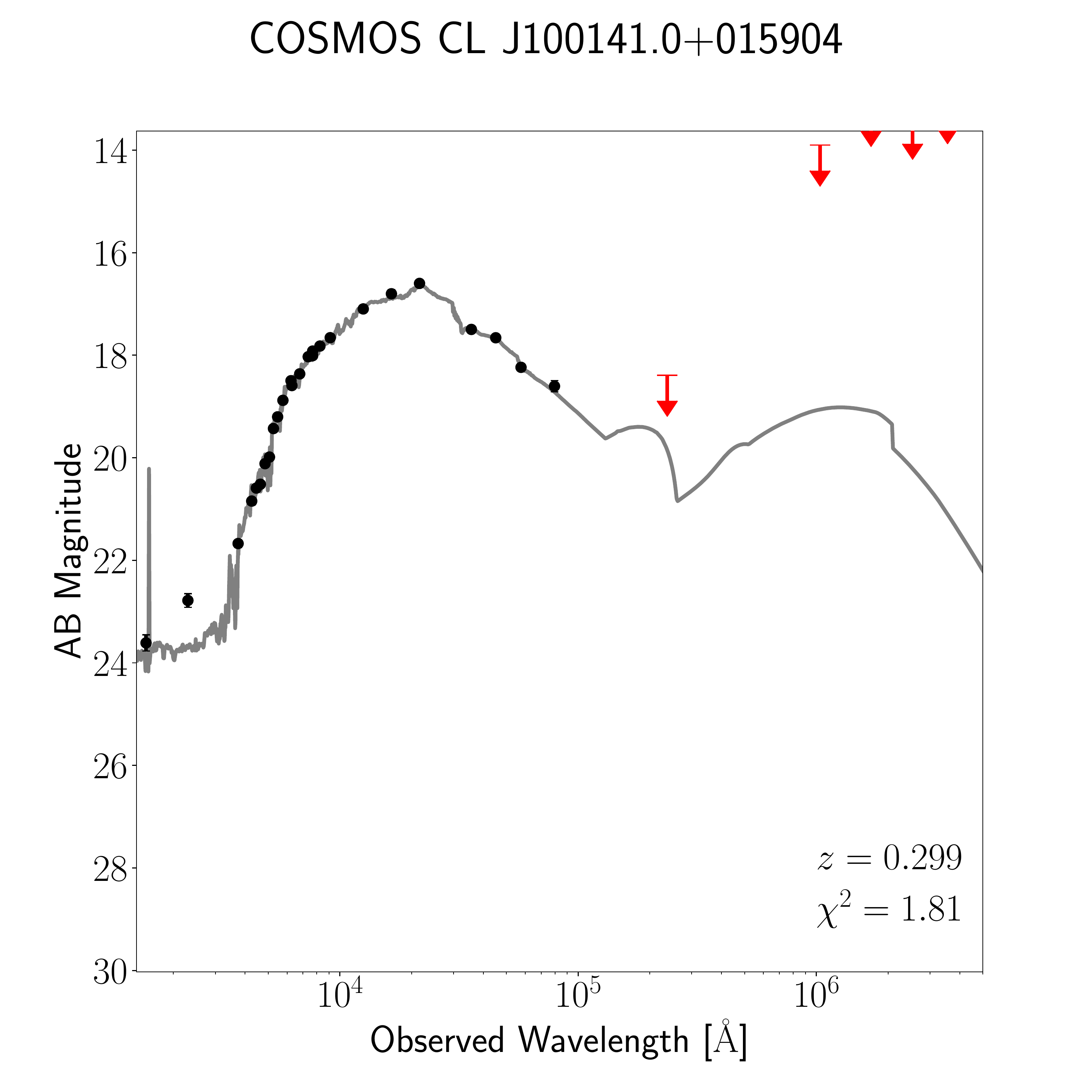,width=4cm,angle=0}
\caption{\textit{Continued}}
\end{figure*}

\renewcommand{\thefigure}{A\arabic{figure}}

\setcounter{figure}{0}

\begin{figure*}
\epsfig{file=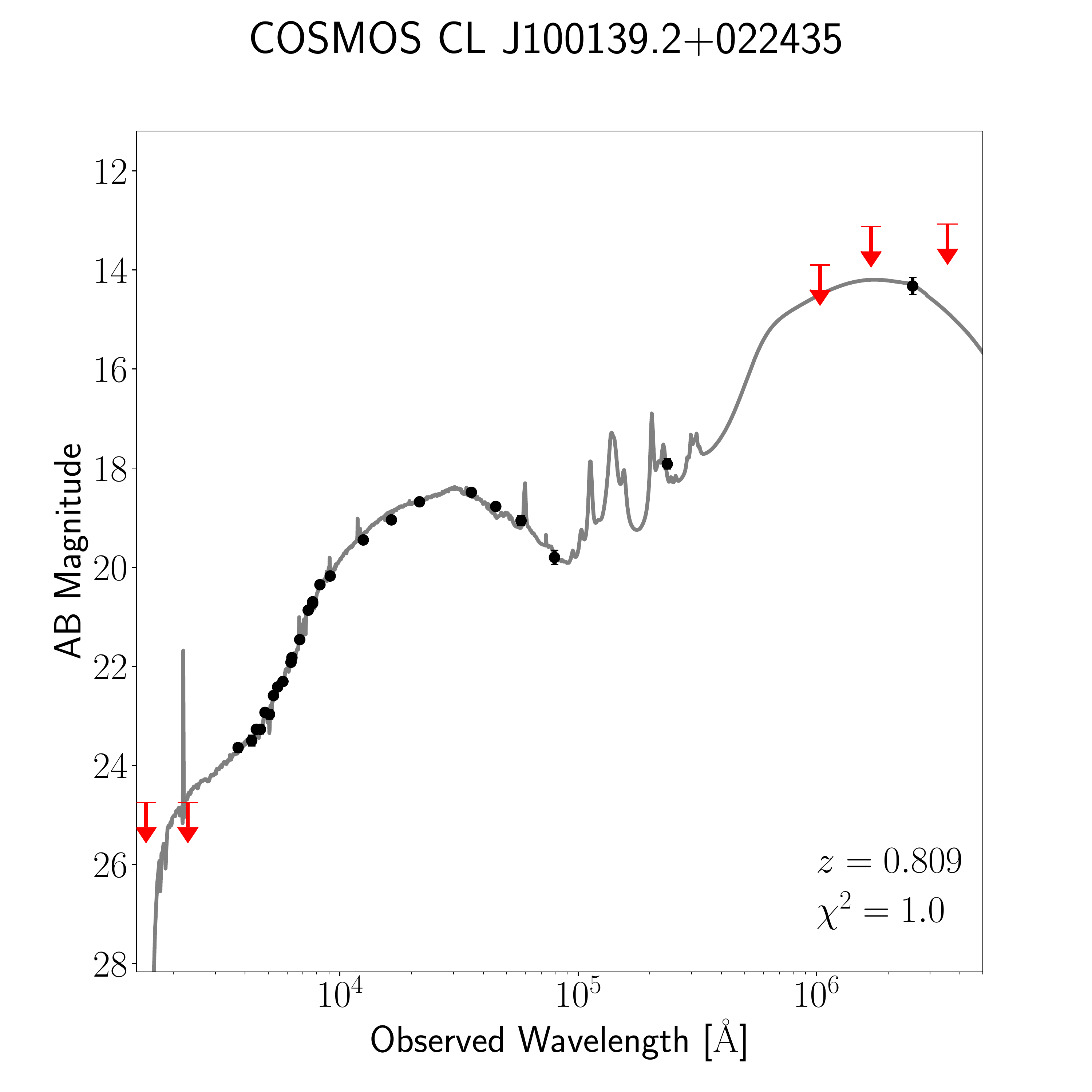,width=4cm,angle=0}
\epsfig{file=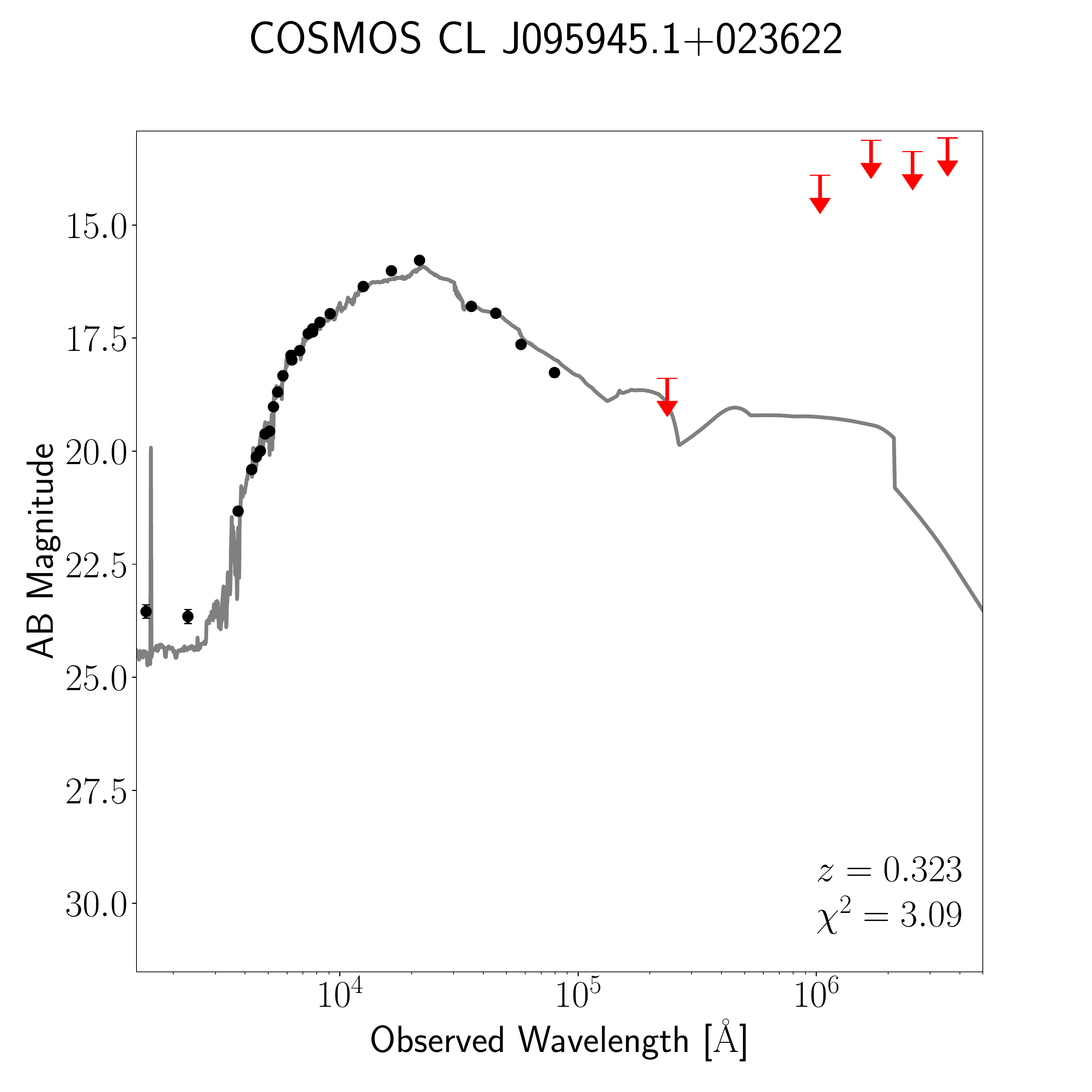,width=4cm,angle=0}
\epsfig{file=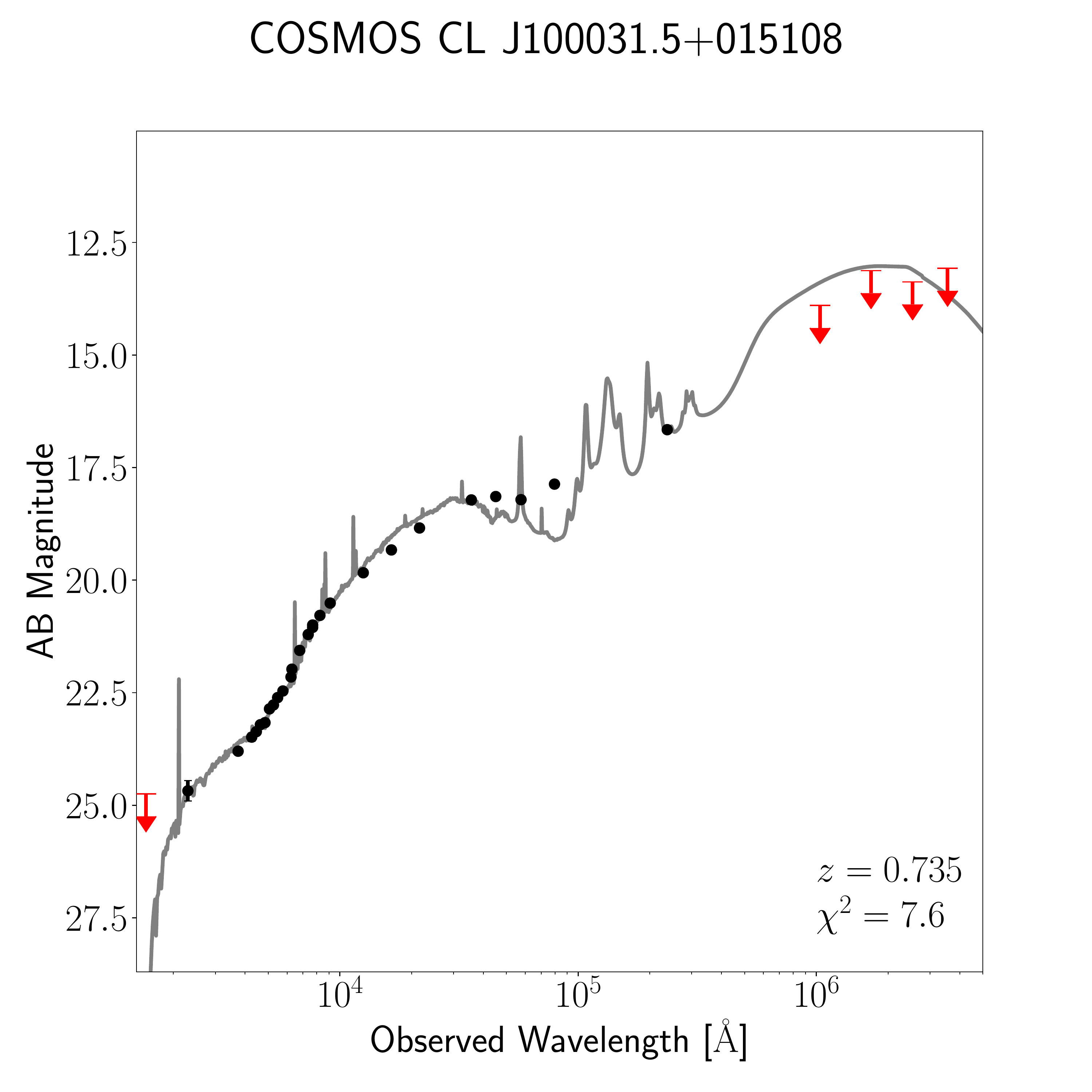,width=4cm,angle=0}
\epsfig{file=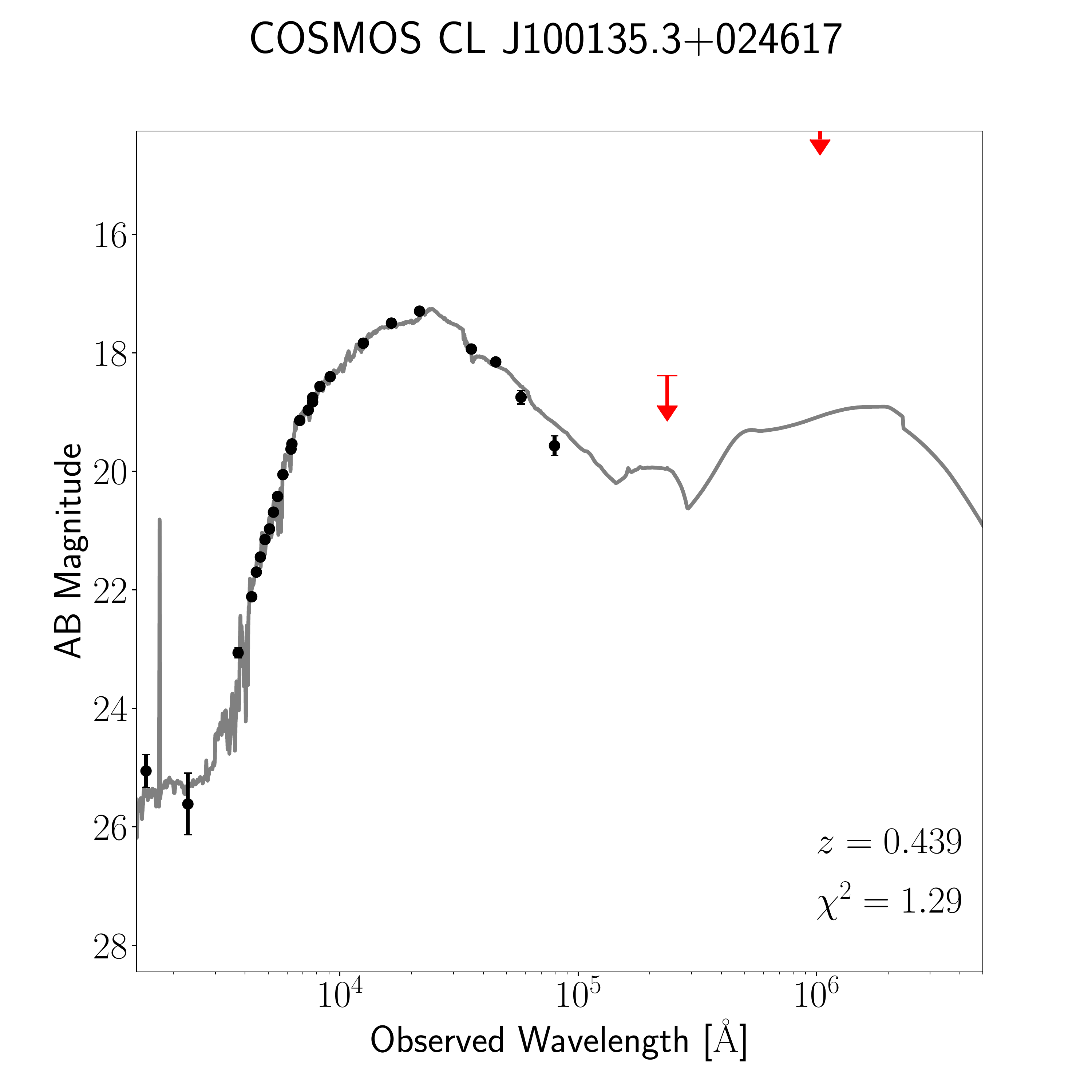,width=4cm,angle=0}
\caption{\textit{Continued}}
\end{figure*}
\end{document}